\documentclass[reprint,superscriptaddress,amssymb,amsmath,aps,prl,longbibliography]{revtex4-1}

\usepackage{graphicx}
\usepackage{dcolumn}
\usepackage{bm}
\usepackage{siunitx}
\usepackage{booktabs}
\usepackage[usenames,dvipsnames]{xcolor}
\usepackage{tcolorbox}
\usepackage{tabularx}
\usepackage{array}
\usepackage{colortbl}
\usepackage{braket}
\usepackage{gensymb}

\usepackage{pdfpages}
\makeatletter
\patchcmd{\@outputpage@head}{\@ifx{\LS@rot\@undefined}{}{\LS@rot}}{}{}{}
\makeatother

\tcbuselibrary{skins}

\newcommand{\units}[1]{\,\mathrm{#1}}
\tcbset{tab2/.style={colback=gray!5!white,colframe=black!50!black,colbacktitle=Gray!40!white,
coltitle=black,center title}}

\begin{document}

\title{Efficient and Low-Backaction Quantum Measurement Using a Chip-Scale Detector}

\author{Eric I. Rosenthal}
\thanks{eric.rosenthal@colorado.edu}
\affiliation{JILA, University of Colorado, Boulder, Colorado 80309, USA}
\affiliation{Department of Physics, University of Colorado, Boulder, Colorado 80309, USA}
\affiliation{National Institute of Standards and Technology, Boulder, Colorado 80305, USA}

\author{Christian M. F. Schneider}
\affiliation{Institute for Quantum Optics and Quantum Information of the Austrian Academy of Sciences, A-6020 Innsbruck, Austria}
\affiliation{Institute for Experimental Physics, University of Innsbruck, A-6020 Innsbruck, Austria}

\author{Maxime Malnou}
\affiliation{Department of Physics, University of Colorado, Boulder, Colorado 80309, USA}
\affiliation{National Institute of Standards and Technology, Boulder, Colorado 80305, USA}

\author{Ziyi Zhao}
\affiliation{JILA, University of Colorado, Boulder, Colorado 80309, USA}
\affiliation{Department of Physics, University of Colorado, Boulder, Colorado 80309, USA}
\affiliation{National Institute of Standards and Technology, Boulder, Colorado 80305, USA}

\author{Felix Leditzky}
\affiliation{Department of Mathematics \& Illinois Quantum Information Science and Technology Center, University of Illinois at Urbana-Champaign, Urbana, IL 61801, USA}
\affiliation{JILA, University of Colorado, Boulder, Colorado 80309, USA}
\affiliation{National Institute of Standards and Technology, Boulder, Colorado 80305, USA}
\affiliation{Center for Theory of Quantum Matter, University of Colorado, Boulder, Colorado 80309, USA}

\author{Benjamin J. Chapman}
\affiliation{Department of Applied Physics, Yale University, New Haven, Connecticut, 06511, USA.}

\author{Waltraut Wustmann}
\affiliation{The Laboratory for Physical Sciences, College Park, Maryland 20740, USA}

\author{Xizheng Ma}
\affiliation{JILA, University of Colorado, Boulder, Colorado 80309, USA}
\affiliation{Department of Physics, University of Colorado, Boulder, Colorado 80309, USA}
\affiliation{National Institute of Standards and Technology, Boulder, Colorado 80305, USA}

\author{Daniel A. Palken}
\affiliation{JILA, University of Colorado, Boulder, Colorado 80309, USA}
\affiliation{Department of Physics, University of Colorado, Boulder, Colorado 80309, USA}
\affiliation{National Institute of Standards and Technology, Boulder, Colorado 80305, USA}

\author{Maximilian F. Zanner}
\affiliation{Institute for Quantum Optics and Quantum Information of the Austrian Academy of Sciences, A-6020 Innsbruck, Austria}
\affiliation{Institute for Experimental Physics, University of Innsbruck, A-6020 Innsbruck, Austria}

\author{Leila R. Vale}
\affiliation{National Institute of Standards and Technology, Boulder, Colorado 80305, USA}

\author{Gene C. Hilton}
\affiliation{National Institute of Standards and Technology, Boulder, Colorado 80305, USA}

\author{Jiansong Gao}
\affiliation{Department of Physics, University of Colorado, Boulder, Colorado 80309, USA}
\affiliation{National Institute of Standards and Technology, Boulder, Colorado 80305, USA}

\author{Graeme Smith}
\affiliation{JILA, University of Colorado, Boulder, Colorado 80309, USA}
\affiliation{Department of Physics, University of Colorado, Boulder, Colorado 80309, USA}
\affiliation{National Institute of Standards and Technology, Boulder, Colorado 80305, USA}
\affiliation{Center for Theory of Quantum Matter, University of Colorado, Boulder, Colorado 80309, USA}

\author{Gerhard Kirchmair}
\affiliation{Institute for Quantum Optics and Quantum Information of the Austrian Academy of Sciences, A-6020 Innsbruck, Austria}
\affiliation{Institute for Experimental Physics, University of Innsbruck, A-6020 Innsbruck, Austria}

\author{K. W. Lehnert}
\affiliation{JILA, University of Colorado, Boulder, Colorado 80309, USA}
\affiliation{Department of Physics, University of Colorado, Boulder, Colorado 80309, USA}
\affiliation{National Institute of Standards and Technology, Boulder, Colorado 80305, USA}

\date{\today}

\begin{abstract}
    Superconducting qubits are a leading platform for scalable quantum computing and quantum error correction. One feature of this platform is the ability to perform projective measurements orders of magnitude more quickly than qubit decoherence times. Such measurements are enabled by the use of quantum-limited parametric amplifiers in conjunction with ferrite circulators --- magnetic devices which provide isolation from noise and decoherence due to amplifier backaction. Because these nonreciprocal elements have limited performance and are not easily integrated on chip, it has been a long-standing goal to replace them with a scalable alternative. Here, we demonstrate a solution to this problem by using a superconducting switch to control the coupling between a qubit and amplifier. Doing so, we measure a transmon qubit using a single, chip-scale device to provide both parametric amplification and isolation from the bulk of amplifier backaction. This measurement is also fast, high fidelity, and has 70\% efficiency, comparable to the best that has been reported in any superconducting qubit measurement. As such, this work constitutes a high-quality platform for the scalable measurement of superconducting qubits.
\end{abstract}

\maketitle

Qubit-specific projective measurement is a requirement for scalable quantum computation and quantum error correction \cite{divincenzo:2000}. In superconducting systems, qubit measurement generally involves scattering a microwave pulse off of a readout cavity dispersively coupled to the qubit \cite{blais:2004,blais:2020b}. This pulse is routed through ferrite circulators and/or isolators to a Josephson-junction-based parametric amplifier \cite{castellanos:2007,yamamoto:2008,bergeal:2010,macklin:2015,frattini:2017}, sent to room temperature, and digitized. This readout scheme can work well \cite{jeffrey:2014}: it is low backaction, quantum nondemolition, and can have infidelity of $10^{-2}$ in less than 100 ns \cite{walter:2017}, with the best reported infidelity of less than $10^{-4}$ \cite{elder:2020}. 

Challenges arise, however, as the scale and requirements of superconducting quantum systems increase. In particular, ferrite circulators are bulky and their requisite number scales linearly with the number of measurement channels. Fitting enough circulators at the base temperature stage of a cryostat is one eventual bottleneck associated with building a scalable quantum computer. Furthermore, circulators are both lossy and provide finite isolation from amplifier noise. Isolation can be improved using multiple isolators in series, but at the cost of increased resistive loss and impedance mismatches, which necessitate a stronger readout pulse in order to make a projective qubit measurement. This can be just as detrimental as amplifier backaction; both have the potential to drive higher-level state transitions which can cause readout errors, and reduce the extent to which a measurement is quantum nondemolition \cite{slichter:2012,sank:2016}.

In recognition of these problems, it has been a long-standing goal to replace ferrite circulators and isolators with a chip-scale, higher-performance alternative. Efforts to do so have often involved parametrically coupling high-$Q$ resonant modes \cite{abdo:2014,ranzani:2015,sliwa:2015,lecocq:2017,abdo:2019,lecocq:2020,abdo:2020,lecocq:2020b} or concatenating frequency conversion and delay operations \cite{kerckhoff:2015,rosenthal:2017,chapman:2017b,chapman:2019}. Such technologies show promise but have yet to supplant ferrites. Performance specifications such as isolation and bandwidth must still be improved, and multiple high-frequency control tones per device are undesirable from the perspective of scalability. An alternate approach is to simply remove any nonreciprocal components between the qubit and first, Josephson-junction-based amplifier \cite{opremcak:2018,eddins:2019,opremcak:2020}. This allows for high efficiency but at the cost of significant exposure to amplifier backaction. 

\begin{figure}[t] 
\begin{center}
\includegraphics[width=1.0\columnwidth]{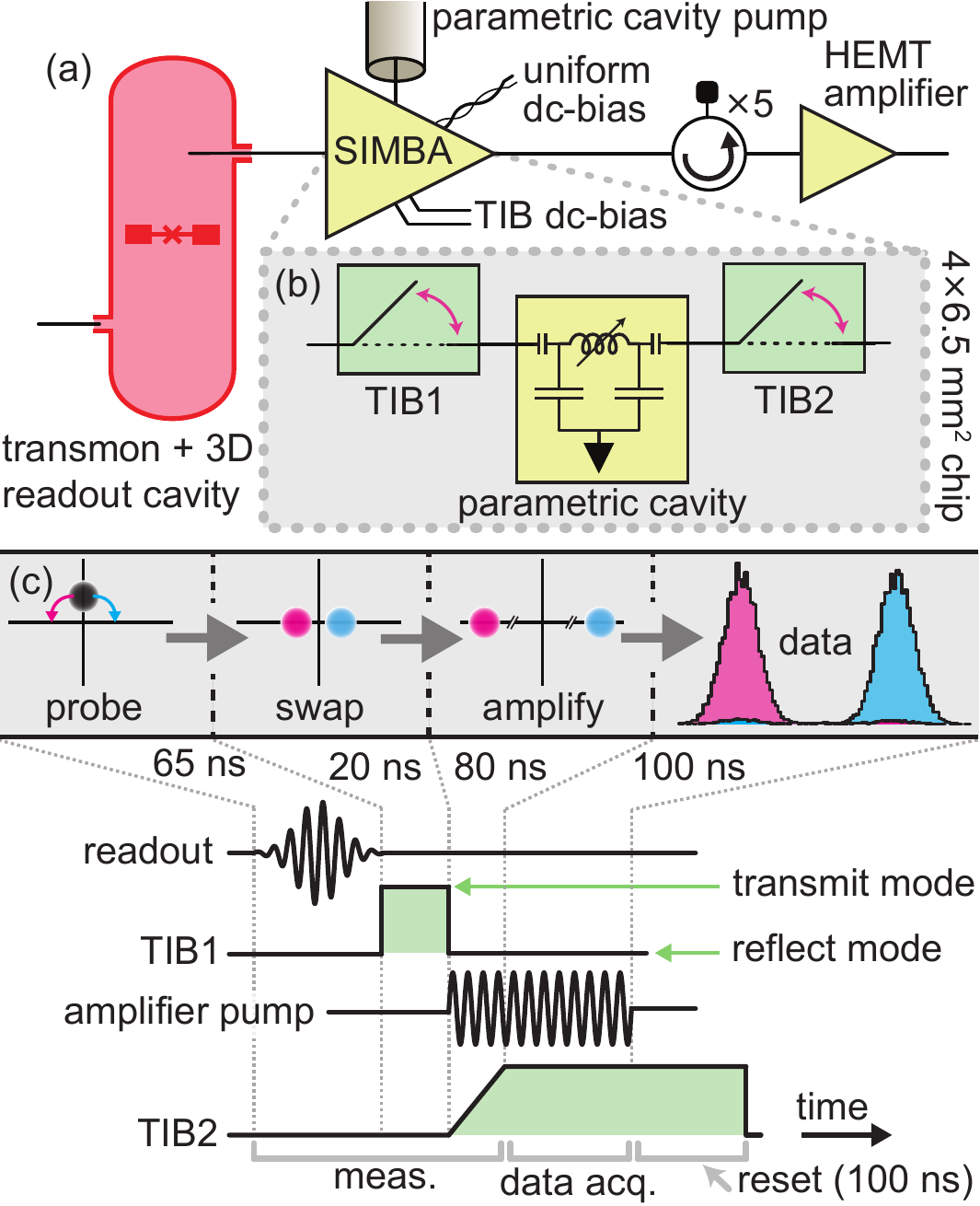}
\caption {Procedure. (a) A transmon qubit is measured using a `superconducting isolating modular bifurcation amplifier' (SIMBA). (b) The SIMBA is composed of a two-port parametric cavity with a tunable-inductor-bridge (TIB) style coupler on each port. (c) To measure the qubit, a probe tone is sent into the readout cavity, swapped into the parametric cavity, and then amplified. The amplified state is then coupled to a standard cryogenic measurement chain and digitized. Cyan (pink) histograms correspond to single-shot measurements when the qubit has been prepared in the ground (excited) state.
}
\label{fig:Intro}
\end{center}
\end{figure}

Here, we instead engineer a replacement for ferrites based on the coordinated operation of superconducting switches. These switches, realized by an improvement upon the design in Refs.~\cite{chapman:2016,chapman:2017}, are integrated into a single, chip-scale device we call a `superconducting isolating modular bifurcation amplifier' (SIMBA), Fig.~\ref{fig:Intro}. 
The SIMBA consists of a two-port parametric cavity (a Josephson parametric amplifier, based on the devices in Refs.~\cite{castellanos:2007,yamamoto:2008,malnou:2019}) with fast, low-loss and high on-off ratio superconducting switches placed on both ports. Importantly, these switches are dc actuated, requiring no microwave control tones. Pulsed, unidirectional gain is realized by the sequential operation of these switches combined with resonant delay, and parametric gain, in the parametric cavity. We use this procedure to demonstrate efficient, high-quality readout of a superconducting qubit while simultaneously isolating it from the bulk of amplifier backaction. We emphasize that this \textit{procedure} is the novel idea in this work, which in the future may be implemented using a wide class of devices.

Central to the SIMBA is a flux-pumped parametric cavity: a lumped-element inductor-capacitor circuit where approximately half the inductance comes from an array of superconducting quantum interference devices (SQUIDs). The parametric cavity resonant frequency can be tuned between 4 and 7.1 GHz by applying an external magnetic flux (see Supplementary Material, Sec~III.D). When flux through these SQUIDs is modulated at twice the cavity resonance frequency, the cavity state undergoes phase-sensitive parametric amplification via three-wave mixing. 

The external coupling of the parametric cavity is controlled by superconducting switches constructed using a `tunable inductor bridge' (TIB) \cite{naaman:2016switch,chapman:2016}. TIB transmission is tuned by a dc signal which changes the balance of a Wheatstone bridge of SQUID arrays. In this experiment, the speed at which transmission can be tuned is limited by off-chip, low-pass filters with a 350-MHz cutoff frequency placed on the TIB bias lines. Tested in isolation, the TIB has an on/off ratio greater than 50 dB tunable between 4 and 7.3 GHz (see Supplementary Material Sec.~III, which includes Refs.~\cite{marchand:1944,mates:2008,reed:2010,malnou:2018}). This overlaps with the range over which the parametric cavity can be tuned, allowing the SIMBA itself to be tuned to operate over several GHz. The TIB 1-dB compression point is approximately $-98$ dBm, which crucially allows the TIB to function effectively while the state in the parametric cavity is amplified. 

\begin{figure}[t] 
\begin{center}
\includegraphics[width=1.0\columnwidth]{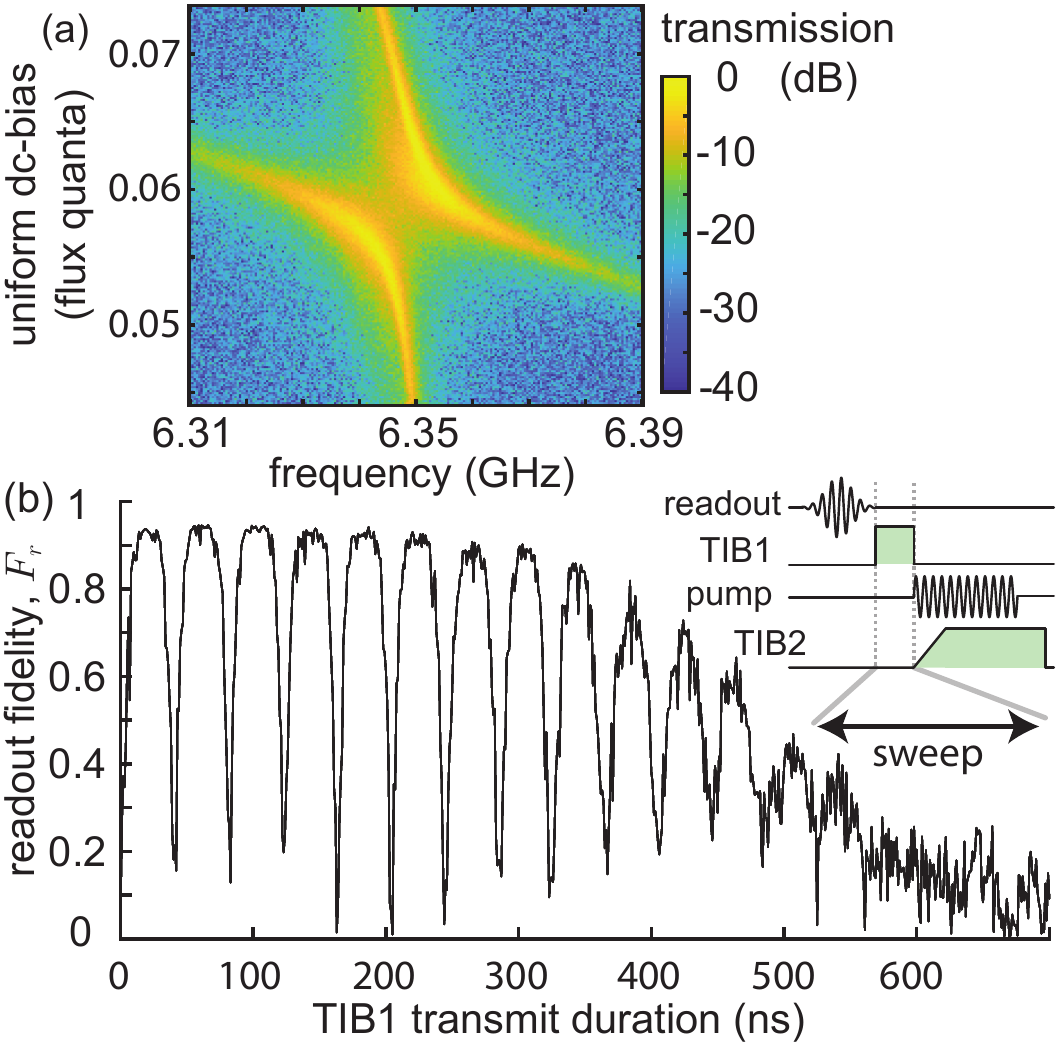}
\caption {Calibration. (a) A uniform external flux is swept while probing the readout cavity in transmission with both TIBs in transmit mode. The avoided crossing shows the parametric cavity tuning through the readout cavity. To operate a SIMBA, this uniform flux bias is set so that the readout and parametric cavities are minimally detuned. (b) Readout fidelity $F_r$ (the ability of a measurement to distinguish the qubit eigenstate \cite{gambetta:2007}) is plotted versus the duration for which TIB1 is set to transmit mode within the measurement sequence. Oscillations with a period of 40 ns indicate coherent swapping of a readout pulse between the readout and parametric cavities.
}
\label{fig:TuneUp}
\end{center}
\end{figure}

We use the SIMBA to measure a transmon qubit dispersively coupled to a readout cavity. As in conventional dispersive readout \cite{blais:2004,blais:2020b}, a pulse is first sent into the weakly coupled port of a two-port readout cavity, where it acquires a qubit-state-dependent phase shift. TIB1 is then set to transmit mode for a duration (20 ns), chosen to fully swap this pulse into the parametric cavity, which has previously been tuned near resonance, Fig.~\ref{fig:TuneUp}. We then strongly flux pump the parametric cavity into the bistable regime \cite{wustmann:2013,krantz:2013,wustmann:2019}: a nonunitary process in which the cavity latches into one of two bistable states with opposite phase but large, equal amplitudes (see Supplementary Material Sec.~IV, which includes Refs.~\cite{vanduzer:1981,siddiqi:2004,manucharyan:2007}). Readout is achieved by seeding the parametric cavity state with the probe tone, such that the postmeasurement qubit state is correlated with the latched state of the parametric cavity \cite{lin:2014,krantz:2016}. We choose to thus discretize and store the measurement result within the cryostat as a step toward implementing rapid and hardware efficient feed-forward protocols \cite{andersen:2016}. To learn the measurement result outside of the cryostat, TIB2 is set to transmit mode, coupling this state to a standard cryogenic microwave measurement chain. 

We focus on three figures of merit to describe the success of this readout: excess backaction $n_b$, measurement efficiency $\eta$, and maximum readout fidelity $F_0$. To characterize these quantities we use the framework of measurement-induced dephasing \cite{clerk:2010} (see Supplementary Material Sec.~II, which includes Refs.~\cite{kraus:1971,loudon:2000,gerry:2004,helstrom:1976,caves:1982,loudon:2000,clerk:2003,gerry:2004,korotkov:2016,han:2018}). 

\begin{figure*}[t] 
\begin{center}
\includegraphics[width=2.0\columnwidth]{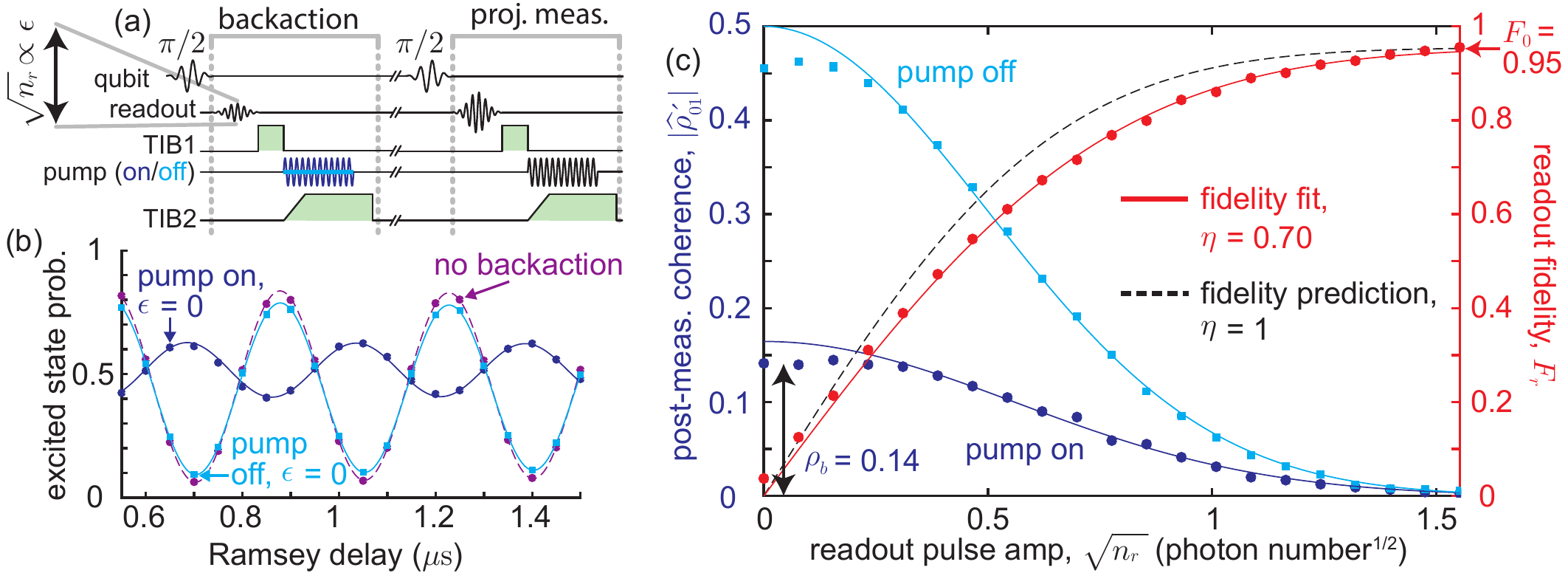}
\caption {Characterization. 
(a) Postmeasurement qubit coherence $|\hat{\rho}_{01}'|$ is obtained by inserting a variable measurement into a Ramsey sequence, exposing the qubit to backaction. 
The ratio of the amplitude of the measured Ramsey fringes to the amplitude of those measured without this backaction (nothing inserted into the Ramsey sequence) equals $2|\hat{\rho}_{01}'|$.
(b) Excess backaction is determined by inserting a ``measurement'' with zero readout amplitude. Postmeasurement coherence after excess backaction with the parametric cavity pump on (indigo) and off (cyan), are compared to a case with no backaction (no readout pulse, pump, or TIB switching inserted in the Ramsey sequence, violet). (c) Postmeasurement coherence $|\hat{\rho}_{01}'|$ (left $y$ axis) and readout fidelity $F_r$ (right $y$ axis, red data points) are measured while sweeping the readout amplitude $\sqrt{n_r}$ of a variable strength measurement. As in (b), $|\hat{\rho}_{01}'|$ is measured both with the parametric pump turned on or off during the variable measurement sequence (indigo or cyan data points, respectively). Postmeasurement coherence with the parametric pump turned on, but in the absence of readout photons, is specified by $\rho_b = |\hat{\rho}_{01}'\left(\sqrt{n_r}=0\right)|$ and determines the excess backaction $n_b = -\log\left(2 \rho_b\right)/2$. Measurement efficiency $\eta$ is determined by a comparison between measurement-induced dephasing and readout fidelity while sweeping readout amplitude.
}
\label{fig:Efficiency}
\end{center}
\end{figure*}

Ideally, measurement-induced dephasing of the qubit comes \textit{only} from a readout pulse. Consider a qubit prepared in a superposition state $\left(\ket{0} + \ket{1}\right)/\sqrt{2}$; a readout pulse at the appropriate frequency interacts with this qubit to create the entangled state $\left(\ket{0} \ket{\alpha_0} + \ket{1} \ket{\alpha_1} \right)/\sqrt{2}$.
Here $\ket{\alpha_0}$ and $\ket{\alpha_1}$ are coherent states both of amplitude $|\alpha|$, 
separated in phase space by the angle $2\theta = 2\arctan\left(2\chi/\kappa_r\right)$, 
where the readout cavity frequency shifts by $\pm\chi/2\pi$ dependent on the qubit state, and $\kappa_r/2\pi$ is the loss rate of the readout cavity \cite{blais:2004,blais:2020b}. After measurement, the off-diagonal element of the qubit density matrix becomes $|\rho_{01}'| = \frac{1}{2}\left|\braket{\alpha_0|\alpha_1}\right| = \frac{1}{2}e^{-2n_{r}}$, 
where $n_{r} = \left(|\alpha|\sin\theta\right)^2$ is the effective photon number of the readout pulse, corresponding to the square of half the separation in phase space between $\ket{\alpha_0}$ and $\ket{\alpha_1}$ (see Supplementary Material Sec.~II.A). Here, $n_r$ is nearly equal to the readout pulse photon number $|\alpha|^2$ because $2\chi/2\pi = 1.93 \units{MHz}$ and $\kappa_r/2\pi = 440 \units{kHz}$, so that $n_r = 0.95|\alpha|^2$. 

In practice, measurement may include ``excess backaction'' or \textit{additional} dephasing. This is modeled as an additional pulse with an effective photon number,
\begin{equation}
    n_b = -\frac{1}{2} \log\left( 2\rho_b \right),
    \label{eqn:Backaction}
\end{equation}
such that the coherence of a superposition state is reduced to $|\hat{\rho}_{01}'| = \frac{1}{2}e^{-2\left(n_b + n_r\right)} = \rho_b e^{-2n_r}$, where $ 0 \leq \rho_b \leq 1/2$ is the postmeasurement coherence in the absence of readout photons. The effective photon number $n_r$ in a given readout pulse is not \textit{a priori} known, but is related to its amplitude expressed in experimental units, $\epsilon \propto \sqrt{n_r}$. The measurement-induced dephasing can therefore be expressed as
\begin{equation}
    |\hat{\rho}_{01}'|= \rho_b e^{-2 \left(\sqrt{n_r}\right)^2} = \rho_b e^{-\epsilon^2 / 2 \sigma^2},
    \label{eqn:MeasInducedDephasingVsAlpha}
\end{equation}
where $\sqrt{n_r} = \epsilon/2\sigma$ and, physically, the constant $\sigma$ calibrates the readout pulse amplitude in units of $\mathrm{\left(photon~number\right)^{1/2}}$.

A dephased qubit indicates that information about its energy eigenstate may be learned by a detector. This information may be quantified by a readout fidelity \cite{gambetta:2007},
\begin{equation}
    F_r = 1 - P(e|0) - P(g|\pi),
    \label{eqn:FidelityDef}
\end{equation}
where $P(e|0)$ and $P(g|\pi)$ are the probability of incorrect assignment when the qubit is prepared in the ground or excited state, respectively.
 
For dispersive readout using a thresholded measurement (see Supplementary Material Sec.~II.B), readout fidelity is
\begin{equation}
    F_r= F_0 \mathrm{erf}\left[\sqrt{2\eta n_r}\right] = F_0 \mathrm{erf}\left[\nu \epsilon\right].
    \label{eqn:FidelityVsAlpha}
\end{equation}
Here $F_0$ is the maximum readout fidelity, and $\eta = \eta_\mathrm{loss}\eta_\mathrm{amp}$ is the measurement efficiency \cite{clerk:2010}, defined here such that $1-\eta_\mathrm{loss}$ is the fraction of readout pulse energy which has been lost before the pulse undergoes parametric amplification, which is assumed to be noiseless such that $\eta_\mathrm{amp}=1$. The constant $\nu = \sqrt{2\eta n_r}/\epsilon$ characterizes how quickly $F_r$ increases with $\epsilon$.

The relationship between $\nu$ and $\sigma$ gives the convenient formula, \begin{equation}
    \eta = 2 \sigma^2 \nu^2.
    \label{eqn:Eta}
\end{equation}
Intuitively, measurement efficiency $\eta$ is determined by the readout fidelity of a weak measurement (quantified by $\nu$), compared to its backaction (quantified by $\sigma$) \cite{bultink:2018}. 

To experimentally determine the figures of merit $n_b$, $\eta$, and $F_0$, we measure readout fidelity and postmeasurement coherence, both as functions of the experimental readout amplitude $\epsilon$. 
Readout fidelity $F_r$ is simply computed by measuring $P(e|0)$ and $P(g|\pi)$, and using Eq.~\ref{eqn:FidelityDef}. To measure $|\hat{\rho}_{01}'|$, the qubit is prepared in a superposition state, exposed to backaction from a variable strength measurement with readout pulse amplitude $\epsilon \propto \sqrt{n_r}$, and then projectively measured after a variable Ramsey delay and a second $\pi/2$ pulse, Fig.~\ref{fig:Efficiency}a. 
We first characterize the backaction from a ``measurement'' of zero readout amplitude, $\epsilon = 0$, meaning backaction solely due to actuating the TIBs (leftmost point in the ``pump off'' data, cyan, Fig.~\ref{fig:Efficiency}c), and the combination of actuating the TIBs and pumping the parametric cavity (leftmost data point, ``pump on'' data, indigo).
We then repeat this sweep over the variable amplitude $\epsilon$, both with the parametric pump turned off (cyan) and on (indigo) during the variable measurement. For comparison, qubit coherence is also measured \textit{without} exposure to any backaction, meaning no variable measurement inserted into the Ramsey delay (e.g., violet data, Fig.~\ref{fig:Efficiency}b). The ratio of the Ramsey fringe amplitudes with or without exposure to backaction gives $2|\hat{\rho}_{01}'|$, with the ratio taken to correct for readout infidelity. 

This characterization determines that our readout is low backaction, high fidelity, and high efficiency.
Excess backaction is found from $\rho_b = 0.141 \pm 0.002$ (leftmost data point, pump on data, Fig.~\ref{fig:Efficiency}c; uncertainty represents $\pm1$ standard deviation). Using Eq.~\ref{eqn:Backaction}, this corresponds to $n_b = 0.63 \pm 0.01$ effective photons of excess backaction: about one-quarter of the $n_r^\mathrm{proj} = 2.4$ effective photons used in a projective measurement (the maximum value on the $x$ axis of Fig.~\ref{fig:Efficiency}c), and far less than the $\sim150$ photons in the pumped state of the parametric cavity (see Supplementary Material Sec.~IV.E). Next, we find $\nu$ and the maximum fidelity $F_0 = 95.5\% \pm 0.3\%$ by fitting $F_r$ versus readout amplitude (red data, Fig.~\ref{fig:Efficiency}c) to Eq.~\ref{eqn:FidelityVsAlpha}. Finally, we obtain $\sigma$ from a fit of the pump off data (cyan) to Eq.~\ref{eqn:MeasInducedDephasingVsAlpha}, and therefore determine $\eta = 70.4\% \pm 0.9\%$ using Eq.~\ref{eqn:Eta}. This fit excludes the first four data points, which level off more quickly than predicted such that excess backaction includes $0.05 \pm 0.01$ effective photons caused solely by actuating the TIBs \footnote{For $\sqrt{n_r} \gtrsim 0.2$ our models for both $|\hat{\rho}_{01}'|$ and $F_r$ generally fall within the 95\% confidence interval of the measurement, and so only these points are used to obtain efficiency. This choice conservatively affects the reported efficiency: including the first four points, or instead fitting the `pump on' data, returns a larger value for $\eta$. We note that in this experiment, $\eta$ must be limited to 78\% or less based on an independent measurement of loss in the parametric cavity and a model for how this loss affects the efficiency (see Supplementary Material Sec. VI.A).}. This dephasing process is not captured by our model, and may result from a noise source on the parametric cavity side of TIB1 (see supplementary material Section~V.B).

The limitations on $n_b$, $\eta$ and $F_0$ are understood and their values may be improved upon (see Supplementary Material Sec.~VI, which includes Refs.~\cite{walls:1994,braginsky:1996,lupascu:2007,gambetta:2008,oconnell:2008,reed:2010b,paik:2011,fowler:2012,reagor:2013,yan:2016,kurpiers:2017,calusine:2018,yan:2018}). Excess backaction primarily results from the $-26 \units{dB}$ of transmission through TIB1 when in reflect mode. This transmission is higher than the $-50 \units{dB}$ of transmission measured in a single TIB in isolation, a discrepancy which may result from the solvable problems of a spurious transmission path within the chip or sample box, or the pumped parametric cavity state approaching the power handling capability of the TIB. Maximum readout fidelity is limited by qubit decay and state preparation error including a $\sim2\%$ thermal population, errors which do not represent limitations of the SIMBA itself. Finally, efficiency is limited primarily by the $4.0 \units{MHz} \pm 0.2 \units{MHz}$ loss rate of the parametric cavity. The dominant contributions to this loss are the nonzero transmission through TIB2 when in reflect mode, on-chip dissipation, and coupling to cable modes: effects which may all be mitigated in future designs.

\begin{table}[h!]
\caption{Readout performance summary.} 
  \begin{center}
    \begin{tabular}{ p{4.0cm}|p{4.0cm}  }
    \hline
    Parameter & Value \\ \hline
    Measurement efficiency & $\eta = 70.4\% \pm 0.9\%$ \\
    Excess backaction & $n_b = 0.66 \pm 0.01$ photons\\
    Maximum readout fidelity & $F_0 = 95.5\% \pm 0.3\%$ \\    
    Measurement time & 265 ns \\
    \end{tabular}
  \end{center}
  \label{tab:PerformanceSummary}
\end{table}

In conclusion, we measure a transmon qubit using a chip-scale, pulsed directional amplifier. The qubit is isolated from amplifier backaction using a superconducting switch to control the coupling between a readout and parametric cavity. Simultaneously demonstrated metrics for this readout are given in Table~\ref{tab:PerformanceSummary}. With reasonable changes to the SIMBA and experimental setup, we estimate it is possible to achieve $\eta > 90\%$ with $F_0 > 99\%$, $n_b \leq 0.02$ and a measurement time of less than 100 ns (see Supplementary Material Sec.~VI). 

This demonstration combines state-of-the-art measurement efficiency \textit{and} considerable isolation from amplifier backaction such that $n_b \sim n_r^\mathrm{proj}/4$. The measurement efficiency of previous superconducting qubit readout schemes have been limited to $\eta = 80\%$ \cite{eddins:2019}, and less when providing any isolation before a parametric amplifier \cite{walter:2017,malnou:2019,touzard:2019,lecocq:2020b} (see Supplementary Material Sec.~I, which includes Refs.~\cite{hatridge:2013,heinsoo:2018,eddins:2018,thorbeck:2017,andersen:2019,andersen:2020,peronnin:2020}, for a broader comparison to other works). Near-unit measurement efficiency after future improvements would allow for near-\textit{complete} access to the information extracted from a quantum system. Additionally, the SIMBA is chip scale, compatible with scalable fabrication procedures including the use of through-silicon vias \cite{rosenberg:2017}, and requires only one microwave control tone to operate. The SIMBA is therefore a favorable choice for high-quality and scalable superconducting qubit measurement.

\vspace{0.1in}
\textit{Acknowledgments} --- The authors thank Florent Lecocq, Alexandre Blais, Jonathan Gross and K. D. Osborn for helpful discussions, and thank James Uhrich, Calvin Schwadron, and Kim Hagen for help in the design and fabrication of the mechanical parts used in this experiment. This work is partially supported by the U.S. Air Force Office of Scientific Research Multidisciplinary Research Program of the University Research Initiative (AFOSR MURI) underGrant No. FA9550-15-1-0015, the U.S. Army Research Office (ARO) under Contract No. W911NF-14-1-0079, and the National Science Foundation under Grant No. 1734006. E.I.R acknowledges support from the ARO QuaCGR fellowship and C.M.F.S. and M.F.Z. acknowledge support from the Austrian Science Fund FWF within the DK-ALM (W1259-N27).

\bibliography{main}

\begin{thebibliography}{79}%
\makeatletter
\providecommand \@ifxundefined [1]{%
 \@ifx{#1\undefined}
}%
\providecommand \@ifnum [1]{%
 \ifnum #1\expandafter \@firstoftwo
 \else \expandafter \@secondoftwo
 \fi
}%
\providecommand \@ifx [1]{%
 \ifx #1\expandafter \@firstoftwo
 \else \expandafter \@secondoftwo
 \fi
}%
\providecommand \natexlab [1]{#1}%
\providecommand \enquote  [1]{``#1''}%
\providecommand \bibnamefont  [1]{#1}%
\providecommand \bibfnamefont [1]{#1}%
\providecommand \citenamefont [1]{#1}%
\providecommand \href@noop [0]{\@secondoftwo}%
\providecommand \href [0]{\begingroup \@sanitize@url \@href}%
\providecommand \@href[1]{\@@startlink{#1}\@@href}%
\providecommand \@@href[1]{\endgroup#1\@@endlink}%
\providecommand \@sanitize@url [0]{\catcode `\\12\catcode `\$12\catcode
  `\&12\catcode `\#12\catcode `\^12\catcode `\_12\catcode `\%12\relax}%
\providecommand \@@startlink[1]{}%
\providecommand \@@endlink[0]{}%
\providecommand \url  [0]{\begingroup\@sanitize@url \@url }%
\providecommand \@url [1]{\endgroup\@href {#1}{\urlprefix }}%
\providecommand \urlprefix  [0]{URL }%
\providecommand \Eprint [0]{\href }%
\providecommand \doibase [0]{http://dx.doi.org/}%
\providecommand \selectlanguage [0]{\@gobble}%
\providecommand \bibinfo  [0]{\@secondoftwo}%
\providecommand \bibfield  [0]{\@secondoftwo}%
\providecommand \translation [1]{[#1]}%
\providecommand \BibitemOpen [0]{}%
\providecommand \bibitemStop [0]{}%
\providecommand \bibitemNoStop [0]{.\EOS\space}%
\providecommand \EOS [0]{\spacefactor3000\relax}%
\providecommand \BibitemShut  [1]{\csname bibitem#1\endcsname}%
\let\auto@bib@innerbib\@empty
\bibitem [{\citenamefont {DiVincenzo}(2000)}]{divincenzo:2000}%
  \BibitemOpen
  \bibfield  {author} {\bibinfo {author} {\bibfnamefont {David~P.}\
  \bibnamefont {DiVincenzo}},\ }\bibfield  {title} {\enquote {\bibinfo {title}
  {The physical implementation of quantum computation},}\ }\href@noop {}
  {\bibfield  {journal} {\bibinfo  {journal} {Fortschr. Phys.}\ }\textbf
  {\bibinfo {volume} {48}},\ \bibinfo {pages} {771--783} (\bibinfo {year}
  {2000})}\BibitemShut {NoStop}%
\bibitem [{\citenamefont {Blais}\ \emph {et~al.}(2004)\citenamefont {Blais},
  \citenamefont {Huang}, \citenamefont {Wallraff}, \citenamefont {Girvin},\
  and\ \citenamefont {Schoelkopf}}]{blais:2004}%
  \BibitemOpen
  \bibfield  {author} {\bibinfo {author} {\bibfnamefont {Alexandre}\
  \bibnamefont {Blais}}, \bibinfo {author} {\bibfnamefont {Ren-Shou}\
  \bibnamefont {Huang}}, \bibinfo {author} {\bibfnamefont {Andreas}\
  \bibnamefont {Wallraff}}, \bibinfo {author} {\bibfnamefont {S.~M.}\
  \bibnamefont {Girvin}}, \ and\ \bibinfo {author} {\bibfnamefont {R.~J.}\
  \bibnamefont {Schoelkopf}},\ }\bibfield  {title} {\enquote {\bibinfo {title}
  {Cavity quantum electrodynamics for superconducting electrical circuits: An
  architecture for quantum computation},}\ }\href@noop {} {\bibfield  {journal}
  {\bibinfo  {journal} {Physical Review A}\ }\textbf {\bibinfo {volume} {69}},\
  \bibinfo {pages} {062320} (\bibinfo {year} {2004})}\BibitemShut {NoStop}%
\bibitem [{\citenamefont {Blais}\ \emph {et~al.}(2020)\citenamefont {Blais},
  \citenamefont {Grimsmo},\ and\ \citenamefont {Wallraff}}]{blais:2020b}%
  \BibitemOpen
  \bibfield  {author} {\bibinfo {author} {\bibfnamefont {Alexandre}\
  \bibnamefont {Blais}}, \bibinfo {author} {\bibfnamefont {Arne~L.}\
  \bibnamefont {Grimsmo}}, \ and\ \bibinfo {author} {\bibfnamefont {Andreas}\
  \bibnamefont {Wallraff}},\ }\bibfield  {title} {\enquote {\bibinfo {title}
  {Circuit quantum electrodynamics},}\ }\href@noop {} {\bibfield  {journal}
  {\bibinfo  {journal} {arXiv:2005.12667}\ } (\bibinfo {year}
  {2020})}\BibitemShut {NoStop}%
\bibitem [{\citenamefont {Castellanos-Beltran}\ and\ \citenamefont
  {Lehnert}(2007)}]{castellanos:2007}%
  \BibitemOpen
  \bibfield  {author} {\bibinfo {author} {\bibfnamefont {M.~A.}\ \bibnamefont
  {Castellanos-Beltran}}\ and\ \bibinfo {author} {\bibfnamefont {K.~W.}\
  \bibnamefont {Lehnert}},\ }\bibfield  {title} {\enquote {\bibinfo {title}
  {Widely tunable parametric amplifier based on a superconducting quantum
  interference device array resonator},}\ }\href@noop {} {\bibfield  {journal}
  {\bibinfo  {journal} {Applied Physics Letters}\ }\textbf {\bibinfo {volume}
  {91}},\ \bibinfo {pages} {083509} (\bibinfo {year} {2007})}\BibitemShut
  {NoStop}%
\bibitem [{\citenamefont {Yamamoto}\ \emph {et~al.}(2008)\citenamefont
  {Yamamoto}, \citenamefont {Inomata}, \citenamefont {Watanabe}, \citenamefont
  {Matsuba}, \citenamefont {Miyazaki}, \citenamefont {Oliver}, \citenamefont
  {Nakamura},\ and\ \citenamefont {Tsai}}]{yamamoto:2008}%
  \BibitemOpen
  \bibfield  {author} {\bibinfo {author} {\bibfnamefont {T.}~\bibnamefont
  {Yamamoto}}, \bibinfo {author} {\bibfnamefont {K.}~\bibnamefont {Inomata}},
  \bibinfo {author} {\bibfnamefont {M.}~\bibnamefont {Watanabe}}, \bibinfo
  {author} {\bibfnamefont {K.}~\bibnamefont {Matsuba}}, \bibinfo {author}
  {\bibfnamefont {T.}~\bibnamefont {Miyazaki}}, \bibinfo {author}
  {\bibfnamefont {W.~D.}\ \bibnamefont {Oliver}}, \bibinfo {author}
  {\bibfnamefont {Y.}~\bibnamefont {Nakamura}}, \ and\ \bibinfo {author}
  {\bibfnamefont {J.~S.}\ \bibnamefont {Tsai}},\ }\bibfield  {title} {\enquote
  {\bibinfo {title} {Flux-driven josephson parametric amplifier},}\ }\href@noop
  {} {\bibfield  {journal} {\bibinfo  {journal} {Applied Physics Letters}\
  }\textbf {\bibinfo {volume} {93}},\ \bibinfo {pages} {042510} (\bibinfo
  {year} {2008})}\BibitemShut {NoStop}%
\bibitem [{\citenamefont {Bergeal}\ \emph {et~al.}(2010)\citenamefont
  {Bergeal}, \citenamefont {Vijay}, \citenamefont {Manucharyan}, \citenamefont
  {Siddiqi}, \citenamefont {Schoelkopf}, \citenamefont {Girvin},\ and\
  \citenamefont {Devoret}}]{bergeal:2010}%
  \BibitemOpen
  \bibfield  {author} {\bibinfo {author} {\bibfnamefont {N.}~\bibnamefont
  {Bergeal}}, \bibinfo {author} {\bibfnamefont {R.}~\bibnamefont {Vijay}},
  \bibinfo {author} {\bibfnamefont {V.~E.}\ \bibnamefont {Manucharyan}},
  \bibinfo {author} {\bibfnamefont {I.}~\bibnamefont {Siddiqi}}, \bibinfo
  {author} {\bibfnamefont {R.~J.}\ \bibnamefont {Schoelkopf}}, \bibinfo
  {author} {\bibfnamefont {S.~M.}\ \bibnamefont {Girvin}}, \ and\ \bibinfo
  {author} {\bibfnamefont {M.~H.}\ \bibnamefont {Devoret}},\ }\bibfield
  {title} {\enquote {\bibinfo {title} {Analog information processing at the
  quantum limit with a {J}osephson ring modulator},}\ }\href@noop {} {\bibfield
   {journal} {\bibinfo  {journal} {Nature Physics}\ }\textbf {\bibinfo {volume}
  {6}},\ \bibinfo {pages} {296--302} (\bibinfo {year} {2010})}\BibitemShut
  {NoStop}%
\bibitem [{\citenamefont {Macklin}\ \emph {et~al.}(2015)\citenamefont
  {Macklin}, \citenamefont {O'Brien}, \citenamefont {Hover}, \citenamefont
  {Schwartz}, \citenamefont {Bolkhovsky}, \citenamefont {Zhang}, \citenamefont
  {Oliver},\ and\ \citenamefont {Siddiqi}}]{macklin:2015}%
  \BibitemOpen
  \bibfield  {author} {\bibinfo {author} {\bibfnamefont {C.}~\bibnamefont
  {Macklin}}, \bibinfo {author} {\bibfnamefont {K.}~\bibnamefont {O'Brien}},
  \bibinfo {author} {\bibfnamefont {D.}~\bibnamefont {Hover}}, \bibinfo
  {author} {\bibfnamefont {M.~E.}\ \bibnamefont {Schwartz}}, \bibinfo {author}
  {\bibfnamefont {V.}~\bibnamefont {Bolkhovsky}}, \bibinfo {author}
  {\bibfnamefont {X.}~\bibnamefont {Zhang}}, \bibinfo {author} {\bibfnamefont
  {W.~D.}\ \bibnamefont {Oliver}}, \ and\ \bibinfo {author} {\bibfnamefont
  {I.}~\bibnamefont {Siddiqi}},\ }\bibfield  {title} {\enquote {\bibinfo
  {title} {A near--quantum-limited {J}osephson traveling-wave parametric
  amplifier},}\ }\href@noop {} {\bibfield  {journal} {\bibinfo  {journal}
  {Science}\ }\textbf {\bibinfo {volume} {350}},\ \bibinfo {pages} {307--310}
  (\bibinfo {year} {2015})}\BibitemShut {NoStop}%
\bibitem [{\citenamefont {Frattini}\ \emph {et~al.}(2017)\citenamefont
  {Frattini}, \citenamefont {Vool}, \citenamefont {Shankar}, \citenamefont
  {Narla}, \citenamefont {Sliwa},\ and\ \citenamefont
  {Devoret}}]{frattini:2017}%
  \BibitemOpen
  \bibfield  {author} {\bibinfo {author} {\bibfnamefont {N.~E.}\ \bibnamefont
  {Frattini}}, \bibinfo {author} {\bibfnamefont {U.}~\bibnamefont {Vool}},
  \bibinfo {author} {\bibfnamefont {S.}~\bibnamefont {Shankar}}, \bibinfo
  {author} {\bibfnamefont {A.}~\bibnamefont {Narla}}, \bibinfo {author}
  {\bibfnamefont {K.~M.}\ \bibnamefont {Sliwa}}, \ and\ \bibinfo {author}
  {\bibfnamefont {M.~H.}\ \bibnamefont {Devoret}},\ }\bibfield  {title}
  {\enquote {\bibinfo {title} {3-wave mixing josephson dipole element},}\
  }\href {\doibase https://doi.org/10.1063/1.4984142} {\bibfield  {journal}
  {\bibinfo  {journal} {Applied Physics Letters}\ }\textbf {\bibinfo {volume}
  {110}},\ \bibinfo {pages} {222603} (\bibinfo {year} {2017})}\BibitemShut
  {NoStop}%
\bibitem [{\citenamefont {Jeffrey}\ \emph {et~al.}(2014)\citenamefont
  {Jeffrey}, \citenamefont {Sank}, \citenamefont {Mutus}, \citenamefont
  {White}, \citenamefont {Kelly}, \citenamefont {Barends}, \citenamefont
  {Chen}, \citenamefont {Chen}, \citenamefont {Chiaro}, \citenamefont
  {Dunsworth}, \citenamefont {Megrant}, \citenamefont {O'Malley}, \citenamefont
  {Neill}, \citenamefont {Roushan}, \citenamefont {Vainsencher}, \citenamefont
  {Wenner}, \citenamefont {Cleland},\ and\ \citenamefont
  {Martinis}}]{jeffrey:2014}%
  \BibitemOpen
  \bibfield  {author} {\bibinfo {author} {\bibfnamefont {Evan}\ \bibnamefont
  {Jeffrey}}, \bibinfo {author} {\bibfnamefont {Daniel}\ \bibnamefont {Sank}},
  \bibinfo {author} {\bibfnamefont {J.~Y.}\ \bibnamefont {Mutus}}, \bibinfo
  {author} {\bibfnamefont {T.~C.}\ \bibnamefont {White}}, \bibinfo {author}
  {\bibfnamefont {J.}~\bibnamefont {Kelly}}, \bibinfo {author} {\bibfnamefont
  {R.}~\bibnamefont {Barends}}, \bibinfo {author} {\bibfnamefont
  {Y.}~\bibnamefont {Chen}}, \bibinfo {author} {\bibfnamefont {Z.}~\bibnamefont
  {Chen}}, \bibinfo {author} {\bibfnamefont {B.}~\bibnamefont {Chiaro}},
  \bibinfo {author} {\bibfnamefont {A.}~\bibnamefont {Dunsworth}}, \bibinfo
  {author} {\bibfnamefont {A.}~\bibnamefont {Megrant}}, \bibinfo {author}
  {\bibfnamefont {P.~J.~J.}\ \bibnamefont {O'Malley}}, \bibinfo {author}
  {\bibfnamefont {C.}~\bibnamefont {Neill}}, \bibinfo {author} {\bibfnamefont
  {P.}~\bibnamefont {Roushan}}, \bibinfo {author} {\bibfnamefont
  {A.}~\bibnamefont {Vainsencher}}, \bibinfo {author} {\bibfnamefont
  {J.}~\bibnamefont {Wenner}}, \bibinfo {author} {\bibfnamefont {A.~N.}\
  \bibnamefont {Cleland}}, \ and\ \bibinfo {author} {\bibfnamefont {John~M.}\
  \bibnamefont {Martinis}},\ }\bibfield  {title} {\enquote {\bibinfo {title}
  {Fast accurate state measurement with superconducting qubits},}\ }\href
  {\doibase 10.1103/PhysRevLett.112.190504} {\bibfield  {journal} {\bibinfo
  {journal} {Phys. Rev. Lett.}\ }\textbf {\bibinfo {volume} {112}},\ \bibinfo
  {pages} {190504} (\bibinfo {year} {2014})}\BibitemShut {NoStop}%
\bibitem [{\citenamefont {Walter}\ \emph {et~al.}(2017)\citenamefont {Walter},
  \citenamefont {Kurpiers}, \citenamefont {Gasparinetti}, \citenamefont
  {Magnard}, \citenamefont {Poto\ifmmode~\check{c}\else \v{c}\fi{}nik},
  \citenamefont {Salath\'e}, \citenamefont {Pechal}, \citenamefont {Mondal},
  \citenamefont {Oppliger}, \citenamefont {Eichler},\ and\ \citenamefont
  {Wallraff}}]{walter:2017}%
  \BibitemOpen
  \bibfield  {author} {\bibinfo {author} {\bibfnamefont {T.}~\bibnamefont
  {Walter}}, \bibinfo {author} {\bibfnamefont {P.}~\bibnamefont {Kurpiers}},
  \bibinfo {author} {\bibfnamefont {S.}~\bibnamefont {Gasparinetti}}, \bibinfo
  {author} {\bibfnamefont {P.}~\bibnamefont {Magnard}}, \bibinfo {author}
  {\bibfnamefont {A.}~\bibnamefont {Poto\ifmmode~\check{c}\else
  \v{c}\fi{}nik}}, \bibinfo {author} {\bibfnamefont {Y.}~\bibnamefont
  {Salath\'e}}, \bibinfo {author} {\bibfnamefont {M.}~\bibnamefont {Pechal}},
  \bibinfo {author} {\bibfnamefont {M.}~\bibnamefont {Mondal}}, \bibinfo
  {author} {\bibfnamefont {M.}~\bibnamefont {Oppliger}}, \bibinfo {author}
  {\bibfnamefont {C.}~\bibnamefont {Eichler}}, \ and\ \bibinfo {author}
  {\bibfnamefont {A.}~\bibnamefont {Wallraff}},\ }\bibfield  {title} {\enquote
  {\bibinfo {title} {Rapid high-fidelity single-shot dispersive readout of
  superconducting qubits},}\ }\href {\doibase 10.1103/PhysRevApplied.7.054020}
  {\bibfield  {journal} {\bibinfo  {journal} {Phys. Rev. Applied}\ }\textbf
  {\bibinfo {volume} {7}},\ \bibinfo {pages} {054020} (\bibinfo {year}
  {2017})}\BibitemShut {NoStop}%
\bibitem [{\citenamefont {Elder}\ \emph {et~al.}(2020)\citenamefont {Elder},
  \citenamefont {Wang}, \citenamefont {Reinhold}, \citenamefont {Hann},
  \citenamefont {Chou}, \citenamefont {Lester}, \citenamefont {Rosenblum},
  \citenamefont {Frunzio}, \citenamefont {Jiang},\ and\ \citenamefont
  {Schoelkopf}}]{elder:2020}%
  \BibitemOpen
  \bibfield  {author} {\bibinfo {author} {\bibfnamefont {Salvatore~S.}\
  \bibnamefont {Elder}}, \bibinfo {author} {\bibfnamefont {Christopher~S.}\
  \bibnamefont {Wang}}, \bibinfo {author} {\bibfnamefont {Philip}\ \bibnamefont
  {Reinhold}}, \bibinfo {author} {\bibfnamefont {Connor~T.}\ \bibnamefont
  {Hann}}, \bibinfo {author} {\bibfnamefont {Kevin~S.}\ \bibnamefont {Chou}},
  \bibinfo {author} {\bibfnamefont {Brian~J.}\ \bibnamefont {Lester}}, \bibinfo
  {author} {\bibfnamefont {Serge}\ \bibnamefont {Rosenblum}}, \bibinfo {author}
  {\bibfnamefont {Luigi}\ \bibnamefont {Frunzio}}, \bibinfo {author}
  {\bibfnamefont {Liang}\ \bibnamefont {Jiang}}, \ and\ \bibinfo {author}
  {\bibfnamefont {Robert~J.}\ \bibnamefont {Schoelkopf}},\ }\bibfield  {title}
  {\enquote {\bibinfo {title} {High-fidelity measurement of qubits encoded in
  multilevel superconducting circuits},}\ }\href {\doibase
  10.1103/PhysRevX.10.011001} {\bibfield  {journal} {\bibinfo  {journal} {Phys.
  Rev. X}\ }\textbf {\bibinfo {volume} {10}},\ \bibinfo {pages} {011001}
  (\bibinfo {year} {2020})}\BibitemShut {NoStop}%
\bibitem [{\citenamefont {Slichter}\ \emph {et~al.}(2012)\citenamefont
  {Slichter}, \citenamefont {Vijay}, \citenamefont {Weber}, \citenamefont
  {Boutin}, \citenamefont {Boissonneault}, \citenamefont {Gambetta},
  \citenamefont {Blais},\ and\ \citenamefont {Siddiqi}}]{slichter:2012}%
  \BibitemOpen
  \bibfield  {author} {\bibinfo {author} {\bibfnamefont {D.~H.}\ \bibnamefont
  {Slichter}}, \bibinfo {author} {\bibfnamefont {R.}~\bibnamefont {Vijay}},
  \bibinfo {author} {\bibfnamefont {S.~J.}\ \bibnamefont {Weber}}, \bibinfo
  {author} {\bibfnamefont {S.}~\bibnamefont {Boutin}}, \bibinfo {author}
  {\bibfnamefont {M.}~\bibnamefont {Boissonneault}}, \bibinfo {author}
  {\bibfnamefont {J.~M.}\ \bibnamefont {Gambetta}}, \bibinfo {author}
  {\bibfnamefont {A.}~\bibnamefont {Blais}}, \ and\ \bibinfo {author}
  {\bibfnamefont {I.}~\bibnamefont {Siddiqi}},\ }\bibfield  {title} {\enquote
  {\bibinfo {title} {Measurement-induced qubit state mixing in circuit qed from
  up-converted dephasing noise},}\ }\href {\doibase
  10.1103/PhysRevLett.109.153601} {\bibfield  {journal} {\bibinfo  {journal}
  {Phys. Rev. Lett.}\ }\textbf {\bibinfo {volume} {109}},\ \bibinfo {pages}
  {153601} (\bibinfo {year} {2012})}\BibitemShut {NoStop}%
\bibitem [{\citenamefont {Sank}\ \emph {et~al.}(2016)\citenamefont {Sank},
  \citenamefont {Chen}, \citenamefont {Khezri}, \citenamefont {Kelly},
  \citenamefont {Barends}, \citenamefont {Campbell}, \citenamefont {Chen},
  \citenamefont {Chiaro}, \citenamefont {Dunsworth}, \citenamefont {Fowler},
  \citenamefont {Jeffrey}, \citenamefont {Lucero}, \citenamefont {Megrant},
  \citenamefont {Mutus}, \citenamefont {Neeley}, \citenamefont {Neill},
  \citenamefont {O'Malley}, \citenamefont {Quintana}, \citenamefont {Roushan},
  \citenamefont {Vainsencher}, \citenamefont {White}, \citenamefont {Wenner},
  \citenamefont {Korotkov},\ and\ \citenamefont {Martinis}}]{sank:2016}%
  \BibitemOpen
  \bibfield  {author} {\bibinfo {author} {\bibfnamefont {Daniel}\ \bibnamefont
  {Sank}}, \bibinfo {author} {\bibfnamefont {Zijun}\ \bibnamefont {Chen}},
  \bibinfo {author} {\bibfnamefont {Mostafa}\ \bibnamefont {Khezri}}, \bibinfo
  {author} {\bibfnamefont {J.}~\bibnamefont {Kelly}}, \bibinfo {author}
  {\bibfnamefont {R.}~\bibnamefont {Barends}}, \bibinfo {author} {\bibfnamefont
  {B.}~\bibnamefont {Campbell}}, \bibinfo {author} {\bibfnamefont
  {Y.}~\bibnamefont {Chen}}, \bibinfo {author} {\bibfnamefont {B.}~\bibnamefont
  {Chiaro}}, \bibinfo {author} {\bibfnamefont {A.}~\bibnamefont {Dunsworth}},
  \bibinfo {author} {\bibfnamefont {A.}~\bibnamefont {Fowler}}, \bibinfo
  {author} {\bibfnamefont {E.}~\bibnamefont {Jeffrey}}, \bibinfo {author}
  {\bibfnamefont {E.}~\bibnamefont {Lucero}}, \bibinfo {author} {\bibfnamefont
  {A.}~\bibnamefont {Megrant}}, \bibinfo {author} {\bibfnamefont
  {J.}~\bibnamefont {Mutus}}, \bibinfo {author} {\bibfnamefont
  {M.}~\bibnamefont {Neeley}}, \bibinfo {author} {\bibfnamefont
  {C.}~\bibnamefont {Neill}}, \bibinfo {author} {\bibfnamefont {P.~J.~J.}\
  \bibnamefont {O'Malley}}, \bibinfo {author} {\bibfnamefont {C.}~\bibnamefont
  {Quintana}}, \bibinfo {author} {\bibfnamefont {P.}~\bibnamefont {Roushan}},
  \bibinfo {author} {\bibfnamefont {A.}~\bibnamefont {Vainsencher}}, \bibinfo
  {author} {\bibfnamefont {T.}~\bibnamefont {White}}, \bibinfo {author}
  {\bibfnamefont {J.}~\bibnamefont {Wenner}}, \bibinfo {author} {\bibfnamefont
  {Alexander~N.}\ \bibnamefont {Korotkov}}, \ and\ \bibinfo {author}
  {\bibfnamefont {John~M.}\ \bibnamefont {Martinis}},\ }\bibfield  {title}
  {\enquote {\bibinfo {title} {Measurement-induced state transitions in a
  superconducting qubit: Beyond the rotating wave approximation},}\ }\href
  {\doibase 10.1103/PhysRevLett.117.190503} {\bibfield  {journal} {\bibinfo
  {journal} {Phys. Rev. Lett.}\ }\textbf {\bibinfo {volume} {117}},\ \bibinfo
  {pages} {190503} (\bibinfo {year} {2016})}\BibitemShut {NoStop}%
\bibitem [{\citenamefont {Abdo}\ \emph {et~al.}(2014)\citenamefont {Abdo},
  \citenamefont {Sliwa}, \citenamefont {Shankar}, \citenamefont {Hatridge},
  \citenamefont {Frunzio}, \citenamefont {Schoelkopf},\ and\ \citenamefont
  {Devoret}}]{abdo:2014}%
  \BibitemOpen
  \bibfield  {author} {\bibinfo {author} {\bibfnamefont {Baleegh}\ \bibnamefont
  {Abdo}}, \bibinfo {author} {\bibfnamefont {Katrina}\ \bibnamefont {Sliwa}},
  \bibinfo {author} {\bibfnamefont {S}~\bibnamefont {Shankar}}, \bibinfo
  {author} {\bibfnamefont {Michael}\ \bibnamefont {Hatridge}}, \bibinfo
  {author} {\bibfnamefont {Luigi}\ \bibnamefont {Frunzio}}, \bibinfo {author}
  {\bibfnamefont {Robert~J.}\ \bibnamefont {Schoelkopf}}, \ and\ \bibinfo
  {author} {\bibfnamefont {Michel~H.}\ \bibnamefont {Devoret}},\ }\bibfield
  {title} {\enquote {\bibinfo {title} {Josephson directional amplifier for
  quantum measurement of superconducting circuits},}\ }\href@noop {} {\bibfield
   {journal} {\bibinfo  {journal} {Physical review letters}\ }\textbf {\bibinfo
  {volume} {112}},\ \bibinfo {pages} {167701} (\bibinfo {year}
  {2014})}\BibitemShut {NoStop}%
\bibitem [{\citenamefont {Ranzani}\ and\ \citenamefont
  {Aumentado}(2015)}]{ranzani:2015}%
  \BibitemOpen
  \bibfield  {author} {\bibinfo {author} {\bibfnamefont {Leonardo}\
  \bibnamefont {Ranzani}}\ and\ \bibinfo {author} {\bibfnamefont {Jos{\'e}}\
  \bibnamefont {Aumentado}},\ }\bibfield  {title} {\enquote {\bibinfo {title}
  {Graph-based analysis of nonreciprocity in coupled-mode systems},}\
  }\href@noop {} {\bibfield  {journal} {\bibinfo  {journal} {New Journal of
  Physics}\ }\textbf {\bibinfo {volume} {17}},\ \bibinfo {pages} {023024}
  (\bibinfo {year} {2015})}\BibitemShut {NoStop}%
\bibitem [{\citenamefont {Sliwa}\ \emph {et~al.}(2015)\citenamefont {Sliwa},
  \citenamefont {Hatridge}, \citenamefont {Narla}, \citenamefont {Shankar},
  \citenamefont {Frunzio}, \citenamefont {Schoelkopf},\ and\ \citenamefont
  {Devoret}}]{sliwa:2015}%
  \BibitemOpen
  \bibfield  {author} {\bibinfo {author} {\bibfnamefont {K.~M.}\ \bibnamefont
  {Sliwa}}, \bibinfo {author} {\bibfnamefont {M.}~\bibnamefont {Hatridge}},
  \bibinfo {author} {\bibfnamefont {A.}~\bibnamefont {Narla}}, \bibinfo
  {author} {\bibfnamefont {S.}~\bibnamefont {Shankar}}, \bibinfo {author}
  {\bibfnamefont {L.}~\bibnamefont {Frunzio}}, \bibinfo {author} {\bibfnamefont
  {R.~J.}\ \bibnamefont {Schoelkopf}}, \ and\ \bibinfo {author} {\bibfnamefont
  {M.~H.}\ \bibnamefont {Devoret}},\ }\bibfield  {title} {\enquote {\bibinfo
  {title} {Reconfigurable {J}osephson circulator/directional amplifier},}\
  }\href@noop {} {\bibfield  {journal} {\bibinfo  {journal} {Phys. Rev. X}\
  }\textbf {\bibinfo {volume} {5}},\ \bibinfo {pages} {041020} (\bibinfo {year}
  {2015})}\BibitemShut {NoStop}%
\bibitem [{\citenamefont {Lecocq}\ \emph {et~al.}(2017)\citenamefont {Lecocq},
  \citenamefont {Ranzani}, \citenamefont {Peterson}, \citenamefont {Cicak},
  \citenamefont {Simmonds}, \citenamefont {Teufel},\ and\ \citenamefont
  {Aumentado}}]{lecocq:2017}%
  \BibitemOpen
  \bibfield  {author} {\bibinfo {author} {\bibfnamefont {F.}~\bibnamefont
  {Lecocq}}, \bibinfo {author} {\bibfnamefont {L.}~\bibnamefont {Ranzani}},
  \bibinfo {author} {\bibfnamefont {G.~A.}\ \bibnamefont {Peterson}}, \bibinfo
  {author} {\bibfnamefont {K.}~\bibnamefont {Cicak}}, \bibinfo {author}
  {\bibfnamefont {R.~W.}\ \bibnamefont {Simmonds}}, \bibinfo {author}
  {\bibfnamefont {J.~D.}\ \bibnamefont {Teufel}}, \ and\ \bibinfo {author}
  {\bibfnamefont {J.}~\bibnamefont {Aumentado}},\ }\bibfield  {title} {\enquote
  {\bibinfo {title} {Nonreciprocal microwave signal processing with a
  field-programmable {J}osephson amplifier},}\ }\href {\doibase
  10.1103/PhysRevApplied.7.024028} {\bibfield  {journal} {\bibinfo  {journal}
  {Phys. Rev. Applied}\ }\textbf {\bibinfo {volume} {7}},\ \bibinfo {pages}
  {024028} (\bibinfo {year} {2017})}\BibitemShut {NoStop}%
\bibitem [{\citenamefont {Abdo}\ \emph {et~al.}(2019)\citenamefont {Abdo},
  \citenamefont {Bronn}, \citenamefont {Jinka}, \citenamefont {Olivadese},
  \citenamefont {C\'orcoles}, \citenamefont {Adiga}, \citenamefont {Brink},
  \citenamefont {Lake}, \citenamefont {Wu}, \citenamefont {Pappas},\ and\
  \citenamefont {Chow}}]{abdo:2019}%
  \BibitemOpen
  \bibfield  {author} {\bibinfo {author} {\bibfnamefont {B.}~\bibnamefont
  {Abdo}}, \bibinfo {author} {\bibfnamefont {N.~T.}\ \bibnamefont {Bronn}},
  \bibinfo {author} {\bibfnamefont {O.}~\bibnamefont {Jinka}}, \bibinfo
  {author} {\bibfnamefont {S.}~\bibnamefont {Olivadese}}, \bibinfo {author}
  {\bibfnamefont {A.~D.}\ \bibnamefont {C\'orcoles}}, \bibinfo {author}
  {\bibfnamefont {V.~P.}\ \bibnamefont {Adiga}}, \bibinfo {author}
  {\bibfnamefont {M.}~\bibnamefont {Brink}}, \bibinfo {author} {\bibfnamefont
  {R.~E.}\ \bibnamefont {Lake}}, \bibinfo {author} {\bibfnamefont
  {X.}~\bibnamefont {Wu}}, \bibinfo {author} {\bibfnamefont {D.~P.}\
  \bibnamefont {Pappas}}, \ and\ \bibinfo {author} {\bibfnamefont {J.~M.}\
  \bibnamefont {Chow}},\ }\bibfield  {title} {\enquote {\bibinfo {title}
  {Active protection of a superconducting qubit with an interferometric
  josephson isolator},}\ }\href@noop {} {\bibfield  {journal} {\bibinfo
  {journal} {Nature communications}\ }\textbf {\bibinfo {volume} {10}},\
  \bibinfo {pages} {3154} (\bibinfo {year} {2019})}\BibitemShut {NoStop}%
\bibitem [{\citenamefont {Lecocq}\ \emph {et~al.}(2020)\citenamefont {Lecocq},
  \citenamefont {Ranzani}, \citenamefont {Peterson}, \citenamefont {Cicak},
  \citenamefont {Metelmann}, \citenamefont {Kotler}, \citenamefont {Simmonds},
  \citenamefont {Teufel},\ and\ \citenamefont {Aumentado}}]{lecocq:2020}%
  \BibitemOpen
  \bibfield  {author} {\bibinfo {author} {\bibfnamefont {F.}~\bibnamefont
  {Lecocq}}, \bibinfo {author} {\bibfnamefont {L.}~\bibnamefont {Ranzani}},
  \bibinfo {author} {\bibfnamefont {G.~A.}\ \bibnamefont {Peterson}}, \bibinfo
  {author} {\bibfnamefont {K.}~\bibnamefont {Cicak}}, \bibinfo {author}
  {\bibfnamefont {A.}~\bibnamefont {Metelmann}}, \bibinfo {author}
  {\bibfnamefont {S.}~\bibnamefont {Kotler}}, \bibinfo {author} {\bibfnamefont
  {R.~W.}\ \bibnamefont {Simmonds}}, \bibinfo {author} {\bibfnamefont {J.~D.}\
  \bibnamefont {Teufel}}, \ and\ \bibinfo {author} {\bibfnamefont
  {J.}~\bibnamefont {Aumentado}},\ }\bibfield  {title} {\enquote {\bibinfo
  {title} {Microwave measurement beyond the quantum limit with a nonreciprocal
  amplifier},}\ }\href {\doibase 10.1103/PhysRevApplied.13.044005} {\bibfield
  {journal} {\bibinfo  {journal} {Phys. Rev. Applied}\ }\textbf {\bibinfo
  {volume} {13}},\ \bibinfo {pages} {044005} (\bibinfo {year}
  {2020})}\BibitemShut {NoStop}%
\bibitem [{\citenamefont {Abdo}\ \emph {et~al.}(2020)\citenamefont {Abdo},
  \citenamefont {Jinka}, \citenamefont {Bronn}, \citenamefont {Olivadese},\
  and\ \citenamefont {Brink}}]{abdo:2020}%
  \BibitemOpen
  \bibfield  {author} {\bibinfo {author} {\bibfnamefont {Baleegh}\ \bibnamefont
  {Abdo}}, \bibinfo {author} {\bibfnamefont {Oblesh}\ \bibnamefont {Jinka}},
  \bibinfo {author} {\bibfnamefont {Nicholas~T.}\ \bibnamefont {Bronn}},
  \bibinfo {author} {\bibfnamefont {Salvatore}\ \bibnamefont {Olivadese}}, \
  and\ \bibinfo {author} {\bibfnamefont {Markus}\ \bibnamefont {Brink}},\
  }\bibfield  {title} {\enquote {\bibinfo {title} {On-chip single-pump
  interferometric josephson isolator for quantum measurements},}\ }\href@noop
  {} {\bibfield  {journal} {\bibinfo  {journal} {arXiv preprint
  arXiv:2006.01918}\ } (\bibinfo {year} {2020})}\BibitemShut {NoStop}%
\bibitem [{\citenamefont {Lecocq}\ \emph {et~al.}(2021)\citenamefont {Lecocq},
  \citenamefont {Ranzani}, \citenamefont {Peterson}, \citenamefont {Cicak},
  \citenamefont {Jin}, \citenamefont {Simmonds}, \citenamefont {Teufel},\ and\
  \citenamefont {Aumentado}}]{lecocq:2020b}%
  \BibitemOpen
  \bibfield  {author} {\bibinfo {author} {\bibfnamefont {F.}~\bibnamefont
  {Lecocq}}, \bibinfo {author} {\bibfnamefont {L.}~\bibnamefont {Ranzani}},
  \bibinfo {author} {\bibfnamefont {G.~A.}\ \bibnamefont {Peterson}}, \bibinfo
  {author} {\bibfnamefont {K.}~\bibnamefont {Cicak}}, \bibinfo {author}
  {\bibfnamefont {X.~Y.}\ \bibnamefont {Jin}}, \bibinfo {author} {\bibfnamefont
  {R.~W.}\ \bibnamefont {Simmonds}}, \bibinfo {author} {\bibfnamefont {J.~D.}\
  \bibnamefont {Teufel}}, \ and\ \bibinfo {author} {\bibfnamefont
  {J.}~\bibnamefont {Aumentado}},\ }\bibfield  {title} {\enquote {\bibinfo
  {title} {Efficient qubit measurement with a nonreciprocal microwave
  amplifier},}\ }\href {\doibase 10.1103/PhysRevLett.126.020502} {\bibfield
  {journal} {\bibinfo  {journal} {Phys. Rev. Lett.}\ }\textbf {\bibinfo
  {volume} {126}},\ \bibinfo {pages} {020502} (\bibinfo {year}
  {2021})}\BibitemShut {NoStop}%
\bibitem [{\citenamefont {Kerckhoff}\ \emph {et~al.}(2015)\citenamefont
  {Kerckhoff}, \citenamefont {Lalumi\`ere}, \citenamefont {Chapman},
  \citenamefont {Blais},\ and\ \citenamefont {Lehnert}}]{kerckhoff:2015}%
  \BibitemOpen
  \bibfield  {author} {\bibinfo {author} {\bibfnamefont {Joseph}\ \bibnamefont
  {Kerckhoff}}, \bibinfo {author} {\bibfnamefont {Kevin}\ \bibnamefont
  {Lalumi\`ere}}, \bibinfo {author} {\bibfnamefont {Benjamin~J.}\ \bibnamefont
  {Chapman}}, \bibinfo {author} {\bibfnamefont {Alexandre}\ \bibnamefont
  {Blais}}, \ and\ \bibinfo {author} {\bibfnamefont {K.~W.}\ \bibnamefont
  {Lehnert}},\ }\bibfield  {title} {\enquote {\bibinfo {title} {On-chip
  superconducting microwave circulator from synthetic rotation},}\ }\href
  {\doibase 10.1103/PhysRevApplied.4.034002} {\bibfield  {journal} {\bibinfo
  {journal} {Phys. Rev. Applied}\ }\textbf {\bibinfo {volume} {4}},\ \bibinfo
  {pages} {034002} (\bibinfo {year} {2015})}\BibitemShut {NoStop}%
\bibitem [{\citenamefont {Rosenthal}\ \emph {et~al.}(2017)\citenamefont
  {Rosenthal}, \citenamefont {Chapman}, \citenamefont {Higginbotham},
  \citenamefont {Kerckhoff},\ and\ \citenamefont {Lehnert}}]{rosenthal:2017}%
  \BibitemOpen
  \bibfield  {author} {\bibinfo {author} {\bibfnamefont {Eric~I.}\ \bibnamefont
  {Rosenthal}}, \bibinfo {author} {\bibfnamefont {Benjamin~J.}\ \bibnamefont
  {Chapman}}, \bibinfo {author} {\bibfnamefont {Andrew~P.}\ \bibnamefont
  {Higginbotham}}, \bibinfo {author} {\bibfnamefont {Joseph}\ \bibnamefont
  {Kerckhoff}}, \ and\ \bibinfo {author} {\bibfnamefont {K.~W.}\ \bibnamefont
  {Lehnert}},\ }\bibfield  {title} {\enquote {\bibinfo {title} {Breaking
  {L}orentz reciprocity with frequency conversion and delay},}\ }\href
  {\doibase 10.1103/PhysRevLett.119.147703} {\bibfield  {journal} {\bibinfo
  {journal} {Phys. Rev. Lett.}\ }\textbf {\bibinfo {volume} {119}},\ \bibinfo
  {pages} {147703} (\bibinfo {year} {2017})}\BibitemShut {NoStop}%
\bibitem [{\citenamefont {Chapman}\ \emph
  {et~al.}(2017{\natexlab{a}})\citenamefont {Chapman}, \citenamefont
  {Rosenthal}, \citenamefont {Kerckhoff}, \citenamefont {Moores}, \citenamefont
  {Vale}, \citenamefont {Mates}, \citenamefont {Hilton}, \citenamefont
  {Lalumi\`ere}, \citenamefont {Blais},\ and\ \citenamefont
  {Lehnert}}]{chapman:2017b}%
  \BibitemOpen
  \bibfield  {author} {\bibinfo {author} {\bibfnamefont {Benjamin~J.}\
  \bibnamefont {Chapman}}, \bibinfo {author} {\bibfnamefont {Eric~I.}\
  \bibnamefont {Rosenthal}}, \bibinfo {author} {\bibfnamefont {Joseph}\
  \bibnamefont {Kerckhoff}}, \bibinfo {author} {\bibfnamefont {Bradley~A.}\
  \bibnamefont {Moores}}, \bibinfo {author} {\bibfnamefont {Leila~R.}\
  \bibnamefont {Vale}}, \bibinfo {author} {\bibfnamefont {J.~A.~B.}\
  \bibnamefont {Mates}}, \bibinfo {author} {\bibfnamefont {Gene~C.}\
  \bibnamefont {Hilton}}, \bibinfo {author} {\bibfnamefont {Kevin}\
  \bibnamefont {Lalumi\`ere}}, \bibinfo {author} {\bibfnamefont {Alexandre}\
  \bibnamefont {Blais}}, \ and\ \bibinfo {author} {\bibfnamefont {K.~W.}\
  \bibnamefont {Lehnert}},\ }\bibfield  {title} {\enquote {\bibinfo {title}
  {Widely tunable on-chip microwave circulator for superconducting quantum
  circuits},}\ }\href {\doibase 10.1103/PhysRevX.7.041043} {\bibfield
  {journal} {\bibinfo  {journal} {Phys. Rev. X}\ }\textbf {\bibinfo {volume}
  {7}},\ \bibinfo {pages} {041043} (\bibinfo {year}
  {2017}{\natexlab{a}})}\BibitemShut {NoStop}%
\bibitem [{\citenamefont {Chapman}\ \emph {et~al.}(2019)\citenamefont
  {Chapman}, \citenamefont {Rosenthal},\ and\ \citenamefont
  {Lehnert}}]{chapman:2019}%
  \BibitemOpen
  \bibfield  {author} {\bibinfo {author} {\bibfnamefont {Benjamin~J.}\
  \bibnamefont {Chapman}}, \bibinfo {author} {\bibfnamefont {Eric~I.}\
  \bibnamefont {Rosenthal}}, \ and\ \bibinfo {author} {\bibfnamefont {K.~W.}\
  \bibnamefont {Lehnert}},\ }\bibfield  {title} {\enquote {\bibinfo {title}
  {Design of an on-chip superconducting microwave circulator with octave
  bandwidth},}\ }\href {\doibase 10.1103/PhysRevApplied.11.044048} {\bibfield
  {journal} {\bibinfo  {journal} {Phys. Rev. Applied}\ }\textbf {\bibinfo
  {volume} {11}},\ \bibinfo {pages} {044048} (\bibinfo {year}
  {2019})}\BibitemShut {NoStop}%
\bibitem [{\citenamefont {Opremcak}\ \emph {et~al.}(2018)\citenamefont
  {Opremcak}, \citenamefont {Pechenezhskiy}, \citenamefont {Howington},
  \citenamefont {Christensen}, \citenamefont {Beck}, \citenamefont
  {Leonard~Jr.}, \citenamefont {Suttle}, \citenamefont {Wilen}, \citenamefont
  {Nesterov}, \citenamefont {Ribeill}, \citenamefont {Thorbeck}, \citenamefont
  {Schlenker}, \citenamefont {Vavilov}, \citenamefont {Plourde},\ and\
  \citenamefont {McDermott}}]{opremcak:2018}%
  \BibitemOpen
  \bibfield  {author} {\bibinfo {author} {\bibfnamefont {A.}~\bibnamefont
  {Opremcak}}, \bibinfo {author} {\bibfnamefont {I.~V.}\ \bibnamefont
  {Pechenezhskiy}}, \bibinfo {author} {\bibfnamefont {C.}~\bibnamefont
  {Howington}}, \bibinfo {author} {\bibfnamefont {B.~G.}\ \bibnamefont
  {Christensen}}, \bibinfo {author} {\bibfnamefont {M.~A.}\ \bibnamefont
  {Beck}}, \bibinfo {author} {\bibfnamefont {E.}~\bibnamefont {Leonard~Jr.}},
  \bibinfo {author} {\bibfnamefont {J.}~\bibnamefont {Suttle}}, \bibinfo
  {author} {\bibfnamefont {C.}~\bibnamefont {Wilen}}, \bibinfo {author}
  {\bibfnamefont {K.~N.}\ \bibnamefont {Nesterov}}, \bibinfo {author}
  {\bibfnamefont {G.~J.}\ \bibnamefont {Ribeill}}, \bibinfo {author}
  {\bibfnamefont {T.}~\bibnamefont {Thorbeck}}, \bibinfo {author}
  {\bibfnamefont {F.}~\bibnamefont {Schlenker}}, \bibinfo {author}
  {\bibfnamefont {M.~G.}\ \bibnamefont {Vavilov}}, \bibinfo {author}
  {\bibfnamefont {B.~L.~T.}\ \bibnamefont {Plourde}}, \ and\ \bibinfo {author}
  {\bibfnamefont {R.}~\bibnamefont {McDermott}},\ }\bibfield  {title} {\enquote
  {\bibinfo {title} {Measurement of a superconducting qubit with a microwave
  photon counter},}\ }\href@noop {} {\bibfield  {journal} {\bibinfo  {journal}
  {Science}\ }\textbf {\bibinfo {volume} {361}},\ \bibinfo {pages}
  {1239--–1242} (\bibinfo {year} {2018})}\BibitemShut {NoStop}%
\bibitem [{\citenamefont {Eddins}\ \emph {et~al.}(2019)\citenamefont {Eddins},
  \citenamefont {Kreikebaum}, \citenamefont {Toyli}, \citenamefont
  {Levenson-Falk}, \citenamefont {Dove}, \citenamefont {Livingston},
  \citenamefont {Levitan}, \citenamefont {Govia}, \citenamefont {Clerk},\ and\
  \citenamefont {Siddiqi}}]{eddins:2019}%
  \BibitemOpen
  \bibfield  {author} {\bibinfo {author} {\bibfnamefont {A.}~\bibnamefont
  {Eddins}}, \bibinfo {author} {\bibfnamefont {J.~M.}\ \bibnamefont
  {Kreikebaum}}, \bibinfo {author} {\bibfnamefont {D.~M.}\ \bibnamefont
  {Toyli}}, \bibinfo {author} {\bibfnamefont {E.~M.}\ \bibnamefont
  {Levenson-Falk}}, \bibinfo {author} {\bibfnamefont {A.}~\bibnamefont {Dove}},
  \bibinfo {author} {\bibfnamefont {W.~P.}\ \bibnamefont {Livingston}},
  \bibinfo {author} {\bibfnamefont {B.~A.}\ \bibnamefont {Levitan}}, \bibinfo
  {author} {\bibfnamefont {L.~C.~G.}\ \bibnamefont {Govia}}, \bibinfo {author}
  {\bibfnamefont {A.~A.}\ \bibnamefont {Clerk}}, \ and\ \bibinfo {author}
  {\bibfnamefont {I.}~\bibnamefont {Siddiqi}},\ }\bibfield  {title} {\enquote
  {\bibinfo {title} {High-efficiency measurement of an artificial atom embedded
  in a parametric amplifier},}\ }\href {\doibase 10.1103/PhysRevX.9.011004}
  {\bibfield  {journal} {\bibinfo  {journal} {Phys. Rev. X}\ }\textbf {\bibinfo
  {volume} {9}},\ \bibinfo {pages} {011004} (\bibinfo {year}
  {2019})}\BibitemShut {NoStop}%
\bibitem [{\citenamefont {Opremcak}\ \emph {et~al.}(2021)\citenamefont
  {Opremcak}, \citenamefont {Liu}, \citenamefont {Wilen}, \citenamefont
  {Okubo}, \citenamefont {Christensen}, \citenamefont {Sank}, \citenamefont
  {White}, \citenamefont {Vainsencher}, \citenamefont {Giustina}, \citenamefont
  {Megrant}, \citenamefont {Burkett}, \citenamefont {Plourde},\ and\
  \citenamefont {McDermott}}]{opremcak:2020}%
  \BibitemOpen
  \bibfield  {author} {\bibinfo {author} {\bibfnamefont {A.}~\bibnamefont
  {Opremcak}}, \bibinfo {author} {\bibfnamefont {C.~H.}\ \bibnamefont {Liu}},
  \bibinfo {author} {\bibfnamefont {C.}~\bibnamefont {Wilen}}, \bibinfo
  {author} {\bibfnamefont {K.}~\bibnamefont {Okubo}}, \bibinfo {author}
  {\bibfnamefont {B.~G.}\ \bibnamefont {Christensen}}, \bibinfo {author}
  {\bibfnamefont {D.}~\bibnamefont {Sank}}, \bibinfo {author} {\bibfnamefont
  {T.~C.}\ \bibnamefont {White}}, \bibinfo {author} {\bibfnamefont
  {A.}~\bibnamefont {Vainsencher}}, \bibinfo {author} {\bibfnamefont
  {M.}~\bibnamefont {Giustina}}, \bibinfo {author} {\bibfnamefont
  {A.}~\bibnamefont {Megrant}}, \bibinfo {author} {\bibfnamefont
  {B.}~\bibnamefont {Burkett}}, \bibinfo {author} {\bibfnamefont {B.~L.~T.}\
  \bibnamefont {Plourde}}, \ and\ \bibinfo {author} {\bibfnamefont
  {R.}~\bibnamefont {McDermott}},\ }\bibfield  {title} {\enquote {\bibinfo
  {title} {High-fidelity measurement of a superconducting qubit using an
  on-chip microwave photon counter},}\ }\href {\doibase
  10.1103/PhysRevX.11.011027} {\bibfield  {journal} {\bibinfo  {journal} {Phys.
  Rev. X}\ }\textbf {\bibinfo {volume} {11}},\ \bibinfo {pages} {011027}
  (\bibinfo {year} {2021})}\BibitemShut {NoStop}%
\bibitem [{\citenamefont {Chapman}\ \emph {et~al.}(2016)\citenamefont
  {Chapman}, \citenamefont {Moores}, \citenamefont {Rosenthal}, \citenamefont
  {Kerckhoff},\ and\ \citenamefont {Lehnert}}]{chapman:2016}%
  \BibitemOpen
  \bibfield  {author} {\bibinfo {author} {\bibfnamefont {Benjamin~J.}\
  \bibnamefont {Chapman}}, \bibinfo {author} {\bibfnamefont {Bradley~A.}\
  \bibnamefont {Moores}}, \bibinfo {author} {\bibfnamefont {Eric~I.}\
  \bibnamefont {Rosenthal}}, \bibinfo {author} {\bibfnamefont {Joseph}\
  \bibnamefont {Kerckhoff}}, \ and\ \bibinfo {author} {\bibfnamefont {K.~W.}\
  \bibnamefont {Lehnert}},\ }\bibfield  {title} {\enquote {\bibinfo {title}
  {General purpose multiplexing device for cryogenic microwave systems},}\
  }\href@noop {} {\bibfield  {journal} {\bibinfo  {journal} {Applied Physics
  Letters}\ }\textbf {\bibinfo {volume} {108}},\ \bibinfo {pages} {222602}
  (\bibinfo {year} {2016})}\BibitemShut {NoStop}%
\bibitem [{\citenamefont {Chapman}\ \emph
  {et~al.}(2017{\natexlab{b}})\citenamefont {Chapman}, \citenamefont
  {Rosenthal}, \citenamefont {Kerckhoff}, \citenamefont {Vale}, \citenamefont
  {Hilton},\ and\ \citenamefont {Lehnert}}]{chapman:2017}%
  \BibitemOpen
  \bibfield  {author} {\bibinfo {author} {\bibfnamefont {Benjamin~J.}\
  \bibnamefont {Chapman}}, \bibinfo {author} {\bibfnamefont {Eric~I.}\
  \bibnamefont {Rosenthal}}, \bibinfo {author} {\bibfnamefont {Joseph}\
  \bibnamefont {Kerckhoff}}, \bibinfo {author} {\bibfnamefont {Leila~R.}\
  \bibnamefont {Vale}}, \bibinfo {author} {\bibfnamefont {Gene~C.}\
  \bibnamefont {Hilton}}, \ and\ \bibinfo {author} {\bibfnamefont {K.~W.}\
  \bibnamefont {Lehnert}},\ }\bibfield  {title} {\enquote {\bibinfo {title}
  {Single-sideband modulator for frequency domain multiplexing of
  superconducting qubit readout},}\ }\href@noop {} {\bibfield  {journal}
  {\bibinfo  {journal} {Applied Physics Letters}\ }\textbf {\bibinfo {volume}
  {110}},\ \bibinfo {pages} {162601} (\bibinfo {year}
  {2017}{\natexlab{b}})}\BibitemShut {NoStop}%
\bibitem [{\citenamefont {Malnou}\ \emph {et~al.}(2019)\citenamefont {Malnou},
  \citenamefont {Palken}, \citenamefont {Brubaker}, \citenamefont {Vale},
  \citenamefont {Hilton},\ and\ \citenamefont {Lehnert}}]{malnou:2019}%
  \BibitemOpen
  \bibfield  {author} {\bibinfo {author} {\bibfnamefont {M.}~\bibnamefont
  {Malnou}}, \bibinfo {author} {\bibfnamefont {D.~A.}\ \bibnamefont {Palken}},
  \bibinfo {author} {\bibfnamefont {B.~M.}\ \bibnamefont {Brubaker}}, \bibinfo
  {author} {\bibfnamefont {Leila~R.}\ \bibnamefont {Vale}}, \bibinfo {author}
  {\bibfnamefont {Gene~C.}\ \bibnamefont {Hilton}}, \ and\ \bibinfo {author}
  {\bibfnamefont {K.~W.}\ \bibnamefont {Lehnert}},\ }\bibfield  {title}
  {\enquote {\bibinfo {title} {Squeezed vacuum used to accelerate the search
  for a weak classical signal},}\ }\href {\doibase 10.1103/PhysRevX.9.021023}
  {\bibfield  {journal} {\bibinfo  {journal} {Phys. Rev. X}\ }\textbf {\bibinfo
  {volume} {9}},\ \bibinfo {pages} {021023} (\bibinfo {year}
  {2019})}\BibitemShut {NoStop}%
\bibitem [{\citenamefont {Naaman}\ \emph {et~al.}(2016)\citenamefont {Naaman},
  \citenamefont {Abutaleb}, \citenamefont {Kirby},\ and\ \citenamefont
  {Rennie}}]{naaman:2016switch}%
  \BibitemOpen
  \bibfield  {author} {\bibinfo {author} {\bibfnamefont {Ofer}\ \bibnamefont
  {Naaman}}, \bibinfo {author} {\bibfnamefont {M.~O.}\ \bibnamefont
  {Abutaleb}}, \bibinfo {author} {\bibfnamefont {Chris}\ \bibnamefont {Kirby}},
  \ and\ \bibinfo {author} {\bibfnamefont {Michael}\ \bibnamefont {Rennie}},\
  }\bibfield  {title} {\enquote {\bibinfo {title} {On-chip {J}osephson junction
  microwave switch},}\ }\href@noop {} {\bibfield  {journal} {\bibinfo
  {journal} {Applied Physics Letters}\ }\textbf {\bibinfo {volume} {108}},\
  \bibinfo {pages} {112601} (\bibinfo {year} {2016})}\BibitemShut {NoStop}%
\bibitem [{\citenamefont {Marchand}(1944)}]{marchand:1944}%
  \BibitemOpen
  \bibfield  {author} {\bibinfo {author} {\bibfnamefont {Nathan}\ \bibnamefont
  {Marchand}},\ }\bibfield  {title} {\enquote {\bibinfo {title}
  {Transmission-line conversion transformers},}\ }\href@noop {} {\bibfield
  {journal} {\bibinfo  {journal} {Electronics}\ }\textbf {\bibinfo {volume}
  {17}},\ \bibinfo {pages} {142--145} (\bibinfo {year} {1944})}\BibitemShut
  {NoStop}%
\bibitem [{\citenamefont {Mates}\ \emph {et~al.}(2008)\citenamefont {Mates},
  \citenamefont {Hilton}, \citenamefont {Irwin}, \citenamefont {Vale},\ and\
  \citenamefont {Lehnert}}]{mates:2008}%
  \BibitemOpen
  \bibfield  {author} {\bibinfo {author} {\bibfnamefont {J.~A.~B.}\
  \bibnamefont {Mates}}, \bibinfo {author} {\bibfnamefont {G.~C.}\ \bibnamefont
  {Hilton}}, \bibinfo {author} {\bibfnamefont {K.~D.}\ \bibnamefont {Irwin}},
  \bibinfo {author} {\bibfnamefont {L.~R.}\ \bibnamefont {Vale}}, \ and\
  \bibinfo {author} {\bibfnamefont {K.~W.}\ \bibnamefont {Lehnert}},\
  }\bibfield  {title} {\enquote {\bibinfo {title} {Demonstration of a
  multiplexer of dissipationless superconducting quantum interference
  devices},}\ }\href@noop {} {\bibfield  {journal} {\bibinfo  {journal}
  {Applied Physics Letters}\ }\textbf {\bibinfo {volume} {92}},\ \bibinfo
  {pages} {023514} (\bibinfo {year} {2008})}\BibitemShut {NoStop}%
\bibitem [{\citenamefont {Reed}\ \emph
  {et~al.}(2010{\natexlab{a}})\citenamefont {Reed}, \citenamefont {DiCarlo},
  \citenamefont {Johnson}, \citenamefont {Sun}, \citenamefont {Schuster},
  \citenamefont {Frunzio},\ and\ \citenamefont {Schoelkopf}}]{reed:2010}%
  \BibitemOpen
  \bibfield  {author} {\bibinfo {author} {\bibfnamefont {M.~D.}\ \bibnamefont
  {Reed}}, \bibinfo {author} {\bibfnamefont {L.}~\bibnamefont {DiCarlo}},
  \bibinfo {author} {\bibfnamefont {B.~R.}\ \bibnamefont {Johnson}}, \bibinfo
  {author} {\bibfnamefont {L.}~\bibnamefont {Sun}}, \bibinfo {author}
  {\bibfnamefont {D.~I.}\ \bibnamefont {Schuster}}, \bibinfo {author}
  {\bibfnamefont {L.}~\bibnamefont {Frunzio}}, \ and\ \bibinfo {author}
  {\bibfnamefont {R.~J.}\ \bibnamefont {Schoelkopf}},\ }\bibfield  {title}
  {\enquote {\bibinfo {title} {High-fidelity readout in circuit quantum
  electrodynamics using the jaynes-cummings nonlinearity},}\ }\href {\doibase
  10.1103/PhysRevLett.105.173601} {\bibfield  {journal} {\bibinfo  {journal}
  {Phys. Rev. Lett.}\ }\textbf {\bibinfo {volume} {105}},\ \bibinfo {pages}
  {173601} (\bibinfo {year} {2010}{\natexlab{a}})}\BibitemShut {NoStop}%
\bibitem [{\citenamefont {Malnou}\ \emph {et~al.}(2018)\citenamefont {Malnou},
  \citenamefont {Palken}, \citenamefont {Vale}, \citenamefont {Hilton},\ and\
  \citenamefont {Lehnert}}]{malnou:2018}%
  \BibitemOpen
  \bibfield  {author} {\bibinfo {author} {\bibfnamefont {M.}~\bibnamefont
  {Malnou}}, \bibinfo {author} {\bibfnamefont {D.~A.}\ \bibnamefont {Palken}},
  \bibinfo {author} {\bibfnamefont {Leila~R.}\ \bibnamefont {Vale}}, \bibinfo
  {author} {\bibfnamefont {Gene~C.}\ \bibnamefont {Hilton}}, \ and\ \bibinfo
  {author} {\bibfnamefont {K.~W.}\ \bibnamefont {Lehnert}},\ }\bibfield
  {title} {\enquote {\bibinfo {title} {Optimal operation of a josephson
  parametric amplifier for vacuum squeezing},}\ }\href {\doibase
  10.1103/PhysRevApplied.9.044023} {\bibfield  {journal} {\bibinfo  {journal}
  {Phys. Rev. Applied}\ }\textbf {\bibinfo {volume} {9}},\ \bibinfo {pages}
  {044023} (\bibinfo {year} {2018})}\BibitemShut {NoStop}%
\bibitem [{\citenamefont {Gambetta}\ \emph {et~al.}(2007)\citenamefont
  {Gambetta}, \citenamefont {Braff}, \citenamefont {Wallraff}, \citenamefont
  {Girvin},\ and\ \citenamefont {Schoelkopf}}]{gambetta:2007}%
  \BibitemOpen
  \bibfield  {author} {\bibinfo {author} {\bibfnamefont {Jay}\ \bibnamefont
  {Gambetta}}, \bibinfo {author} {\bibfnamefont {W.~A.}\ \bibnamefont {Braff}},
  \bibinfo {author} {\bibfnamefont {A.}~\bibnamefont {Wallraff}}, \bibinfo
  {author} {\bibfnamefont {S.~M.}\ \bibnamefont {Girvin}}, \ and\ \bibinfo
  {author} {\bibfnamefont {R.~J.}\ \bibnamefont {Schoelkopf}},\ }\bibfield
  {title} {\enquote {\bibinfo {title} {Protocols for optimal readout of qubits
  using a continuous quantum nondemolition measurement},}\ }\href {\doibase
  10.1103/PhysRevA.76.012325} {\bibfield  {journal} {\bibinfo  {journal} {Phys.
  Rev. A}\ }\textbf {\bibinfo {volume} {76}},\ \bibinfo {pages} {012325}
  (\bibinfo {year} {2007})}\BibitemShut {NoStop}%
\bibitem [{\citenamefont {Wustmann}\ and\ \citenamefont
  {Shumeiko}(2013)}]{wustmann:2013}%
  \BibitemOpen
  \bibfield  {author} {\bibinfo {author} {\bibfnamefont {Waltraut}\
  \bibnamefont {Wustmann}}\ and\ \bibinfo {author} {\bibfnamefont {Vitaly}\
  \bibnamefont {Shumeiko}},\ }\bibfield  {title} {\enquote {\bibinfo {title}
  {Parametric resonance in tunable superconducting cavities},}\ }\href
  {\doibase 10.1103/PhysRevB.87.184501} {\bibfield  {journal} {\bibinfo
  {journal} {Phys. Rev. B}\ }\textbf {\bibinfo {volume} {87}},\ \bibinfo
  {pages} {184501} (\bibinfo {year} {2013})}\BibitemShut {NoStop}%
\bibitem [{\citenamefont {Krantz}\ \emph {et~al.}(2013)\citenamefont {Krantz},
  \citenamefont {Reshitnyk}, \citenamefont {Wustmann}, \citenamefont
  {Bylander}, \citenamefont {Gustavsson}, \citenamefont {Oliver}, \citenamefont
  {Duty}, \citenamefont {Shumeiko},\ and\ \citenamefont
  {Delsing}}]{krantz:2013}%
  \BibitemOpen
  \bibfield  {author} {\bibinfo {author} {\bibfnamefont {P.}~\bibnamefont
  {Krantz}}, \bibinfo {author} {\bibfnamefont {Y.}~\bibnamefont {Reshitnyk}},
  \bibinfo {author} {\bibfnamefont {W.}~\bibnamefont {Wustmann}}, \bibinfo
  {author} {\bibfnamefont {J.}~\bibnamefont {Bylander}}, \bibinfo {author}
  {\bibfnamefont {S.}~\bibnamefont {Gustavsson}}, \bibinfo {author}
  {\bibfnamefont {W.~D.}\ \bibnamefont {Oliver}}, \bibinfo {author}
  {\bibfnamefont {T.}~\bibnamefont {Duty}}, \bibinfo {author} {\bibfnamefont
  {V.}~\bibnamefont {Shumeiko}}, \ and\ \bibinfo {author} {\bibfnamefont
  {P.}~\bibnamefont {Delsing}},\ }\bibfield  {title} {\enquote {\bibinfo
  {title} {Investigation of nonlinear effects in josephson parametric
  oscillators used in circuit quantum electrodynamics},}\ }\href@noop {}
  {\bibfield  {journal} {\bibinfo  {journal} {New J. Phys.}\ }\textbf {\bibinfo
  {volume} {15}},\ \bibinfo {pages} {105002} (\bibinfo {year}
  {2013})}\BibitemShut {NoStop}%
\bibitem [{\citenamefont {Wustmann}\ and\ \citenamefont
  {Shumeiko}(2019)}]{wustmann:2019}%
  \BibitemOpen
  \bibfield  {author} {\bibinfo {author} {\bibfnamefont {Waltraut}\
  \bibnamefont {Wustmann}}\ and\ \bibinfo {author} {\bibfnamefont {Vitaly}\
  \bibnamefont {Shumeiko}},\ }\bibfield  {title} {\enquote {\bibinfo {title}
  {Parametric effects in circuit quantum electrodynamics},}\ }\href@noop {}
  {\bibfield  {journal} {\bibinfo  {journal} {Low Temperature Physics}\
  }\textbf {\bibinfo {volume} {45}},\ \bibinfo {pages} {848} (\bibinfo {year}
  {2019})}\BibitemShut {NoStop}%
\bibitem [{\citenamefont {Van~Duzer}\ and\ \citenamefont
  {Turner}(1981)}]{vanduzer:1981}%
  \BibitemOpen
  \bibfield  {author} {\bibinfo {author} {\bibfnamefont {Theodore}\
  \bibnamefont {Van~Duzer}}\ and\ \bibinfo {author} {\bibfnamefont
  {Charles~William}\ \bibnamefont {Turner}},\ }\href@noop {} {\emph {\bibinfo
  {title} {Principles of superconductive devices and circuits}}},\ \bibinfo
  {edition} {2nd}\ ed.\ (\bibinfo  {publisher} {Prentice Hall},\ \bibinfo
  {year} {1981})\BibitemShut {NoStop}%
\bibitem [{\citenamefont {Siddiqi}\ \emph {et~al.}(2004)\citenamefont
  {Siddiqi}, \citenamefont {Vijay}, \citenamefont {Pierre}, \citenamefont
  {Wilson}, \citenamefont {Metcalfe}, \citenamefont {Rigetti}, \citenamefont
  {Frunzio},\ and\ \citenamefont {Devoret}}]{siddiqi:2004}%
  \BibitemOpen
  \bibfield  {author} {\bibinfo {author} {\bibfnamefont {I.}~\bibnamefont
  {Siddiqi}}, \bibinfo {author} {\bibfnamefont {R.}~\bibnamefont {Vijay}},
  \bibinfo {author} {\bibfnamefont {F.}~\bibnamefont {Pierre}}, \bibinfo
  {author} {\bibfnamefont {C.~M.}\ \bibnamefont {Wilson}}, \bibinfo {author}
  {\bibfnamefont {M.}~\bibnamefont {Metcalfe}}, \bibinfo {author}
  {\bibfnamefont {C.}~\bibnamefont {Rigetti}}, \bibinfo {author} {\bibfnamefont
  {L.}~\bibnamefont {Frunzio}}, \ and\ \bibinfo {author} {\bibfnamefont
  {M.~H.}\ \bibnamefont {Devoret}},\ }\bibfield  {title} {\enquote {\bibinfo
  {title} {Rf-driven josephson bifurcation amplifier for quantum
  measurement},}\ }\href {\doibase 10.1103/PhysRevLett.93.207002} {\bibfield
  {journal} {\bibinfo  {journal} {Phys. Rev. Lett.}\ }\textbf {\bibinfo
  {volume} {93}},\ \bibinfo {pages} {207002} (\bibinfo {year}
  {2004})}\BibitemShut {NoStop}%
\bibitem [{\citenamefont {Manucharyan}\ \emph {et~al.}(2007)\citenamefont
  {Manucharyan}, \citenamefont {Boaknin}, \citenamefont {Metcalfe},
  \citenamefont {Vijay}, \citenamefont {Siddiqi},\ and\ \citenamefont
  {Devoret}}]{manucharyan:2007}%
  \BibitemOpen
  \bibfield  {author} {\bibinfo {author} {\bibfnamefont {V.~E.}\ \bibnamefont
  {Manucharyan}}, \bibinfo {author} {\bibfnamefont {E.}~\bibnamefont
  {Boaknin}}, \bibinfo {author} {\bibfnamefont {M.}~\bibnamefont {Metcalfe}},
  \bibinfo {author} {\bibfnamefont {R.}~\bibnamefont {Vijay}}, \bibinfo
  {author} {\bibfnamefont {I.}~\bibnamefont {Siddiqi}}, \ and\ \bibinfo
  {author} {\bibfnamefont {M.}~\bibnamefont {Devoret}},\ }\bibfield  {title}
  {\enquote {\bibinfo {title} {Microwave bifurcation of a {J}osephson junction:
  Embedding-circuit requirements},}\ }\href {\doibase
  10.1103/PhysRevB.76.014524} {\bibfield  {journal} {\bibinfo  {journal} {Phys.
  Rev. B}\ }\textbf {\bibinfo {volume} {76}},\ \bibinfo {pages} {014524}
  (\bibinfo {year} {2007})}\BibitemShut {NoStop}%
\bibitem [{\citenamefont {Lin}\ \emph {et~al.}(2014)\citenamefont {Lin},
  \citenamefont {Inomata}, \citenamefont {Koshino}, \citenamefont {Oliver},
  \citenamefont {Nakamura}, \citenamefont {Tsai},\ and\ \citenamefont
  {Yamamoto}}]{lin:2014}%
  \BibitemOpen
  \bibfield  {author} {\bibinfo {author} {\bibfnamefont {Z.~R.}\ \bibnamefont
  {Lin}}, \bibinfo {author} {\bibfnamefont {K.}~\bibnamefont {Inomata}},
  \bibinfo {author} {\bibfnamefont {K.}~\bibnamefont {Koshino}}, \bibinfo
  {author} {\bibfnamefont {W.~D.}\ \bibnamefont {Oliver}}, \bibinfo {author}
  {\bibfnamefont {Y.}~\bibnamefont {Nakamura}}, \bibinfo {author}
  {\bibfnamefont {J.~S.}\ \bibnamefont {Tsai}}, \ and\ \bibinfo {author}
  {\bibfnamefont {T.}~\bibnamefont {Yamamoto}},\ }\bibfield  {title} {\enquote
  {\bibinfo {title} {Josephson parametric phase-locked oscillator and its
  application to dispersive readout of superconducting qubits},}\ }\href@noop
  {} {\bibfield  {journal} {\bibinfo  {journal} {Nature communications}\
  }\textbf {\bibinfo {volume} {5}},\ \bibinfo {pages} {4480} (\bibinfo {year}
  {2014})}\BibitemShut {NoStop}%
\bibitem [{\citenamefont {Krantz}\ \emph {et~al.}(2016)\citenamefont {Krantz},
  \citenamefont {Bengtsson}, \citenamefont {Simoen}, \citenamefont
  {Gustavsson}, \citenamefont {Shumeiko}, \citenamefont {Oliver}, \citenamefont
  {Wilson}, \citenamefont {Delsing},\ and\ \citenamefont
  {Bylander}}]{krantz:2016}%
  \BibitemOpen
  \bibfield  {author} {\bibinfo {author} {\bibfnamefont {P.}~\bibnamefont
  {Krantz}}, \bibinfo {author} {\bibfnamefont {A.}~\bibnamefont {Bengtsson}},
  \bibinfo {author} {\bibfnamefont {M.}~\bibnamefont {Simoen}}, \bibinfo
  {author} {\bibfnamefont {S.}~\bibnamefont {Gustavsson}}, \bibinfo {author}
  {\bibfnamefont {V.}~\bibnamefont {Shumeiko}}, \bibinfo {author}
  {\bibfnamefont {W.~D.}\ \bibnamefont {Oliver}}, \bibinfo {author}
  {\bibfnamefont {C.~M.}\ \bibnamefont {Wilson}}, \bibinfo {author}
  {\bibfnamefont {P.}~\bibnamefont {Delsing}}, \ and\ \bibinfo {author}
  {\bibfnamefont {B.}~\bibnamefont {Bylander}},\ }\bibfield  {title} {\enquote
  {\bibinfo {title} {Single-shot read-out of a superconducting qubit using a
  josephson parametric oscillator},}\ }\href@noop {} {\bibfield  {journal}
  {\bibinfo  {journal} {Nature communications}\ }\textbf {\bibinfo {volume}
  {7}},\ \bibinfo {pages} {11417} (\bibinfo {year} {2016})}\BibitemShut
  {NoStop}%
\bibitem [{\citenamefont {Andersen}\ \emph {et~al.}(2016)\citenamefont
  {Andersen}, \citenamefont {Kerckhoff}, \citenamefont {Lehnert}, \citenamefont
  {Chapman},\ and\ \citenamefont {M\o{}lmer}}]{andersen:2016}%
  \BibitemOpen
  \bibfield  {author} {\bibinfo {author} {\bibfnamefont {Christian~Kraglund}\
  \bibnamefont {Andersen}}, \bibinfo {author} {\bibfnamefont {Joseph}\
  \bibnamefont {Kerckhoff}}, \bibinfo {author} {\bibfnamefont {Konrad~W.}\
  \bibnamefont {Lehnert}}, \bibinfo {author} {\bibfnamefont {Benjamin~J.}\
  \bibnamefont {Chapman}}, \ and\ \bibinfo {author} {\bibfnamefont {Klaus}\
  \bibnamefont {M\o{}lmer}},\ }\bibfield  {title} {\enquote {\bibinfo {title}
  {Closing a quantum feedback loop inside a cryostat: Autonomous state
  preparation and long-time memory of a superconducting qubit},}\ }\href
  {\doibase 10.1103/PhysRevA.93.012346} {\bibfield  {journal} {\bibinfo
  {journal} {Phys. Rev. A}\ }\textbf {\bibinfo {volume} {93}},\ \bibinfo
  {pages} {012346} (\bibinfo {year} {2016})}\BibitemShut {NoStop}%
\bibitem [{\citenamefont {Clerk}\ \emph {et~al.}(2010)\citenamefont {Clerk},
  \citenamefont {Devoret}, \citenamefont {Girvin}, \citenamefont {Marquardt},\
  and\ \citenamefont {Schoelkopf}}]{clerk:2010}%
  \BibitemOpen
  \bibfield  {author} {\bibinfo {author} {\bibfnamefont {A.~A.}\ \bibnamefont
  {Clerk}}, \bibinfo {author} {\bibfnamefont {M.~H.}\ \bibnamefont {Devoret}},
  \bibinfo {author} {\bibfnamefont {S.~M.}\ \bibnamefont {Girvin}}, \bibinfo
  {author} {\bibfnamefont {Florian}\ \bibnamefont {Marquardt}}, \ and\ \bibinfo
  {author} {\bibfnamefont {R.~J.}\ \bibnamefont {Schoelkopf}},\ }\bibfield
  {title} {\enquote {\bibinfo {title} {Introduction to quantum noise,
  measurement, and amplification},}\ }\href@noop {} {\bibfield  {journal}
  {\bibinfo  {journal} {Reviews of Modern Physics}\ }\textbf {\bibinfo {volume}
  {82}},\ \bibinfo {pages} {1155} (\bibinfo {year} {2010})}\BibitemShut
  {NoStop}%
\bibitem [{\citenamefont {Kraus}(1971)}]{kraus:1971}%
  \BibitemOpen
  \bibfield  {author} {\bibinfo {author} {\bibfnamefont {K.}~\bibnamefont
  {Kraus}},\ }\bibfield  {title} {\enquote {\bibinfo {title} {General state
  changes in quantum theory},}\ }\href@noop {} {\bibfield  {journal} {\bibinfo
  {journal} {Ann. Phys.}\ }\textbf {\bibinfo {volume} {64}},\ \bibinfo {pages}
  {311--335} (\bibinfo {year} {1971})}\BibitemShut {NoStop}%
\bibitem [{\citenamefont {Loudon}(2000)}]{loudon:2000}%
  \BibitemOpen
  \bibfield  {author} {\bibinfo {author} {\bibfnamefont {Rodney}\ \bibnamefont
  {Loudon}},\ }\href@noop {} {\emph {\bibinfo {title} {The Quantum Theory of
  Light}}},\ \bibinfo {edition} {3rd}\ ed.\ (\bibinfo  {publisher} {Oxford
  Science Publications},\ \bibinfo {year} {2000})\BibitemShut {NoStop}%
\bibitem [{\citenamefont {Gerry}\ and\ \citenamefont
  {Knight}(2004)}]{gerry:2004}%
  \BibitemOpen
  \bibfield  {author} {\bibinfo {author} {\bibfnamefont {Christopher}\
  \bibnamefont {Gerry}}\ and\ \bibinfo {author} {\bibfnamefont {Peter}\
  \bibnamefont {Knight}},\ }\href@noop {} {\emph {\bibinfo {title}
  {Introductory Quantum Optics}}}\ (\bibinfo  {publisher} {Cambridge University
  Press},\ \bibinfo {year} {2004})\BibitemShut {NoStop}%
\bibitem [{\citenamefont {Helstrom}(1976)}]{helstrom:1976}%
  \BibitemOpen
  \bibfield  {author} {\bibinfo {author} {\bibfnamefont {C.W.}\ \bibnamefont
  {Helstrom}},\ }\href@noop {} {\emph {\bibinfo {title} {Quantum detection and
  estimation theory}}}\ (\bibinfo  {publisher} {Academic Press: New York},\
  \bibinfo {year} {1976})\BibitemShut {NoStop}%
\bibitem [{\citenamefont {Caves}(1982)}]{caves:1982}%
  \BibitemOpen
  \bibfield  {author} {\bibinfo {author} {\bibfnamefont {Carlton~M.}\
  \bibnamefont {Caves}},\ }\bibfield  {title} {\enquote {\bibinfo {title}
  {Quantum limits on noise in linear amplifiers},}\ }\href {\doibase
  10.1103/PhysRevD.26.1817} {\bibfield  {journal} {\bibinfo  {journal} {Phys.
  Rev. D}\ }\textbf {\bibinfo {volume} {26}},\ \bibinfo {pages} {1817--1839}
  (\bibinfo {year} {1982})}\BibitemShut {NoStop}%
\bibitem [{\citenamefont {Clerk}\ \emph {et~al.}(2003)\citenamefont {Clerk},
  \citenamefont {Girvin},\ and\ \citenamefont {Stone}}]{clerk:2003}%
  \BibitemOpen
  \bibfield  {author} {\bibinfo {author} {\bibfnamefont {A.~A.}\ \bibnamefont
  {Clerk}}, \bibinfo {author} {\bibfnamefont {S.~M.}\ \bibnamefont {Girvin}}, \
  and\ \bibinfo {author} {\bibfnamefont {A.~D.}\ \bibnamefont {Stone}},\
  }\bibfield  {title} {\enquote {\bibinfo {title} {Quantum-limited measurement
  and information in mesoscopic detectors},}\ }\href {\doibase
  10.1103/PhysRevB.67.165324} {\bibfield  {journal} {\bibinfo  {journal} {Phys.
  Rev. B}\ }\textbf {\bibinfo {volume} {67}},\ \bibinfo {pages} {165324}
  (\bibinfo {year} {2003})}\BibitemShut {NoStop}%
\bibitem [{\citenamefont {Korotkov}(2016)}]{korotkov:2016}%
  \BibitemOpen
  \bibfield  {author} {\bibinfo {author} {\bibfnamefont {Alexander~N.}\
  \bibnamefont {Korotkov}},\ }\bibfield  {title} {\enquote {\bibinfo {title}
  {Quantum bayesian approach to circuit qed measurement with moderate
  bandwidth},}\ }\href {\doibase 10.1103/PhysRevA.94.042326} {\bibfield
  {journal} {\bibinfo  {journal} {Phys. Rev. A}\ }\textbf {\bibinfo {volume}
  {94}},\ \bibinfo {pages} {042326} (\bibinfo {year} {2016})}\BibitemShut
  {NoStop}%
\bibitem [{\citenamefont {Han}\ \emph {et~al.}(2018)\citenamefont {Han},
  \citenamefont {Leuchs},\ and\ \citenamefont {Grassl}}]{han:2018}%
  \BibitemOpen
  \bibfield  {author} {\bibinfo {author} {\bibfnamefont {Rui}\ \bibnamefont
  {Han}}, \bibinfo {author} {\bibfnamefont {Gerd}\ \bibnamefont {Leuchs}}, \
  and\ \bibinfo {author} {\bibfnamefont {Markus}\ \bibnamefont {Grassl}},\
  }\bibfield  {title} {\enquote {\bibinfo {title} {Residual and destroyed
  accessible information after measurements},}\ }\href {\doibase
  10.1103/PhysRevLett.120.160501} {\bibfield  {journal} {\bibinfo  {journal}
  {Phys. Rev. Lett.}\ }\textbf {\bibinfo {volume} {120}},\ \bibinfo {pages}
  {160501} (\bibinfo {year} {2018})}\BibitemShut {NoStop}%
\bibitem [{\citenamefont {Bultink}\ \emph {et~al.}(2018)\citenamefont
  {Bultink}, \citenamefont {Tarasinski}, \citenamefont {Haandb{\ae}k},
  \citenamefont {Poletto}, \citenamefont {Haider}, \citenamefont {Michalak},
  \citenamefont {Bruno},\ and\ \citenamefont {DiCarlo}}]{bultink:2018}%
  \BibitemOpen
  \bibfield  {author} {\bibinfo {author} {\bibfnamefont {C.~C.}\ \bibnamefont
  {Bultink}}, \bibinfo {author} {\bibfnamefont {B.}~\bibnamefont {Tarasinski}},
  \bibinfo {author} {\bibfnamefont {N.}~\bibnamefont {Haandb{\ae}k}}, \bibinfo
  {author} {\bibfnamefont {S.}~\bibnamefont {Poletto}}, \bibinfo {author}
  {\bibfnamefont {N.}~\bibnamefont {Haider}}, \bibinfo {author} {\bibfnamefont
  {D.~J.}\ \bibnamefont {Michalak}}, \bibinfo {author} {\bibfnamefont
  {A.}~\bibnamefont {Bruno}}, \ and\ \bibinfo {author} {\bibfnamefont
  {L.}~\bibnamefont {DiCarlo}},\ }\bibfield  {title} {\enquote {\bibinfo
  {title} {General method for extracting the quantum efficiency of dispersive
  qubit readout in circuit qed},}\ }\href@noop {} {\bibfield  {journal}
  {\bibinfo  {journal} {Applied Physics Letters}\ }\textbf {\bibinfo {volume}
  {112}},\ \bibinfo {pages} {092601} (\bibinfo {year} {2018})}\BibitemShut
  {NoStop}%
\bibitem [{Note1()}]{Note1}%
  \BibitemOpen
  \bibinfo {note} {For $\protect \sqrt {n_r} \gtrsim 0.2$ our models for both
  $|\protect \hat {\rho }_{01}'|$ and $F_r$ generally fall within the 95\%
  confidence interval of the measurement, and so only these points are used to
  obtain efficiency. This choice conservatively affects the reported
  efficiency: including the first four points, or instead fitting the `pump on'
  data, returns a larger value for $\eta $. We note that in this experiment,
  $\eta $ must be limited to 78\% or less based on an independent measurement
  of loss in the parametric cavity and a model for how this loss affects the
  efficiency (see Supplementary Material Sec. VI.A).}\BibitemShut {Stop}%
\bibitem [{\citenamefont {Walls}\ and\ \citenamefont
  {Milburn}(1994)}]{walls:1994}%
  \BibitemOpen
  \bibfield  {author} {\bibinfo {author} {\bibfnamefont {D.~F.}\ \bibnamefont
  {Walls}}\ and\ \bibinfo {author} {\bibfnamefont {G.~J.}\ \bibnamefont
  {Milburn}},\ }\href@noop {} {\emph {\bibinfo {title} {Quantum Optics}}}\
  (\bibinfo  {publisher} {Springer, Berlin},\ \bibinfo {year}
  {1994})\BibitemShut {NoStop}%
\bibitem [{\citenamefont {Braginsky}\ and\ \citenamefont
  {Khalili}(1996)}]{braginsky:1996}%
  \BibitemOpen
  \bibfield  {author} {\bibinfo {author} {\bibfnamefont {V.~B.}\ \bibnamefont
  {Braginsky}}\ and\ \bibinfo {author} {\bibfnamefont {F.~Ya.}\ \bibnamefont
  {Khalili}},\ }\bibfield  {title} {\enquote {\bibinfo {title} {Quantum
  nondemolition measurements: the route from toys to tools},}\ }\href {\doibase
  10.1103/RevModPhys.68.1} {\bibfield  {journal} {\bibinfo  {journal} {Rev.
  Mod. Phys.}\ }\textbf {\bibinfo {volume} {68}},\ \bibinfo {pages} {1--11}
  (\bibinfo {year} {1996})}\BibitemShut {NoStop}%
\bibitem [{\citenamefont {Lupa\c{s}cu}\ \emph {et~al.}(2007)\citenamefont
  {Lupa\c{s}cu}, \citenamefont {Saito}, \citenamefont {Picot}, \citenamefont
  {De~Groot}, \citenamefont {Harmans},\ and\ \citenamefont
  {Mooij}}]{lupascu:2007}%
  \BibitemOpen
  \bibfield  {author} {\bibinfo {author} {\bibfnamefont {A.}~\bibnamefont
  {Lupa\c{s}cu}}, \bibinfo {author} {\bibfnamefont {S.}~\bibnamefont {Saito}},
  \bibinfo {author} {\bibfnamefont {T.}~\bibnamefont {Picot}}, \bibinfo
  {author} {\bibfnamefont {P.~C.}\ \bibnamefont {De~Groot}}, \bibinfo {author}
  {\bibfnamefont {C.~J. P.~M.}\ \bibnamefont {Harmans}}, \ and\ \bibinfo
  {author} {\bibfnamefont {J.~E.}\ \bibnamefont {Mooij}},\ }\bibfield  {title}
  {\enquote {\bibinfo {title} {Quantum non-demolition measurement of a
  superconducting two-level system},}\ }\href@noop {} {\bibfield  {journal}
  {\bibinfo  {journal} {Nature Physics}\ }\textbf {\bibinfo {volume} {3}},\
  \bibinfo {pages} {119--123} (\bibinfo {year} {2007})}\BibitemShut {NoStop}%
\bibitem [{\citenamefont {Gambetta}\ \emph {et~al.}(2008)\citenamefont
  {Gambetta}, \citenamefont {Blais}, \citenamefont {Boissonneault},
  \citenamefont {Houck}, \citenamefont {Schuster},\ and\ \citenamefont
  {Girvin}}]{gambetta:2008}%
  \BibitemOpen
  \bibfield  {author} {\bibinfo {author} {\bibfnamefont {Jay}\ \bibnamefont
  {Gambetta}}, \bibinfo {author} {\bibfnamefont {Alexandre}\ \bibnamefont
  {Blais}}, \bibinfo {author} {\bibfnamefont {M.}~\bibnamefont
  {Boissonneault}}, \bibinfo {author} {\bibfnamefont {A.~A.}\ \bibnamefont
  {Houck}}, \bibinfo {author} {\bibfnamefont {D.~I.}\ \bibnamefont {Schuster}},
  \ and\ \bibinfo {author} {\bibfnamefont {S.~M.}\ \bibnamefont {Girvin}},\
  }\bibfield  {title} {\enquote {\bibinfo {title} {Quantum trajectory approach
  to circuit qed: Quantum jumps and the zeno effect},}\ }\href {\doibase
  10.1103/PhysRevA.77.012112} {\bibfield  {journal} {\bibinfo  {journal} {Phys.
  Rev. A}\ }\textbf {\bibinfo {volume} {77}},\ \bibinfo {pages} {012112}
  (\bibinfo {year} {2008})}\BibitemShut {NoStop}%
\bibitem [{\citenamefont {O'Connell}\ \emph {et~al.}(2008)\citenamefont
  {O'Connell}, \citenamefont {Ansmann}, \citenamefont {Bialczak}, \citenamefont
  {Hofheinz}, \citenamefont {Katz}, \citenamefont {Lucero}, \citenamefont
  {McKenney}, \citenamefont {Neeley}, \citenamefont {Wang}, \citenamefont
  {Weig}, \citenamefont {Cleland},\ and\ \citenamefont
  {Martinis}}]{oconnell:2008}%
  \BibitemOpen
  \bibfield  {author} {\bibinfo {author} {\bibfnamefont {Aaron~D.}\
  \bibnamefont {O'Connell}}, \bibinfo {author} {\bibfnamefont {M.}~\bibnamefont
  {Ansmann}}, \bibinfo {author} {\bibfnamefont {R.~C.}\ \bibnamefont
  {Bialczak}}, \bibinfo {author} {\bibfnamefont {M.}~\bibnamefont {Hofheinz}},
  \bibinfo {author} {\bibfnamefont {N.}~\bibnamefont {Katz}}, \bibinfo {author}
  {\bibfnamefont {Erik}\ \bibnamefont {Lucero}}, \bibinfo {author}
  {\bibfnamefont {C.}~\bibnamefont {McKenney}}, \bibinfo {author}
  {\bibfnamefont {M.}~\bibnamefont {Neeley}}, \bibinfo {author} {\bibfnamefont
  {H.}~\bibnamefont {Wang}}, \bibinfo {author} {\bibfnamefont {E.~M.}\
  \bibnamefont {Weig}}, \bibinfo {author} {\bibfnamefont {A.~N.}\ \bibnamefont
  {Cleland}}, \ and\ \bibinfo {author} {\bibfnamefont {J.~M.}\ \bibnamefont
  {Martinis}},\ }\bibfield  {title} {\enquote {\bibinfo {title} {Microwave
  dielectric loss at single photon energies and millikelvin temperatures},}\
  }\href@noop {} {\bibfield  {journal} {\bibinfo  {journal} {Applied Physics
  Letters}\ }\textbf {\bibinfo {volume} {92}},\ \bibinfo {pages} {112903}
  (\bibinfo {year} {2008})}\BibitemShut {NoStop}%
\bibitem [{\citenamefont {Reed}\ \emph
  {et~al.}(2010{\natexlab{b}})\citenamefont {Reed}, \citenamefont {Johnson},
  \citenamefont {Houck}, \citenamefont {DiCarlo}, \citenamefont {M.},
  \citenamefont {Schuster}, \citenamefont {Frunzio},\ and\ \citenamefont
  {Schoelkopf}}]{reed:2010b}%
  \BibitemOpen
  \bibfield  {author} {\bibinfo {author} {\bibfnamefont {M.~D.}\ \bibnamefont
  {Reed}}, \bibinfo {author} {\bibfnamefont {B.~R.}\ \bibnamefont {Johnson}},
  \bibinfo {author} {\bibfnamefont {A.~A.}\ \bibnamefont {Houck}}, \bibinfo
  {author} {\bibfnamefont {L.}~\bibnamefont {DiCarlo}}, \bibinfo {author}
  {\bibfnamefont {Chow~J.}\ \bibnamefont {M.}}, \bibinfo {author}
  {\bibfnamefont {D.~I.}\ \bibnamefont {Schuster}}, \bibinfo {author}
  {\bibfnamefont {L.}~\bibnamefont {Frunzio}}, \ and\ \bibinfo {author}
  {\bibfnamefont {R.~J.}\ \bibnamefont {Schoelkopf}},\ }\bibfield  {title}
  {\enquote {\bibinfo {title} {Fast reset and suppressing spontaneous emission
  of a superconducting qubit},}\ }\href@noop {} {\bibfield  {journal} {\bibinfo
   {journal} {Applied Physics Letters}\ }\textbf {\bibinfo {volume} {96}},\
  \bibinfo {pages} {203110} (\bibinfo {year} {2010}{\natexlab{b}})}\BibitemShut
  {NoStop}%
\bibitem [{\citenamefont {Paik}\ \emph {et~al.}(2011)\citenamefont {Paik},
  \citenamefont {Schuster}, \citenamefont {Bishop}, \citenamefont {Kirchmair},
  \citenamefont {Catelani}, \citenamefont {Sears}, \citenamefont {Johnson},
  \citenamefont {Reagor}, \citenamefont {Frunzio}, \citenamefont {Glazman},
  \citenamefont {Girvin}, \citenamefont {Devoret},\ and\ \citenamefont
  {Schoelkopf}}]{paik:2011}%
  \BibitemOpen
  \bibfield  {author} {\bibinfo {author} {\bibfnamefont {Hanhee}\ \bibnamefont
  {Paik}}, \bibinfo {author} {\bibfnamefont {D.~I.}\ \bibnamefont {Schuster}},
  \bibinfo {author} {\bibfnamefont {Lev~S.}\ \bibnamefont {Bishop}}, \bibinfo
  {author} {\bibfnamefont {G.}~\bibnamefont {Kirchmair}}, \bibinfo {author}
  {\bibfnamefont {G.}~\bibnamefont {Catelani}}, \bibinfo {author}
  {\bibfnamefont {A.~P.}\ \bibnamefont {Sears}}, \bibinfo {author}
  {\bibfnamefont {B.~R.}\ \bibnamefont {Johnson}}, \bibinfo {author}
  {\bibfnamefont {M.~J.}\ \bibnamefont {Reagor}}, \bibinfo {author}
  {\bibfnamefont {L.}~\bibnamefont {Frunzio}}, \bibinfo {author} {\bibfnamefont
  {L.~I.}\ \bibnamefont {Glazman}}, \bibinfo {author} {\bibfnamefont {S.~M.}\
  \bibnamefont {Girvin}}, \bibinfo {author} {\bibfnamefont {M.~H.}\
  \bibnamefont {Devoret}}, \ and\ \bibinfo {author} {\bibfnamefont {R.~J.}\
  \bibnamefont {Schoelkopf}},\ }\bibfield  {title} {\enquote {\bibinfo {title}
  {Observation of high coherence in josephson junction qubits measured in a
  three-dimensional circuit qed architecture},}\ }\href {\doibase
  10.1103/PhysRevLett.107.240501} {\bibfield  {journal} {\bibinfo  {journal}
  {Phys. Rev. Lett.}\ }\textbf {\bibinfo {volume} {107}},\ \bibinfo {pages}
  {240501} (\bibinfo {year} {2011})}\BibitemShut {NoStop}%
\bibitem [{\citenamefont {Fowler}\ \emph {et~al.}(2012)\citenamefont {Fowler},
  \citenamefont {Mariantoni}, \citenamefont {Martinis},\ and\ \citenamefont
  {Cleland}}]{fowler:2012}%
  \BibitemOpen
  \bibfield  {author} {\bibinfo {author} {\bibfnamefont {Austin~G.}\
  \bibnamefont {Fowler}}, \bibinfo {author} {\bibfnamefont {Matteo}\
  \bibnamefont {Mariantoni}}, \bibinfo {author} {\bibfnamefont {John~M.}\
  \bibnamefont {Martinis}}, \ and\ \bibinfo {author} {\bibfnamefont
  {Andrew~N.}\ \bibnamefont {Cleland}},\ }\bibfield  {title} {\enquote
  {\bibinfo {title} {Surface codes: Towards practical large-scale quantum
  computation},}\ }\href {\doibase 10.1103/PhysRevA.86.032324} {\bibfield
  {journal} {\bibinfo  {journal} {Phys. Rev. A}\ }\textbf {\bibinfo {volume}
  {86}},\ \bibinfo {pages} {032324} (\bibinfo {year} {2012})}\BibitemShut
  {NoStop}%
\bibitem [{\citenamefont {Reagor}\ \emph {et~al.}(2013)\citenamefont {Reagor},
  \citenamefont {Paik}, \citenamefont {Catelani}, \citenamefont {Sun},
  \citenamefont {Axline}, \citenamefont {Holland}, \citenamefont {Pop},
  \citenamefont {Masluk}, \citenamefont {Brecht}, \citenamefont {Frunzio},
  \citenamefont {Devoret}, \citenamefont {Glazman},\ and\ \citenamefont
  {Schoelkopf}}]{reagor:2013}%
  \BibitemOpen
  \bibfield  {author} {\bibinfo {author} {\bibfnamefont {Matthew}\ \bibnamefont
  {Reagor}}, \bibinfo {author} {\bibfnamefont {Hanhee}\ \bibnamefont {Paik}},
  \bibinfo {author} {\bibfnamefont {Gianluigi}\ \bibnamefont {Catelani}},
  \bibinfo {author} {\bibfnamefont {Luyan}\ \bibnamefont {Sun}}, \bibinfo
  {author} {\bibfnamefont {Christopher}\ \bibnamefont {Axline}}, \bibinfo
  {author} {\bibfnamefont {Eric}\ \bibnamefont {Holland}}, \bibinfo {author}
  {\bibfnamefont {Ioan~M.}\ \bibnamefont {Pop}}, \bibinfo {author}
  {\bibfnamefont {Nicholas~A.}\ \bibnamefont {Masluk}}, \bibinfo {author}
  {\bibfnamefont {Teresa}\ \bibnamefont {Brecht}}, \bibinfo {author}
  {\bibfnamefont {Luigi}\ \bibnamefont {Frunzio}}, \bibinfo {author}
  {\bibfnamefont {Michel~H.}\ \bibnamefont {Devoret}}, \bibinfo {author}
  {\bibfnamefont {Leonid}\ \bibnamefont {Glazman}}, \ and\ \bibinfo {author}
  {\bibfnamefont {Robert~J.}\ \bibnamefont {Schoelkopf}},\ }\bibfield  {title}
  {\enquote {\bibinfo {title} {Reaching 10 ms single photon lifetimes for
  superconducting aluminum cavities},}\ }\href@noop {} {\bibfield  {journal}
  {\bibinfo  {journal} {Applied Physics Letters}\ }\textbf {\bibinfo {volume}
  {102}},\ \bibinfo {pages} {192604} (\bibinfo {year} {2013})}\BibitemShut
  {NoStop}%
\bibitem [{\citenamefont {Yan}\ \emph {et~al.}(2016)\citenamefont {Yan},
  \citenamefont {Gustavsson}, \citenamefont {Kamal}, \citenamefont {Birenbaum},
  \citenamefont {Sears}, \citenamefont {Hover}, \citenamefont {Gudmundsen},
  \citenamefont {Rosenberg}, \citenamefont {Samach}, \citenamefont {Weber},
  \citenamefont {Yoder}, \citenamefont {Orlando}, \citenamefont {Clarke},
  \citenamefont {Kerman},\ and\ \citenamefont {Oliver}}]{yan:2016}%
  \BibitemOpen
  \bibfield  {author} {\bibinfo {author} {\bibfnamefont {Fei}\ \bibnamefont
  {Yan}}, \bibinfo {author} {\bibfnamefont {Simon}\ \bibnamefont {Gustavsson}},
  \bibinfo {author} {\bibfnamefont {Archana}\ \bibnamefont {Kamal}}, \bibinfo
  {author} {\bibfnamefont {Jeffrey}\ \bibnamefont {Birenbaum}}, \bibinfo
  {author} {\bibfnamefont {Adam~P.}\ \bibnamefont {Sears}}, \bibinfo {author}
  {\bibfnamefont {David}\ \bibnamefont {Hover}}, \bibinfo {author}
  {\bibfnamefont {Ted~J.}\ \bibnamefont {Gudmundsen}}, \bibinfo {author}
  {\bibfnamefont {Danna}\ \bibnamefont {Rosenberg}}, \bibinfo {author}
  {\bibfnamefont {Gabriel}\ \bibnamefont {Samach}}, \bibinfo {author}
  {\bibfnamefont {S.}~\bibnamefont {Weber}}, \bibinfo {author} {\bibfnamefont
  {Jonilyn~L.}\ \bibnamefont {Yoder}}, \bibinfo {author} {\bibfnamefont
  {Terry~P.}\ \bibnamefont {Orlando}}, \bibinfo {author} {\bibfnamefont {John}\
  \bibnamefont {Clarke}}, \bibinfo {author} {\bibfnamefont {Andrew~J.}\
  \bibnamefont {Kerman}}, \ and\ \bibinfo {author} {\bibfnamefont {William~D.}\
  \bibnamefont {Oliver}},\ }\bibfield  {title} {\enquote {\bibinfo {title} {The
  flux qubit revisited to enhance coherence and reproducibility},}\ }\href@noop
  {} {\bibfield  {journal} {\bibinfo  {journal} {Nature communications}\
  }\textbf {\bibinfo {volume} {7}},\ \bibinfo {pages} {12964} (\bibinfo {year}
  {2016})}\BibitemShut {NoStop}%
\bibitem [{\citenamefont {Kurpiers}\ \emph {et~al.}(2017)\citenamefont
  {Kurpiers}, \citenamefont {Walter}, \citenamefont {Magnard}, \citenamefont
  {Salathe},\ and\ \citenamefont {Wallraff}}]{kurpiers:2017}%
  \BibitemOpen
  \bibfield  {author} {\bibinfo {author} {\bibfnamefont {Philipp}\ \bibnamefont
  {Kurpiers}}, \bibinfo {author} {\bibfnamefont {Theodore}\ \bibnamefont
  {Walter}}, \bibinfo {author} {\bibfnamefont {Paul}\ \bibnamefont {Magnard}},
  \bibinfo {author} {\bibfnamefont {Yves}\ \bibnamefont {Salathe}}, \ and\
  \bibinfo {author} {\bibfnamefont {Andreas}\ \bibnamefont {Wallraff}},\
  }\bibfield  {title} {\enquote {\bibinfo {title} {Characterizing the
  attenuation of coaxial and rectangular microwave-frequency waveguides at
  cryogenic temperatures},}\ }\href@noop {} {\bibfield  {journal} {\bibinfo
  {journal} {. EPJ Quantum Technology}\ }\textbf {\bibinfo {volume} {4}},\
  \bibinfo {pages} {8} (\bibinfo {year} {2017})}\BibitemShut {NoStop}%
\bibitem [{\citenamefont {Calusine}\ \emph {et~al.}(2018)\citenamefont
  {Calusine}, \citenamefont {Melville}, \citenamefont {Woods}, \citenamefont
  {Das}, \citenamefont {Stull}, \citenamefont {Bolkhovsky}, \citenamefont
  {Braje}, \citenamefont {Hover}, \citenamefont {Kim}, \citenamefont {Miloshi},
  \citenamefont {Rosenberg}, \citenamefont {Sevi}, \citenamefont {Yoder},
  \citenamefont {Dauler},\ and\ \citenamefont {Oliver}}]{calusine:2018}%
  \BibitemOpen
  \bibfield  {author} {\bibinfo {author} {\bibfnamefont {G.}~\bibnamefont
  {Calusine}}, \bibinfo {author} {\bibfnamefont {A.}~\bibnamefont {Melville}},
  \bibinfo {author} {\bibfnamefont {W.}~\bibnamefont {Woods}}, \bibinfo
  {author} {\bibfnamefont {R.}~\bibnamefont {Das}}, \bibinfo {author}
  {\bibfnamefont {C.}~\bibnamefont {Stull}}, \bibinfo {author} {\bibfnamefont
  {V.}~\bibnamefont {Bolkhovsky}}, \bibinfo {author} {\bibfnamefont
  {D.}~\bibnamefont {Braje}}, \bibinfo {author} {\bibfnamefont
  {D.}~\bibnamefont {Hover}}, \bibinfo {author} {\bibfnamefont {D.~K.}\
  \bibnamefont {Kim}}, \bibinfo {author} {\bibfnamefont {X.}~\bibnamefont
  {Miloshi}}, \bibinfo {author} {\bibfnamefont {D.}~\bibnamefont {Rosenberg}},
  \bibinfo {author} {\bibfnamefont {A.}~\bibnamefont {Sevi}}, \bibinfo {author}
  {\bibfnamefont {J.~L.}\ \bibnamefont {Yoder}}, \bibinfo {author}
  {\bibfnamefont {E.}~\bibnamefont {Dauler}}, \ and\ \bibinfo {author}
  {\bibfnamefont {W.~D.}\ \bibnamefont {Oliver}},\ }\bibfield  {title}
  {\enquote {\bibinfo {title} {Analysis and mitigation of interface losses in
  trenched superconducting coplanar waveguide resonators},}\ }\href@noop {}
  {\bibfield  {journal} {\bibinfo  {journal} {Applied Physics Letters}\
  }\textbf {\bibinfo {volume} {112}},\ \bibinfo {pages} {062601} (\bibinfo
  {year} {2018})}\BibitemShut {NoStop}%
\bibitem [{\citenamefont {Yan}\ \emph {et~al.}(2018)\citenamefont {Yan},
  \citenamefont {Campbell}, \citenamefont {Krantz}, \citenamefont {Kjaergaard},
  \citenamefont {Kim}, \citenamefont {Yoder}, \citenamefont {Hover},
  \citenamefont {Sears}, \citenamefont {Kerman}, \citenamefont {Orlando},
  \citenamefont {Gustavsson},\ and\ \citenamefont {Oliver}}]{yan:2018}%
  \BibitemOpen
  \bibfield  {author} {\bibinfo {author} {\bibfnamefont {Fei}\ \bibnamefont
  {Yan}}, \bibinfo {author} {\bibfnamefont {Dan}\ \bibnamefont {Campbell}},
  \bibinfo {author} {\bibfnamefont {Philip}\ \bibnamefont {Krantz}}, \bibinfo
  {author} {\bibfnamefont {Morten}\ \bibnamefont {Kjaergaard}}, \bibinfo
  {author} {\bibfnamefont {David}\ \bibnamefont {Kim}}, \bibinfo {author}
  {\bibfnamefont {Jonilyn~L.}\ \bibnamefont {Yoder}}, \bibinfo {author}
  {\bibfnamefont {David}\ \bibnamefont {Hover}}, \bibinfo {author}
  {\bibfnamefont {Adam}\ \bibnamefont {Sears}}, \bibinfo {author}
  {\bibfnamefont {Andrew~J.}\ \bibnamefont {Kerman}}, \bibinfo {author}
  {\bibfnamefont {Terry~P.}\ \bibnamefont {Orlando}}, \bibinfo {author}
  {\bibfnamefont {Simon}\ \bibnamefont {Gustavsson}}, \ and\ \bibinfo {author}
  {\bibfnamefont {William~D.}\ \bibnamefont {Oliver}},\ }\bibfield  {title}
  {\enquote {\bibinfo {title} {Distinguishing coherent and thermal photon noise
  in a circuit quantum electrodynamical system},}\ }\href {\doibase
  10.1103/PhysRevLett.120.260504} {\bibfield  {journal} {\bibinfo  {journal}
  {Phys. Rev. Lett.}\ }\textbf {\bibinfo {volume} {120}},\ \bibinfo {pages}
  {260504} (\bibinfo {year} {2018})}\BibitemShut {NoStop}%
\bibitem [{\citenamefont {Touzard}\ \emph {et~al.}(2019)\citenamefont
  {Touzard}, \citenamefont {Kou}, \citenamefont {Frattini}, \citenamefont
  {Sivak}, \citenamefont {Puri}, \citenamefont {Grimm}, \citenamefont
  {Frunzio}, \citenamefont {Shankar},\ and\ \citenamefont
  {Devoret}}]{touzard:2019}%
  \BibitemOpen
  \bibfield  {author} {\bibinfo {author} {\bibfnamefont {S.}~\bibnamefont
  {Touzard}}, \bibinfo {author} {\bibfnamefont {A.}~\bibnamefont {Kou}},
  \bibinfo {author} {\bibfnamefont {N.~E.}\ \bibnamefont {Frattini}}, \bibinfo
  {author} {\bibfnamefont {V.~V.}\ \bibnamefont {Sivak}}, \bibinfo {author}
  {\bibfnamefont {S.}~\bibnamefont {Puri}}, \bibinfo {author} {\bibfnamefont
  {A.}~\bibnamefont {Grimm}}, \bibinfo {author} {\bibfnamefont
  {L.}~\bibnamefont {Frunzio}}, \bibinfo {author} {\bibfnamefont
  {S.}~\bibnamefont {Shankar}}, \ and\ \bibinfo {author} {\bibfnamefont
  {M.~H.}\ \bibnamefont {Devoret}},\ }\bibfield  {title} {\enquote {\bibinfo
  {title} {Gated conditional displacement readout of superconducting qubits},}\
  }\href {\doibase 10.1103/PhysRevLett.122.080502} {\bibfield  {journal}
  {\bibinfo  {journal} {Phys. Rev. Lett.}\ }\textbf {\bibinfo {volume} {122}},\
  \bibinfo {pages} {080502} (\bibinfo {year} {2019})}\BibitemShut {NoStop}%
\bibitem [{\citenamefont {Hatridge}\ \emph {et~al.}(2013)\citenamefont
  {Hatridge}, \citenamefont {Shankar}, \citenamefont {Mirrahimi}, \citenamefont
  {Schackert}, \citenamefont {Geerlings}, \citenamefont {Brecht}, \citenamefont
  {Sliwa}, \citenamefont {Abdo}, \citenamefont {Frunzio}, \citenamefont
  {Girvin}, \citenamefont {Schoelkopf},\ and\ \citenamefont
  {Devoret}}]{hatridge:2013}%
  \BibitemOpen
  \bibfield  {author} {\bibinfo {author} {\bibfnamefont {M.}~\bibnamefont
  {Hatridge}}, \bibinfo {author} {\bibfnamefont {S.}~\bibnamefont {Shankar}},
  \bibinfo {author} {\bibfnamefont {M.}~\bibnamefont {Mirrahimi}}, \bibinfo
  {author} {\bibfnamefont {F.}~\bibnamefont {Schackert}}, \bibinfo {author}
  {\bibfnamefont {K.}~\bibnamefont {Geerlings}}, \bibinfo {author}
  {\bibfnamefont {T.}~\bibnamefont {Brecht}}, \bibinfo {author} {\bibfnamefont
  {K.~M.}\ \bibnamefont {Sliwa}}, \bibinfo {author} {\bibfnamefont
  {B.}~\bibnamefont {Abdo}}, \bibinfo {author} {\bibfnamefont {L.}~\bibnamefont
  {Frunzio}}, \bibinfo {author} {\bibfnamefont {S.~M.}\ \bibnamefont {Girvin}},
  \bibinfo {author} {\bibfnamefont {R.~J.}\ \bibnamefont {Schoelkopf}}, \ and\
  \bibinfo {author} {\bibfnamefont {M.~H.}\ \bibnamefont {Devoret}},\
  }\bibfield  {title} {\enquote {\bibinfo {title} {Quantum back-action of an
  individual variable-strength measurement},}\ }\href@noop {} {\bibfield
  {journal} {\bibinfo  {journal} {Science}\ }\textbf {\bibinfo {volume}
  {339}},\ \bibinfo {pages} {178--–181} (\bibinfo {year} {2013})}\BibitemShut
  {NoStop}%
\bibitem [{\citenamefont {Heinsoo}\ \emph {et~al.}(2018)\citenamefont
  {Heinsoo}, \citenamefont {Andersen}, \citenamefont {Remm}, \citenamefont
  {Krinner}, \citenamefont {Walter}, \citenamefont {Salath\'e}, \citenamefont
  {Gasparinetti}, \citenamefont {Besse}, \citenamefont
  {Poto\ifmmode~\check{c}\else \v{c}\fi{}nik}, \citenamefont {Wallraff},\ and\
  \citenamefont {Eichler}}]{heinsoo:2018}%
  \BibitemOpen
  \bibfield  {author} {\bibinfo {author} {\bibfnamefont {Johannes}\
  \bibnamefont {Heinsoo}}, \bibinfo {author} {\bibfnamefont
  {Christian~Kraglund}\ \bibnamefont {Andersen}}, \bibinfo {author}
  {\bibfnamefont {Ants}\ \bibnamefont {Remm}}, \bibinfo {author} {\bibfnamefont
  {Sebastian}\ \bibnamefont {Krinner}}, \bibinfo {author} {\bibfnamefont
  {Theodore}\ \bibnamefont {Walter}}, \bibinfo {author} {\bibfnamefont {Yves}\
  \bibnamefont {Salath\'e}}, \bibinfo {author} {\bibfnamefont {Simone}\
  \bibnamefont {Gasparinetti}}, \bibinfo {author} {\bibfnamefont {Jean-Claude}\
  \bibnamefont {Besse}}, \bibinfo {author} {\bibfnamefont {Anton}\ \bibnamefont
  {Poto\ifmmode~\check{c}\else \v{c}\fi{}nik}}, \bibinfo {author}
  {\bibfnamefont {Andreas}\ \bibnamefont {Wallraff}}, \ and\ \bibinfo {author}
  {\bibfnamefont {Christopher}\ \bibnamefont {Eichler}},\ }\bibfield  {title}
  {\enquote {\bibinfo {title} {Rapid high-fidelity multiplexed readout of
  superconducting qubits},}\ }\href {\doibase 10.1103/PhysRevApplied.10.034040}
  {\bibfield  {journal} {\bibinfo  {journal} {Phys. Rev. Applied}\ }\textbf
  {\bibinfo {volume} {10}},\ \bibinfo {pages} {034040} (\bibinfo {year}
  {2018})}\BibitemShut {NoStop}%
\bibitem [{\citenamefont {Eddins}\ \emph {et~al.}(2018)\citenamefont {Eddins},
  \citenamefont {Schreppler}, \citenamefont {Toyli}, \citenamefont {Martin},
  \citenamefont {Hacohen-Gourgy}, \citenamefont {Govia}, \citenamefont
  {Ribeiro}, \citenamefont {Clerk},\ and\ \citenamefont
  {Siddiqi}}]{eddins:2018}%
  \BibitemOpen
  \bibfield  {author} {\bibinfo {author} {\bibfnamefont {A.}~\bibnamefont
  {Eddins}}, \bibinfo {author} {\bibfnamefont {S.}~\bibnamefont {Schreppler}},
  \bibinfo {author} {\bibfnamefont {D.~M.}\ \bibnamefont {Toyli}}, \bibinfo
  {author} {\bibfnamefont {L.~S.}\ \bibnamefont {Martin}}, \bibinfo {author}
  {\bibfnamefont {S.}~\bibnamefont {Hacohen-Gourgy}}, \bibinfo {author}
  {\bibfnamefont {L.~C.~G.}\ \bibnamefont {Govia}}, \bibinfo {author}
  {\bibfnamefont {H.}~\bibnamefont {Ribeiro}}, \bibinfo {author} {\bibfnamefont
  {A.~A.}\ \bibnamefont {Clerk}}, \ and\ \bibinfo {author} {\bibfnamefont
  {I.}~\bibnamefont {Siddiqi}},\ }\bibfield  {title} {\enquote {\bibinfo
  {title} {Stroboscopic qubit measurement with squeezed illumination},}\ }\href
  {\doibase 10.1103/PhysRevLett.120.040505} {\bibfield  {journal} {\bibinfo
  {journal} {Phys. Rev. Lett.}\ }\textbf {\bibinfo {volume} {120}},\ \bibinfo
  {pages} {040505} (\bibinfo {year} {2018})}\BibitemShut {NoStop}%
\bibitem [{\citenamefont {Thorbeck}\ \emph {et~al.}(2017)\citenamefont
  {Thorbeck}, \citenamefont {Zhu}, \citenamefont {Leonard~Jr.}, \citenamefont
  {Barends}, \citenamefont {Kelly}, \citenamefont {Martinis},\ and\
  \citenamefont {McDermott}}]{thorbeck:2017}%
  \BibitemOpen
  \bibfield  {author} {\bibinfo {author} {\bibfnamefont {T.}~\bibnamefont
  {Thorbeck}}, \bibinfo {author} {\bibfnamefont {S.}~\bibnamefont {Zhu}},
  \bibinfo {author} {\bibfnamefont {E.}~\bibnamefont {Leonard~Jr.}}, \bibinfo
  {author} {\bibfnamefont {R.}~\bibnamefont {Barends}}, \bibinfo {author}
  {\bibfnamefont {J.}~\bibnamefont {Kelly}}, \bibinfo {author} {\bibfnamefont
  {John~M.}\ \bibnamefont {Martinis}}, \ and\ \bibinfo {author} {\bibfnamefont
  {R.}~\bibnamefont {McDermott}},\ }\bibfield  {title} {\enquote {\bibinfo
  {title} {Reverse isolation and backaction of the slug microwave amplifier},}\
  }\href@noop {} {\bibfield  {journal} {\bibinfo  {journal} {Physical Review
  Applied}\ }\textbf {\bibinfo {volume} {8}},\ \bibinfo {pages} {054007}
  (\bibinfo {year} {2017})}\BibitemShut {NoStop}%
\bibitem [{\citenamefont {Andersen}\ \emph {et~al.}(2019)\citenamefont
  {Andersen}, \citenamefont {Remm}, \citenamefont {Lazar}, \citenamefont
  {Krinner}, \citenamefont {Heinsoo}, \citenamefont {Besse}, \citenamefont
  {Gabureac}, \citenamefont {Wallraff},\ and\ \citenamefont
  {Eichler}}]{andersen:2019}%
  \BibitemOpen
  \bibfield  {author} {\bibinfo {author} {\bibfnamefont {Christian~Kraglund}\
  \bibnamefont {Andersen}}, \bibinfo {author} {\bibfnamefont {Ants}\
  \bibnamefont {Remm}}, \bibinfo {author} {\bibfnamefont {Stefania}\
  \bibnamefont {Lazar}}, \bibinfo {author} {\bibfnamefont {Sebastian}\
  \bibnamefont {Krinner}}, \bibinfo {author} {\bibfnamefont {Johannes}\
  \bibnamefont {Heinsoo}}, \bibinfo {author} {\bibfnamefont {Jean-Claude}\
  \bibnamefont {Besse}}, \bibinfo {author} {\bibfnamefont {Mihai}\ \bibnamefont
  {Gabureac}}, \bibinfo {author} {\bibfnamefont {Andreas}\ \bibnamefont
  {Wallraff}}, \ and\ \bibinfo {author} {\bibfnamefont {Christopher}\
  \bibnamefont {Eichler}},\ }\bibfield  {title} {\enquote {\bibinfo {title}
  {Entanglement stabilization using ancilla-based parity detection and
  real-time feedback in superconducting circuits},}\ }\href@noop {} {\bibfield
  {journal} {\bibinfo  {journal} {npj Quantum Information}\ }\textbf {\bibinfo
  {volume} {5}},\ \bibinfo {pages} {69} (\bibinfo {year} {2019})}\BibitemShut
  {NoStop}%
\bibitem [{\citenamefont {Andersen}\ \emph {et~al.}(2020)\citenamefont
  {Andersen}, \citenamefont {Remm}, \citenamefont {Lazar}, \citenamefont
  {Krinner}, \citenamefont {Lacroix}, \citenamefont {Norris}, \citenamefont
  {Gabureac}, \citenamefont {Eichler},\ and\ \citenamefont
  {Wallraff}}]{andersen:2020}%
  \BibitemOpen
  \bibfield  {author} {\bibinfo {author} {\bibfnamefont {Christian~Kraglund}\
  \bibnamefont {Andersen}}, \bibinfo {author} {\bibfnamefont {Ants}\
  \bibnamefont {Remm}}, \bibinfo {author} {\bibfnamefont {Stefania}\
  \bibnamefont {Lazar}}, \bibinfo {author} {\bibfnamefont {Sebastian}\
  \bibnamefont {Krinner}}, \bibinfo {author} {\bibfnamefont {Nathan}\
  \bibnamefont {Lacroix}}, \bibinfo {author} {\bibfnamefont {Graham~J.}\
  \bibnamefont {Norris}}, \bibinfo {author} {\bibfnamefont {Mihai}\
  \bibnamefont {Gabureac}}, \bibinfo {author} {\bibfnamefont {Christopher}\
  \bibnamefont {Eichler}}, \ and\ \bibinfo {author} {\bibfnamefont {Andreas}\
  \bibnamefont {Wallraff}},\ }\bibfield  {title} {\enquote {\bibinfo {title}
  {Repeated quantum error detection in a surface code},}\ }\href@noop {}
  {\bibfield  {journal} {\bibinfo  {journal} {Nature Physics}\ }\textbf
  {\bibinfo {volume} {16}},\ \bibinfo {pages} {875–880} (\bibinfo {year}
  {2020})}\BibitemShut {NoStop}%
\bibitem [{\citenamefont {Peronnin}\ \emph {et~al.}(2020)\citenamefont
  {Peronnin}, \citenamefont {Markovi\ifmmode~\acute{c}\else \'{c}\fi{}},
  \citenamefont {Ficheux},\ and\ \citenamefont {Huard}}]{peronnin:2020}%
  \BibitemOpen
  \bibfield  {author} {\bibinfo {author} {\bibfnamefont {T.}~\bibnamefont
  {Peronnin}}, \bibinfo {author} {\bibfnamefont {D.}~\bibnamefont
  {Markovi\ifmmode~\acute{c}\else \'{c}\fi{}}}, \bibinfo {author}
  {\bibfnamefont {Q.}~\bibnamefont {Ficheux}}, \ and\ \bibinfo {author}
  {\bibfnamefont {B.}~\bibnamefont {Huard}},\ }\bibfield  {title} {\enquote
  {\bibinfo {title} {Sequential dispersive measurement of a superconducting
  qubit},}\ }\href {\doibase 10.1103/PhysRevLett.124.180502} {\bibfield
  {journal} {\bibinfo  {journal} {Phys. Rev. Lett.}\ }\textbf {\bibinfo
  {volume} {124}},\ \bibinfo {pages} {180502} (\bibinfo {year}
  {2020})}\BibitemShut {NoStop}%
\bibitem [{\citenamefont {Rosenberg}\ \emph {et~al.}(2017)\citenamefont
  {Rosenberg}, \citenamefont {Kim}, \citenamefont {Das}, \citenamefont {Yost},
  \citenamefont {Gustavsson}, \citenamefont {Hover}, \citenamefont {Krantz},
  \citenamefont {Melville}, \citenamefont {Racz}, \citenamefont {Samach},
  \citenamefont {Weber}, \citenamefont {Yan}, \citenamefont {Yoder},
  \citenamefont {Kerman},\ and\ \citenamefont {Oliver}}]{rosenberg:2017}%
  \BibitemOpen
  \bibfield  {author} {\bibinfo {author} {\bibfnamefont {D.}~\bibnamefont
  {Rosenberg}}, \bibinfo {author} {\bibfnamefont {D.}~\bibnamefont {Kim}},
  \bibinfo {author} {\bibfnamefont {R.}~\bibnamefont {Das}}, \bibinfo {author}
  {\bibfnamefont {D.}~\bibnamefont {Yost}}, \bibinfo {author} {\bibfnamefont
  {S.}~\bibnamefont {Gustavsson}}, \bibinfo {author} {\bibfnamefont
  {D.}~\bibnamefont {Hover}}, \bibinfo {author} {\bibfnamefont
  {P.}~\bibnamefont {Krantz}}, \bibinfo {author} {\bibfnamefont
  {A.}~\bibnamefont {Melville}}, \bibinfo {author} {\bibfnamefont
  {L.}~\bibnamefont {Racz}}, \bibinfo {author} {\bibfnamefont {G.~O.}\
  \bibnamefont {Samach}}, \bibinfo {author} {\bibfnamefont {S.~J.}\
  \bibnamefont {Weber}}, \bibinfo {author} {\bibfnamefont {F.}~\bibnamefont
  {Yan}}, \bibinfo {author} {\bibfnamefont {J.~L.}\ \bibnamefont {Yoder}},
  \bibinfo {author} {\bibfnamefont {A.~J.}\ \bibnamefont {Kerman}}, \ and\
  \bibinfo {author} {\bibfnamefont {W.~D.}\ \bibnamefont {Oliver}},\ }\bibfield
   {title} {\enquote {\bibinfo {title} {3d integrated superconducting
  qubits},}\ }\href@noop {} {\bibfield  {journal} {\bibinfo  {journal} {npj
  Quantum Inf}\ }\textbf {\bibinfo {volume} {3}},\ \bibinfo {pages} {42}
  (\bibinfo {year} {2017})}\BibitemShut {NoStop}%
\end{thebibliography}%

\clearpage
\newpage



\section*{Supplementary material}

# Supplementary material for "Efficient and Low-Backaction Quantum Measurement Using a Chip-Scale Detector"

(Dated: February 19, 2021)

## CONTENTS

## I. INTRODUCTION

### A. Outline

In the main text we introduce a novel superconducting device: the 'Superconducting Isolating Modular Bifurcation Amplifier' (SIMBA). Using the SIMBA, we demonstrate efficient, low-backaction and high-fidelity measurement of a superconducting qubit. The purpose of this supplementary material is to clarify and support claims made in the main text, to present experimental details, and finally to serve as a resource for the future operation and design of SIMBAs or SIMBA-like devices. The supplementary material is organized as follows:

In Section II, we begin with a theoretical discussion of measurement-induced dephasing of a qubit. We show how this fundamental property of quantum mechanics can be used to calibrate the efficiency of a measurement [1]. We first show this for measurement using a linear amplifier, as previously done in Refs. [2–6]. This method is then extended to the similar case of a bifurcation amplifier. Finally, we examine qubit measurement from an information theoretical standpoint [7, 8], contrasting the use of a linear vs. bifurcation amplifier.

In Section III, we turn to hardware. First, we describe the SIMBA along with its components: two superconducting switches and the Josephson parametric amplifier. Section IV is devoted to a theoretical understanding of our parametric amplifier pumped into the bifurcation regime. This is useful for understanding the parametric amplifier design space. In particular, we build on work in Ref. [9] for the case of a parametric amplifier constructed from an array of SQUIDs. Section V discusses the calibration and performance of qubit readout using a SIMBA. Section VI concludes with a discussion of the limitations of the current SIMBA, along with suggested improvements.

### B. Comparison to previous work

Before an in-depth analysis of readout using a SIMBA, we briefly compare this work to other recent demonstrations of superconducting qubit readout. Doing so is helpful for understanding how this work fits into the superconducting qubit literature. In this comparison, we focus on the different figures of merit reported in Table I of the main text: measurement efficiency $\eta$, excess backaction $n_b$, maximum readout fidelity $F_0$, and the time in which readout is completed [20].

Table. S1 gives a list (not exhaustive) of recent demonstrations of superconducting qubit readout in which the

TABLE S1. **Comparison to related works.** The references in this table are a selection of recent demonstrations of superconducting qubit readout in which the measurement efficiency $\eta$ is characterized. When doing conventional dispersive readout [10] using ferrite circulators or isolators, many recent demonstrations have readout fidelities $F_r$ of 0.95 or greater, readout times of several hundred nanoseconds, and measurement efficiencies between $\eta = 0.1$ and $\eta = 0.6$ (characterized by a comparison of measurement induced dephasing to the information gain of a weak measurement, as in this work), [2, 3, 5, 11–14]. All works in this table use Josephson-junction based parametric amplifiers, which are operated either in a phase-*sensitive* (S) or phase-*preserving* (P) manner. Note that the works in this table which use a phase-preserving amplifier (P) generally use a definition of measurement efficiency which goes to unity when using a phase-preserving amplifier. That convention differs by a factor of $1/2$ from the definition of efficiency used in this work. In other words, $\eta$ in these works refers to the loss between the qubit and amplifier, plus added noise of the amplifier in excess of the required half-quanta.

| Reference | Amplifier | $\eta$ | Backaction | $F_r$ | Readout time |
|---|---|---|---|---|---|
| This work | SIMBA (S) | 0.70 | 0.66 photons | 0.955 | 265 ns |
| Lecocq *et al.* (2021), [6] | FPJA (S) | 0.72 | 0.07 photons | 0.97 | 350 ns |
| Andersen *et al.* (2020), [12] | TWPA (P) | $0.15 - 0.30$ | ferrite-iso. | $0.978 - 0.994$ | $300 - 400$ ns |
| Abdo *et al.* (2020), [15] | JDA (P) | $\sim 0.2^{\text{a}}$ | 0.002 photons | 0.92 | 1000 ns |
| Peronnin *et al.* (2020), [14] | TWPA (P) | 0.11 | ferrite-iso. | $0.95^{\text{b}}$ | 220 ns |
| Andersen *et al.* (2019), [12] | TWPA (P) | 0.24 | ferrite-iso. | $0.987 - 0.992$ | 200 ns |
| Abdo *et al.* (2019), [16] | JPC (P) | $\sim 0.3^{\text{a}}$ | 0.01 photons$^{\text{c}}$ | 0.9 | 200 ns |
| Touzard *et al.* (2019), [5] | SPA (S) | 0.6 | ferrite-iso. | 0.978 | 870 ns |
| Eddins *et al.* (2019), [4] | JPA (S) | 0.80 | no isolation before amp. | not specified | N/A |
| Heinsoo *et al.* (2018), [11] | TWPA (P) | $0.43 - 0.52$ | ferrite-iso. | $0.936 - 0.988$ | 250 ns |
| Bultink *et al.* (2018), [2] | TWPA (P) | 0.165 | ferrite-iso. | not specified | N/A |
| Eddins *et al.* (2018), [3] | JPA (S) | 0.38 | ferrite-iso. | not specified | N/A |
| Walter *et al.* (2017), [17] | JPD (S) | $0.75^{\text{d}}$ | ferrite-iso. | 0.992 | 88 ns |
| Macklin *et al.* (2015), [18] | TWPA (P) | 0.49 | ferrite-iso.$^{\text{e}}$ | 0.967 | 100 ns |

[a] Refs. [15, 16] characterize the measurement efficiency using the method described in Ref. [19].
[b] Ref. [14] reports readout fidelity using a different definition than in this work. Here, we convert their reported maximum readout fidelity to the definition in Eq. 3, main text.
[c] In Ref. [16], isolation between a qubit and amplifier is provided by the combination of a chip-scale superconducting isolator and a ferrite circulator.
[d] In Ref. [17], $\eta$ is estimated by determining the power spectral density of a readout pulse using the ac-Stark shift of the qubit, and comparing this to the power spectral density of a measured signal.
[e] Without any ferrite circulators or isolators at its input, a pumped TWPA will in-principle not cause excess backaction if its pump signal is perfectly matched with the TWPA. In practice, it is difficult to prevent some of this relatively large ($\gg 1$ photon) pump signal from scattering toward the qubit.

measurement efficiency $\eta$ has been characterized. The highest measurement efficiency which has thus-far been reported in a circuit quantum electrodynamics system is $\eta = 0.80$ in Ref. [4]. In Ref. [4], a qubit is dispersively coupled to a readout cavity which shares functionality as a flux-pumped parametric amplifier. There are therefore no non-reciprocal elements between the qubit and amplifier. To the authors' knowledge, the highest measurement efficiencies that have yet been reported *with* considerable ferrite-based isolation between the qubit and amplifier are $\eta = 0.6$ in Ref. [5], and $\eta = 0.75$ in Ref. [17] (where $\eta$ is calibrated without a comparison to measurement-induced dephasing, as done elsewhere in Table. S1).

One advantage of the SIMBA is the elimination of *any* ferrite circulators or isolators between a qubit and parametric amplifier. There have thus-far been relatively few demonstrations of this in the superconducting qubit literature. To the authors' knowledge, Refs. [6, 15, 21] are the only other such demonstrations yet-reported *with* considerable isolation from amplifier backaction [22]. Readout using a SIMBA demonstrates the combination of high efficiency and relatively fast and high-fidelity measurement, while still limiting excess backaction to 0.66 photons, equivalent to isolation of approximately -26 dB. We note that these performance specifications are similar to recent work by Lecocq *et al.* [6], which uses a parametrically modulated superconducting circuit [23] to provide both gain and isolation.

## II. EFFICIENCY

The inherent decoherence associated with quantum measurement can be used as a resource for metrology. As such, we describe how a variable strength quantum measurement is used to precisely measure the *efficiency* of a measurement, a metric which quantifies the loss and added noise of a detector [1].

## A. Measurement-induced dephasing

Here we consider measurement of the state of a qubit via a projective measurement performed on an ancilla system, entangled with the qubit. A quantum measurement on the system as a whole is characterized by a positive operator-valued measure (POVM) $\Pi = \{\Pi_x \colon x \in \mathcal{X}\}$, which consists of a set of positive semidefinite operators $\Pi_x \geq 0$ labeled by the measurement outcome $x$ in some set $\mathcal{X}$, and satisfying $\sum_{x \in \mathcal{X}} \Pi_x = \mathbb{I}$. For each $x$ one may choose Kraus operators $A_x$ such that $\Pi_x = A_x^\dagger A_x$. Note that the choice of Kraus operators is not unique. Given a measurement outcome $x$, the POVM maps a quantum state $\hat{\rho}$ to a new quantum state $\hat{\rho}'_x$ via the formula [24],

$$\hat{\rho}'_x = \frac{A_x \hat{\rho} A_x^\dagger}{\mathrm{Tr}(\Pi_x \hat{\rho})}. \tag{S1}$$

Since the POVM is complete, $\sum_{x \in \mathcal{X}} \Pi_x = \mathbb{I}$, the post-measurement state is then

$$\hat{\rho}' = \sum_x \lambda_x \hat{\rho}'_x \tag{S2}$$

with probabilities $\lambda_x = \mathrm{Tr}(\Pi_x \hat{\rho})$.

Now consider a qubit entangled with a measurement system in the states $|\varphi_0\rangle$ and $|\varphi_1\rangle$,

$$|\psi\rangle = \frac{1}{\sqrt{2}}(|0\rangle|\varphi_0\rangle + |1\rangle|\varphi_1\rangle), \tag{S3}$$

with density matrix $\hat{\sigma} = |\psi\rangle\langle\psi|$ describing the joint state of the qubit and the measurement system. For the POVM element $\Pi_x = \mathbb{I} \otimes |x\rangle\langle x|$, Eq. S1 gives the following state after obtaining the measurement outcome $x$,

$$\hat{\sigma}'_x = \frac{1}{2\lambda_x}\begin{pmatrix} |\langle x|\varphi_0\rangle|^2 & \langle x|\varphi_0\rangle\langle\varphi_1|x\rangle \\ \langle x|\varphi_1\rangle\langle\varphi_0|x\rangle & |\langle x|\varphi_1\rangle|^2 \end{pmatrix} \otimes |x\rangle\langle x|, \tag{S4}$$

where $\lambda_x = \mathrm{Tr}(\Pi_x \hat{\sigma})$. Summing Eq. S4 over all measurement outcomes as in Eq. S2 and taking the partial trace over the second system gives the average density matrix $\hat{\rho}' = \mathrm{Tr}_2(\sum_x \lambda_x \hat{\sigma}'_x)$ of the qubit after measurement [25],

$$\hat{\rho}' = \frac{1}{2}\begin{pmatrix} 1 & \langle\varphi_1|\varphi_0\rangle \\ \langle\varphi_0|\varphi_1\rangle & 1 \end{pmatrix}. \tag{S5}$$

In other words, the projective measurement on the ancilla via POVM element $\Pi_x$ transforms the pure qubit state into the mixed state given by Eq. S5. Qubit coherence (the off-diagonal density matrix element of the qubit) will be reduced by the extent to which $|\varphi_0\rangle$ and $|\varphi_1\rangle$ are orthogonal.

In dispersive readout [10], a qubit is entangled with a coherent state such that $|\varphi_0\rangle = |\alpha_0\rangle$ and $|\varphi_1\rangle = |\alpha_1\rangle$, Fig. S1a. These states have equal amplitudes $|\alpha|$ but are separated by a maximum angle $2\theta = 2\arctan(2\chi/\kappa_r)$ in quadrature space, Fig. S1a. Here, $\kappa_r/2\pi$ is the loss rate of the readout resonator, whose frequency changes by $\pm\chi/2\pi$ depending on the qubit state.

The wavefunction of a normalized coherent state along the $x$-quadrature is defined to be $\langle x|\alpha\rangle = \left(\frac{2}{\pi}\right)^{1/4} e^{-i|\alpha|^2 \sin(2\theta)/2} e^{-[(x-\mathrm{Re}[\alpha])^2 - 2ix\mathrm{Im}[\alpha]]}$, where $\mathrm{Re}[\alpha] = |\alpha|\sin\theta$ and $\mathrm{Im}[\alpha] = |\alpha|\cos\theta$. Note that in this definition, $\langle x|\alpha\rangle$ is a Gaussian function with a standard deviation of $1/\sqrt{2}$, variance of $1/2$, and that $|\langle x|\alpha\rangle|^2 = \sqrt{2/\pi} e^{-2(x-|\alpha|\sin\theta)^2}$ [8, 26, 27]. Therefore $|\langle x|\alpha\rangle|^2$, the modulus-squared of the coherent state wavefunction, has a standard deviation of $1/2$ and a variance of $1/4$, as per the convention in Refs. [26, 27]. From Eq. S5, the qubit after measurement is in a mixed state with coherence $\hat{\rho}'_{01} = \frac{1}{2}\langle\alpha_0|\alpha_1\rangle$. Evaluating this gives,

$$\hat{\rho}' = \frac{1}{2}\begin{pmatrix} 1 & e^{-2|\alpha|^2 \sin^2\theta} \\ e^{-2|\alpha|^2 \sin^2\theta} & 1 \end{pmatrix}. \tag{S6}$$

As $|\alpha| \to \infty$, qubit coherence vanishes and Eq. S6 reduces to the density matrix of a qubit prepared in a superposition state and then projectively measured. Eq. S6 also concludes that dephasing goes to zero in the limit where $\theta \to 0$ such that $\kappa_r \gg 2\chi$. This makes sense: a readout signal much more strongly coupled to the environment than to the qubit will dissipate before becoming entangled with the qubit.

With no readout signal, $|\alpha| = 0$, a qubit may still be dephased by the excess backaction of a detector (for example, due to finite isolation between the qubit and amplifier). Post-measurement qubit coherence is then modified to,

$$|\hat{\rho}'_{01}| = \rho_b e^{-2|\alpha|^2 \sin^2\theta}, \tag{S7}$$

where $0 \leq \rho_b \leq 1/2$ is the dephasing in excess of that caused by the readout pulse.

## B. Measurement efficiency

### 1. Definition

As a qubit is dephased by measurement, classical information about the qubit state may also be gained. As given by Eq. S7, the amount of dephasing — and therefore possible information gain — is a function of the separation in phase space between coherent states $|\alpha_0\rangle$ and $|\alpha_1\rangle$. In practice, loss present between the qubit and detector will diminish this separation before $|\alpha_0\rangle$ and $|\alpha_1\rangle$ can be measured. Loss scales the measured coherent state amplitudes by $|\alpha| \to \sqrt{\eta_{\mathrm{loss}}}|\alpha|$, where $0 \leq 1 - \eta_{\mathrm{loss}} \leq 1$ is the fraction of readout energy remaining upon amplification, Fig. S1a. The total efficiency of a measurement chain, i.e. its *measurement efficiency* [28], is equal to


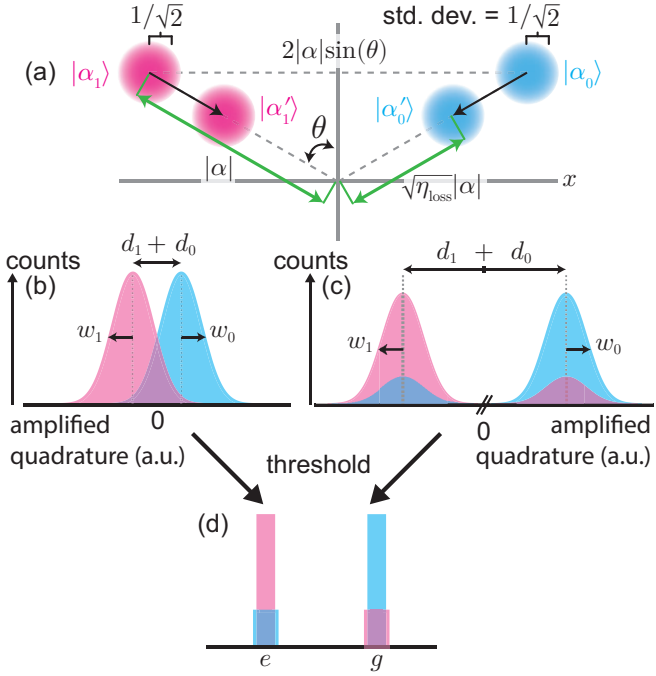

FIG. S1. **Dispersive readout model.** (a) A coherent state of amplitude $|\alpha|$, whose wavefunction $|\langle x|\alpha\rangle\langle\alpha|y\rangle|$ is illustrated by a blob in phase space, is a Gaussian function with a single quadrature standard deviation $1/\sqrt{2}$. This coherent state acquires a phase shift $\pm\theta$ dependent on the qubit state. Loss before a detector scales the amplitude of this coherent state by $\sqrt{\eta_\text{loss}}$ to create $|\alpha'_0\rangle$ and $|\alpha'_1\rangle$, here assuming that $\eta = \eta_\text{loss}$. (b) Linear measurement along the $x$-quadrature yields the cyan (pink) distribution if the qubit was prepared in the ground (excited) state. The height of these distributions are proportional to the modulus squared of the coherent state wavefunctions $|\langle x|\alpha'_0\rangle|^2$ and $|\langle x|\alpha'_1\rangle|^2$, and have standard deviations $w_0$ and $w_1$. After noiseless parametric amplification, the experimentally measured signal-to-noise ratio (SNR) is defined according to Eq. S9. Errors due to finite SNR and preparation infidelity, illustrated in (c), are not distinguishable when a linear measurement result is thresholded as shown in (d). This is true regardless of whether the thresholding is due to use of a bifurcation amplifier, or due to post-processing of a linear measurement.

$\eta = \eta_\text{loss}\eta_\text{amp}$, where $0 \leq \eta_\text{amp} \leq 1$ is the efficiency of the amplifier chain (quantifying its added noise). In this work, we measure $\eta$ but we assume $\eta_\text{amp} = 1$ and therefore $\eta = \eta_\text{loss}$, because we use a phase-sensitive parametric amplifier which in principle adds no noise along the amplified quadrature [29].

### 2. Linear measurement

Loss before measurement transforms $|\alpha_0\rangle$ and $|\alpha_1\rangle$ to the states $|\alpha'_0\rangle$ and $|\alpha'_1\rangle$, respectively, Fig. S1a. The ability of a detector to discriminate $|\alpha'_0\rangle$ and $|\alpha'_1\rangle$ may be quantified by a signal-to-noise ratio (SNR) [1]. We de-

fine SNR [5, 30] as the separation in phase space between these states divided by their single quadrature standard deviation of $1/\sqrt{2}$. The expected SNR of a measurement along the $x$-quadrature is therefore

$$\text{SNR} = \sqrt{8\eta}|\alpha|\sin\theta. \quad (S8)$$

Referring to Fig. S1b,c, this is equal to the experimentally measured value of

$$\text{SNR} = \frac{|d_1 + d_0|}{\sqrt{w_1^2 + w_0^2}}, \quad (S9)$$

where $|d_1 + d_0|$ is the separation between the two measured histograms (corresponding to the qubit in the ground/excited state), and $w_1 = w_0$ are the standard deviations of these histograms (corresponding to the coherent state variance).

Eq. S7 and Eq. S8 can be used together to quantify the efficiency of a linear measurement

$$\eta = \frac{\text{SNR}^2}{-4\log\left(|\hat{\rho}'_{01}|/\rho_b\right)}, \quad (S10)$$

which is a function of the experimentally measurable quantities of $|\hat{\rho}'_{01}|$, $\rho_b$ and SNR. For any readout amplitude $|\alpha|$, measurement of SNR and dephasing $|\hat{\rho}'_{01}|/\rho_b$ thus determines $\eta$. It is expedient to simplify this determination by measuring $|\hat{\rho}'_{01}|$ and SNR at different values of readout amplitude. Consider a readout amplitude in the experimental units $\epsilon$ proportional to $|\alpha|$ (e.g. voltage bias on a mixer). From Eq. S8, the signal-to-noise ratio will increase linearly with $\epsilon$. From Eq. S7, qubit coherence is reduced as a Gaussian function of the readout amplitude $\epsilon$. Therefore,

$$\text{SNR} = a\epsilon, \qquad |\hat{\rho}'_{01}(\epsilon)| = \rho_b e^{-\epsilon^2/2\sigma^2}, \quad (S11)$$

where $a$ is a constant of proportionality and $\sigma$ is a Gaussian standard deviation. Substituting these expressions into Eq. S10 and solving for $\eta$ gives

$$\eta = \frac{a^2\sigma^2}{2}. \quad (S12)$$

Eq. S12 gives the measurement efficiency in terms of $\sigma$ and $a$, experimental quantities which are determined from fits to measurements of $|\hat{\rho}'_{01}|$ and SNR as functions of the readout amplitude $\epsilon$. We reiterate that in this work, including Eq. S12, we define measurement efficiency such that $\eta = 1$ when using an ideal, phase-*sensitive* amplifier, with no other loss or added noise introduced by the detector [31].

### 3. Bifurcated measurement

The measurement discussed in this work, however, does not use a linear amplifier. Pumping a parametric amplifier into bifurcation is a non-unitary process which

destroys information: all possible input states are irreversibly mapped to two output states, illustrated by the distribution in Fig. S1d. Finite SNR and preparation infidelity are indistinguishable after thresholding which prevents a direct measurement of SNR for use in Eq. S10 and thus Eq. S12.

To understand the efficiency of a bifurcated measurement, we instead consider the qubit readout fidelity [32]

$$F_r = 1 - P(e|0) - P(g|\pi), \qquad (S13)$$

where $P(e|0)$ and $P(g|\pi)$ are the error probabilities of measuring the qubit to be in the excited (ground) state, when the qubit is prepared in the ground (excited) state, respectively, such that $0 \leq F_r \leq 1$. Intuitively, readout fidelity is the probability of the detector to correctly discriminate the state that the qubit has been prepared in: $F_r = \frac{1}{2}\left[P(e|\pi) - P(e|0)\right] + \frac{1}{2}\left[P(g|0) - P(g|\pi)\right]$.

Readout fidelity is limited by both finite readout amplitude and preparation infidelity. Preparation infidelities, $p_0$ and $q_0$, are defined as the probability of having prepared the qubit in the excited (ground) state when trying to prepare it in the ground (excited) state, respectively. These infidelities $p_0$ and $q_0$ define a maximum readout fidelity $F_0 = 1 - p_0 - q_0$.

To model the readout fidelity, we model our bifurcated measurement as a linear amplification whose result (a measured value along the $x$-quadrature, Fig. S1) is thresholded by its sign, assigning the qubit state to one of two distinct outcomes. Measurement error is the sum of correct state preparation but incorrect assignment, and incorrect state preparation with correct assignment [32]:

$$P(e|0) = \int_{-\infty}^{0} dx \left[(1-p_0)\left|\langle x|\alpha_0'\rangle\right|^2 + p_0\left|\langle x|\alpha_1'\rangle\right|^2\right], \qquad (S14)$$

$$P(g|\pi) = \int_{0}^{\infty} dx \left[q_0\left|\langle x|\alpha_0'\rangle\right|^2 + (1-q_0)\left|\langle x|\alpha_1'\rangle\right|^2\right]. \qquad (S15)$$

Evaluating these integrals and plugging the results into Eq. S13 yields,

$$F_r = F_0 \mathrm{erf}\left[\sqrt{2\eta}|\alpha|\sin\theta\right]. \qquad (S16)$$

Note that the argument of Eq. S16 is always positive, since $0 \leq \theta \leq \pi/2$ as defined in Fig. S1a. Using Eq. S7, readout fidelity can be expressed in terms of the measurement induced dephasing,

$$F_r = F_0 \mathrm{erf}\left[\sqrt{-\eta\log\left(|\hat{\rho}_{01}'|/\rho_b\right)}\right]. \qquad (S17)$$

In order to finally determine $\eta$, we use Eq. S17 to solve for $\eta$ in a similar manner to how we obtained Eq. S12. Again, we consider readout amplitude in the experimental units of $\epsilon \propto |\alpha|$ (see Eq. 2 in the main text, also). From Eq. S16, readout fidelity will increase as an error function with respect to $\epsilon$, and qubit coherence will still decrease as a Gaussian:

$$F_r(\epsilon) = F_0 \mathrm{erf}\left[\nu\epsilon\right], \qquad \left|\hat{\rho}_{01}'(\epsilon)\right| = \rho_b e^{-\epsilon^2/2\sigma^2}. \qquad (S18)$$

Where $\nu$ and $\sigma$ are obtained from fits of experimental data, e.g. Fig 3c. Using Eq. S17 and Eq. S18 to solve for $\eta$ gives

$$\eta = 2\nu^2\sigma^2. \qquad (S19)$$

Eq. S19 (Eq. 5 in the main text) gives the measurement efficiency $\eta$ in terms of $\sigma$ and $\nu$, quantities which can be experimentally determined from fits to measurements of $|\hat{\rho}_{01}'|$ and $F_r$ as functions of $\epsilon$.

### C. Information efficiency

The methods described in the previous section are specific to either a linear or bifurcated measurement. Alternatively, measurement of a qubit using *any* type of detector can be understood in terms of information theory [7, 8]. This analysis can be useful if the specific mechanism for qubit readout measurement is unclear or does not cleanly follow the models of Eq. S12 or Eq. S19. This analysis will also allow us to quantitatively compare the information gained by linear vs. bifurcated measurements.

#### 1. Accessible information

When a qubit prepared in a superposition state $(|0\rangle + |1\rangle)/\sqrt{2}$ is measured, it is collapsed toward the eigenstate $|0\rangle$ or $|1\rangle$. How projective, or how strong, the measurement is quantifies this collapse. For two quantum states $|\varphi_0\rangle$ and $|\varphi_1\rangle$ of a measurement system which are entangled with the qubit as given by Eq. S3, this collapse is characterized by the 'error probability' $r$ associated with discriminating $|\varphi_0\rangle$ and $|\varphi_1\rangle$. Error here, given by the Helstrom bound [33], is a function of the non-orthogonality of these states. For pure states $|\varphi_0\rangle$ and $|\varphi_1\rangle$ this minimum error probability is $r = \frac{1}{2}\left(1 - \sqrt{1 - |\langle\varphi_0|\varphi_1\rangle|^2}\right)$, which from Eq. S5 equals

$$r = \frac{1}{2} - \frac{1}{2}\sqrt{1 - 4|\hat{\rho}_{01}'|^2}, \qquad (S20)$$

with $0 \leq r \leq \frac{1}{2}$. The Helstrom bound has a simple geometric interpretation: Fig. S2 illustrates a qubit density matrix before and after measurement in the $z$-basis, such that a superposition state is partially projected onto $|0\rangle$, with some remaining phase coherence $|\hat{\rho}_{01}'|$. The error probability $r$ is the remaining uncertainty of the post-measurement qubit state, equal to the length of the green vector.

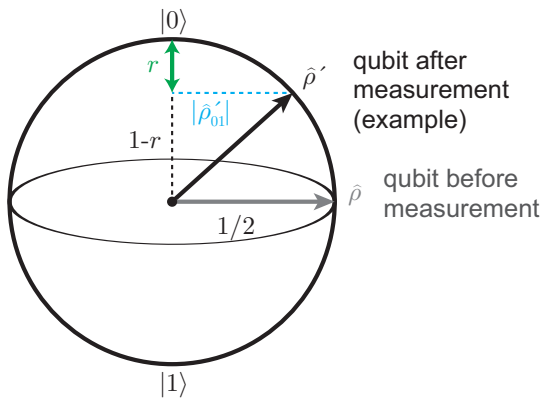

FIG. S2. **The Helstrom bound.** A qubit is prepared in a superposition state $\hat{\rho}$ and non-projectively measured in the $z$-basis. For a given measurement outcome, the post-measurement qubit state is $|\phi\rangle = \sqrt{1-r}\,|0\rangle + \sqrt{r}\,|1\rangle$, with density matrix $\hat{\rho}' = |\phi\rangle\langle\phi|$. The projection of this state along the $z$-axis, $1-r$, quantifies the remaining uncertainty $r$ about the qubit state. In this example a projection toward $|0\rangle$ is illustrated, but a projection toward $|1\rangle$ occurs with equal probability.

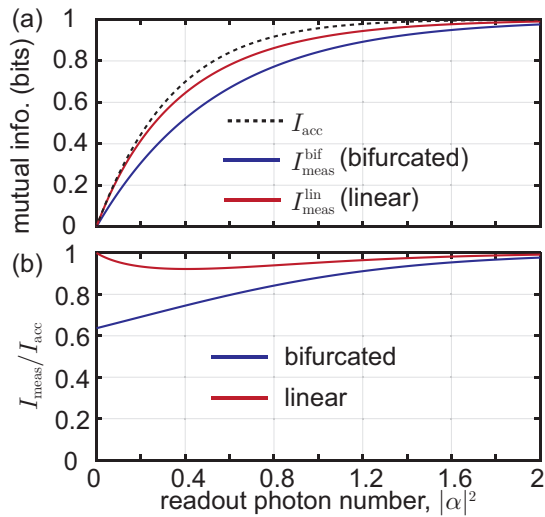

FIG. S3. **Information gained by a qubit measurement.** (a) Accessible information, $I_{\mathrm{acc}}$ from Eq. S23, after the qubit is measured by a coherent state with amplitude $|\alpha|$ (dashed line, and assuming $\theta = \pi/2$). $I_{\mathrm{acc}}$ grows with $|\alpha|$, and (b) is always greater than or equal to the information gain from either a bifurcated (blue) or linear (red) measurement.

Explicitly, consider the states $|\varphi_0\rangle$ and $|\varphi_1\rangle$ to be prepared with equal probability. We take that a measurement of $|\varphi_0\rangle$ occurs with error probability $r_0$, and a measurement of $|\varphi_1\rangle$ with error probability $r_1$. The probability of obtaining the measurement outcomes corresponding to $|\varphi_0\rangle$ or $|\varphi_1\rangle$ are therefore $s_0 = (1-r_0)/2 + r_1/2$ and $s_1 = 1 - s_0$, respectively. The quantities $r_1/2s_0$ and $r_0/2s_1$ are the fractional likelihoods of error, given by Bayes' theorem. This is generalizable to the case where $|\varphi_0\rangle$ and $|\varphi_1\rangle$ are prepared with different probabilities, but here we consider only the case of Eq. S3 for simplicity.

Measurement therefore gains *information* about the qubit state, quantified by the probabilities $r_1/2s_0$ and $r_0/2s_1$. As in Ref. [8], this information is quantified by the mutual information,

$$I = 1 - \left[s_0 H\left(\frac{r_1}{2s_0}\right) + s_1 H\left(\frac{r_0}{2s_1}\right)\right], \quad (S21)$$

where $H(x) = -x\log_2(x) - (1-x)\log_2(1-x)$ denotes the binary entropy.

The accessible information $I_{\mathrm{acc}}$ is defined as the maximum possible mutual information $I$ which may be obtained by a measurement. This occurs when the measurement error probabilities saturate the Helstrom bound, i.e., $r_0 = r_1 = r$ and $s_0 = s_1 = 1/2$. Eq. S21 then simplifies to

$$I_{\mathrm{acc}} = 1 - H(r). \quad (S22)$$

Note that $I_{\mathrm{acc}}$ is a function of *only* the post-measurement qubit density matrix $|\hat{\rho}'_{01}|$, on which $r$ depends via Eq. S20.

### 2. Information gain

We can similarly determine the information attained by measurement. The information $I^{\mathrm{bif}}_{\mathrm{meas}}$ gained from a bifurcated measurement (here meaning two possible outcomes) is:

$$I^{\mathrm{bif}}_{\mathrm{meas}} = 1 - \frac{1}{2}H\left[P(e|0)\right] - \frac{1}{2}H\left[P(g|\pi)\right]. \quad (S23)$$

Here, we take that $|\varphi_0\rangle = |\alpha\rangle$ and $|\varphi_1\rangle = |-\alpha\rangle$ (assuming $2\chi \gg \kappa_r$ such that $\theta = \pi/2$, for simplicity). Also for simplicity, we assume that preparation infidelity is zero, $p_0 = q_0 = 0$. The probability of a measurement error for a bifurcated measurement, Eq. S15, therefore reduces to $P(e|0) = P(g|\pi) = \int_{-\infty}^{0}|\langle\alpha|x\rangle|^2\,dx = \frac{1}{2}\left(1 - \mathrm{erf}\left[\sqrt{2}|\alpha|\right]\right)$. This is plugged into Eq. S23 to model $I^{\mathrm{bif}}_{\mathrm{meas}}$.

Of course, a detector need not only give two measurements but instead can yield a continuum of outcomes. An example of this is a *linear* measurement such as homodyne detection of a coherent state which has been linearly amplified. To quantify information gain from a linear measurement, consider discrimination between the pink and blue Gaussian distributions in Fig. S1b, which are prepared with equal likelihood of $1/2$. If the qubit was prepared in the ground state (cyan distribution, Fig. S1), the probability of error for a measurement returning the value $x$ is $r_x^0 = P(x|\pi) = \sqrt{2/\pi}\,e^{-2(x-|\alpha|)^2}$. And if the pink distribution corresponding to the qubit in the excited state was prepared, the probability of error is $r_x^\pi = P(x|0) = \sqrt{2/\pi}\,e^{-2(x+|\alpha|)^2}$. The probabil-

ity of measuring the value $x$ at all is the sum $s_x = \frac{1}{2}\left[P(x|0) + P(x|\pi)\right] = \frac{1}{2}\left(r_x^0 + r_x^\pi\right)$. At each $x$, the information acquired from measurement is the entropy of the normalized measurement probability, $H\left(r_x^0/s_x\right)$ or $H\left(r_x^\pi/s_x\right)$. The mutual information from measurement is the integral of these conditional entropies over all values $x$, weighted by the likelihood $s_x$ of that measurement outcome occurring [7, 8],

$$I_{\text{meas}}^{\text{lin}} = 1 - \frac{1}{2}\int_{-\infty}^{\infty} dx \left[H\left(\frac{r_x^0}{s_x}\right) + H\left(\frac{r_x^\pi}{s_x}\right)\right] s_x, \quad (S24)$$

where $I_{\text{meas}}^{\text{bif}} < I_{\text{meas}}^{\text{lin}}$ for $|\alpha| < \infty$, seen in Fig. S3a. As $|\alpha| \to \infty$, however, both $I_{\text{meas}}^{\text{bif}}$ and $I_{\text{meas}}^{\text{lin}}$ saturate to unity. This means that thresholding a linear measurement destroys information, but only for the case of a weak or non-projective measurement.

### 3. Analysis of experimental results

The ratio $I_{\text{meas}}/I_{\text{acc}}$ defines an *information efficiency* [8]. In Fig. S3b, we plot $I_{\text{meas}}/I_{\text{acc}}$ for the experimental readout characterization shown in Fig. 3, compared to various models for a bifurcated measurement.

As seen in Fig. S3, choice of a bifurcated detector destroys information but *only* for a 'weak' (non-projective) measurement, i.e. $I_{acc} < 1$. This is seen quantitatively in Fig. S4: the black, dashed diagonal line at $I_{\text{meas}} = I_{\text{acc}}$ bounds any measurement. The pink, dashed line below it gives the limit of a 'perfect' bifurcated measurement (such that $n_b = 0$, $\eta = 1$ and $F_0 = 1$). However in the projective measurement limit, meaning that $I_{\text{acc}} \to 1$, the information gained for a bifurcated measurement also goes to one. There is therefore no disadvantage to using a bifurcated detector, as opposed to a linear detector, when making a projective measurement.

In Fig. S4 we also plot $I_{\text{meas}}$ vs. $I_{\text{acc}}$ for the variable strength measurements shown in Fig. 3c. Both the 'pump on' and 'pump off' data are compared to models (cyan and indigo lines) which use the parameters returned from the fits shown in Fig. 3c. The strong agreement between experiment and these models supports our determination of $n_b$, $\eta$ and $F_0$ in the main text.

## III. DEVICE

In this section, we turn to the SIMBA itself; specifically, the microwave engineering which went into its design. The SIMBA is a relatively complicated multi-layer circuit. Its modular design can be understood, however, by its compartmentalization into two superconducting switches (tunable inductor bridges, TIBs) surrounding a two port Josephson parametric amplifier (JPA).

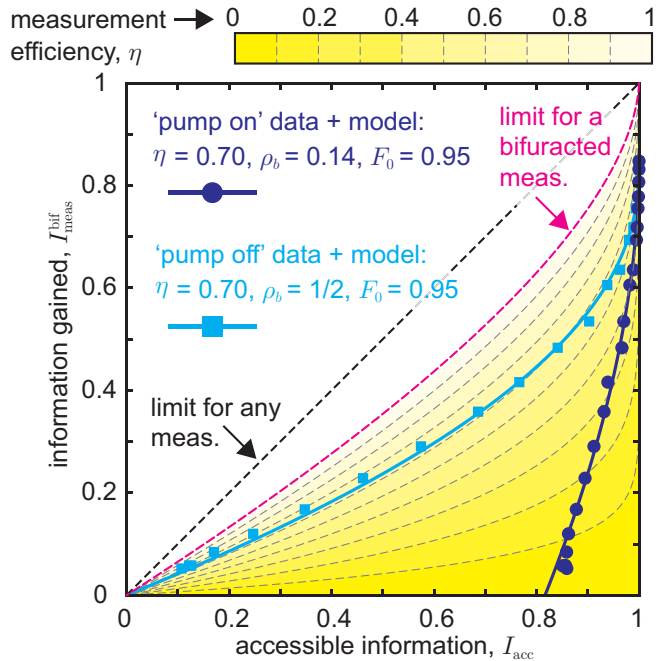

FIG. S4. **Information theory analysis.** A comparison of the information gained $I_{\text{meas}}$ (Eq. S23) to the accessible information $I_{\text{acc}}$ (Eq. S22) for each variable strength measurement in Fig. 3c (data points). Cyan and indigo data points correspond to the parametric cavity pump turned off or on, respectively, within the variable strength measurement sequence whose backaction is being characterized. Data are compared to models (solid lines) which assume thresholded dispersive readout and the parameters obtained from fits to the data in Fig. 3. Dashed gray lines are models for bifurcated measurement with no excess backaction $n_b = 0$, and unit fidelity $F_0 = 1$, but measurement efficiency $\eta < 1$ (and $\eta = 1$ for the pink dashed line).

### A. Design and layout

A layout of the SIMBA chip used in this work is shown in Fig. S5, and an optical micrograph of the device is shown in Fig. S6.

### B. Fabrication

Devices were fabricated at NIST Boulder in a $Nb/AlO_x/Nb$ tri-layer process [34]. The SIMBA fabrication procedure is especially similar to that of the devices used in Refs. [35–39]. A low-loss amorphous silicon dielectric (loss tangent $\delta = 1.5 - 5 \times 10^{-4}$ at mK temperatures) [40] was used in the metal-insulator-metal capacitors within the TIBs.

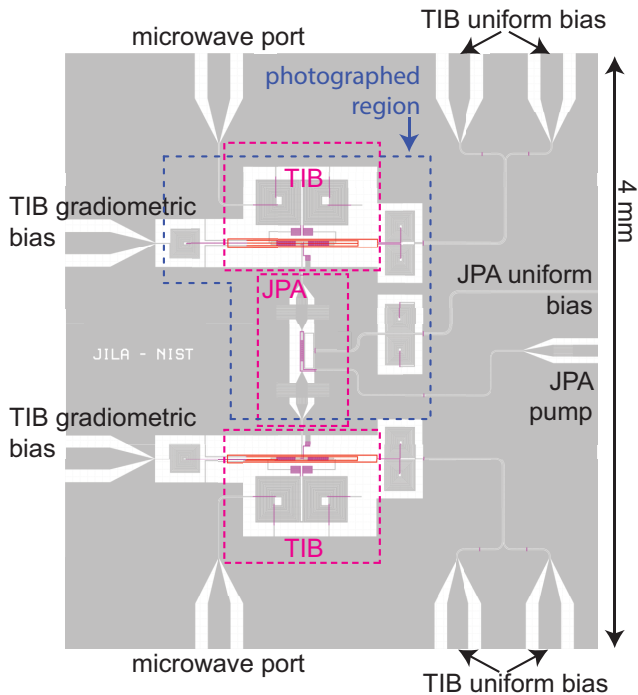

FIG. S5. **SIMBA chip layout**. The SIMBA consists of two TIBs and a JPA. A micrograph of the region within the blue, dashed line is shown in Fig. S6. The SIMBA chip extends 2.5 mm to the right of the region illustrated here, such that the wire bonding pads for the 'JPA uniform bias' and 'JPA pump' lines are not shown.

### C. Superconducting switch

In this section we delve into the internal complexity of the TIB, showing how it has been engineered to function as a simple microwave switch. Before doing so, we note that the TIBs in this work have undergone significant improvements compared to those in our previous work. In particular, the TIBs in Refs. [35, 36] had a chip-mode around 5 GHz (near to the qubit frequency in this SIMBA experiment), and, had greater loss out of their bias lines due to a lack of any on-chip, low-pass filters on these lines. Finally, these bias lines were constructed in an 'unshielded' way such that crosstalk between bias lines on a compact circuit like the SIMBA would likely have presented a problem. Expanding upon techniques developed in Ref. [37], the TIBs in this work have been engineered to eliminate these specific problems. We believe that such improvements were essential to achieving high-quality performance from the integrated SIMBA device.

Conceptually, the TIB can be thought of as a superconducting analog to a microwave mixer, with diodes replaced by SQUID arrays. As with a mixer, the TIB functions as a microwave switching/modulation element where symmetry of a Wheatstone bridge allows for high-performance, broadband operation. In particular, the process of preserving vs. breaking the symmetry of the bridge allows for transmission through the TIB to be tuned by a far greater ratio than its constituent inductors can be tuned.

A lumped-element schematic of a TIB is shown in Fig. S6b: a balun couples the left port of the TIB to the differential voltage across the top and bottom nodes of the bridge (nodes $a$ and $c$). No signal can couple between the two ports when the bridge is balanced, meaning that all four bridge inductors have equal value. To see this, consider an oscillating signal of amplitude $v$ applied at the left port of the lumped element circuit in Fig. S6b. Voltage at the top and bottom nodes of the bridge (nodes $b$ and $d$) will oscillate with amplitude $\pm v'$, respectively, creating an effective ground at the right port and therefore the TIB will reflect (the amplitude $v'$ will in general depend on the operating frequency, choice of capacitors, etc.). If instead the bridge is imbalanced, as drawn, the symmetry of the bridge is broken so that the right port does *not* see an effective ground, and thus transmission can be nonzero. Capacitors are added to match the circuit over a desired frequency range.

In the TIB circuit layout, the Wheatstone bridge is twisted into a figure-eight geometry, Fig. S6c, in order to tune the bridge imbalance with a single bias line while preserving as much symmetry in the circuit as possible. This bias line runs through the center of the figure-eight and puts a gradiometric flux $\pm \Phi_g$ into the SQUID arrays on opposite sides of the bridge. At the same time, all the arrays see an identical uniform background flux $\Phi_u$. The inductance $l_+$ and $l_-$ of the thick and thin inductors in Fig. S6c becomes [35],

$$l_{+/-} = \frac{l_0}{\left|\cos\left(\frac{\Phi_u \pm \Phi_g}{2\phi_0}\right)\right|}. \quad (S25)$$

Where $\phi_0 = \hbar/2e$ is the reduced magnetic flux quantum, with $e$ the charge of an electron. The inductance $l_0 = N\phi_0/2I_c = 0.66\,\mathrm{nH}$ is determined by the Josephson junctions critical currents, $I_c = 5\,\mu\mathrm{A}$ and the number of SQUIDs per array, $N = 20$. As long as $\Phi_u/2\phi_0$ is not a multiple of $\pi/2$, applying a nonzero $\Phi_g$ will imbalance the bridge such that $l_+ \neq l_-$ and transmission will be nonzero. The greatest imbalance is achieved when $\Phi_g/2\phi_0 = \pi/4$ and $\Phi_u/2\phi_0 = \pi/4 + n\pi$, where $n$ is an integer.

The gradiometric bias lines (red, Fig. S6a) contains a low-pass filter (LPF), realized with a $\sim 20\,\mathrm{nH}$ spiral inductor. This filter limits microwave power coupling out of the bias line. A numerical finite-element simulation indicates that this inductor has a self-resonance frequency of 7.9 GHz (note that in this style of inductive filter, a higher inductance will generally lead to a lower self-resonance frequency). Further simulations indicate that transmission from a microwave port of the TIB out the bias port is generally smaller than $-40\,\mathrm{dB}$ between 4 and 8 GHz with inclusion of this LPF, but as high as $-20\,\mathrm{dB}$ without it. At the operational frequency of 6.34 GHz,

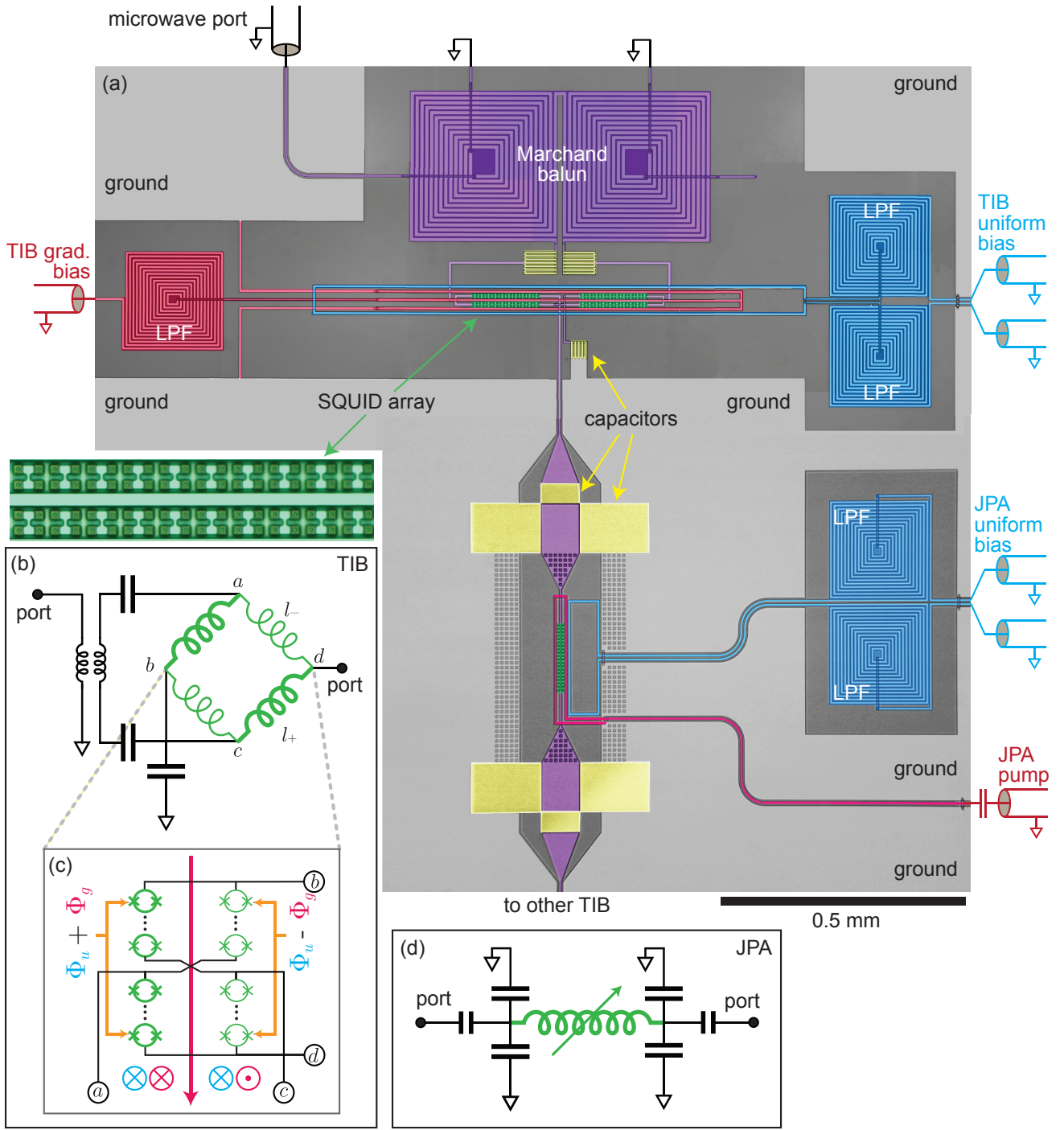

FIG. S6. **SIMBA design**. (a) False color optical micrograph of the region of the SIMBA boxed in Fig. S5. The SIMBA consists of a 2-port parametric cavity (the JPA) with a superconducting switch (a TIB) placed on each port, as in Fig. 1b. The TIB is the structure in the top half of this image, and the JPA is the structure in the bottom half. The bias lines to both devices contain low-pass filters (LPFs) in order to prevent microwave signal in the SIMBA being transmitted out of them. Note that the ground plane around the JPA is waffled in order to pin flux vorticies in place. (b) Lumped element schematic for the TIBs used in this work. The input of the SIMBA (output of the readout cavity), is connected to the port of the TIB containing the balun (left), while the single-ended port (without a balun, right) is connected to the JPA. The TIB is constructed from a Wheatstone bridge of SQUID arrays, with capacitors added to match the circuit. (c) The Wheatstone bridge is constructed by twisting the SQUID arrays into a figure-eight geometry. The flux through the SQUID arrays is a combination of a background, uniform flux $\Phi_u$, and a gradiometric flux $\Phi_g$. The gradiometric flux is controlled by the current in a bias line running through the center of the figure-eight. (d) Lumped element schematic of the JPA. The tunable inductor is also realized with an array of SQUIDs. Flux through the SQUIDs can be controlled by two on-chip bias lines. One of these bias lines is designed for the microwave frequency parametric cavity pump signal, and the other is designed to apply a dc flux to tune the parametric cavity frequency.

transmission from a microwave port of a TIB out of its gradiometric bias port is simulated to be between $-39\,\mathrm{dB}$ and $-47\,\mathrm{dB}$ (the exact value changes slightly depending on which of the two microwave ports is used, and whether the TIB is in transmit or reflect mode). The TIBs used in this work contain a second on-chip bias line (blue, Fig. S6a), for applying a dc uniform flux $\Phi_u$. This bias line also contains two LPFs.

### 1. On/off ratio

To test this design, we have fabricated chips which contain a single TIB by itself. In Fig. S7, we show a measurement of transmission through a single TIB when sweeping its gradiometric bias flux $\Phi_g$. Near $\Phi_g = 0$ the TIB is in 'reflect mode', meaning transmission is near zero. As we change to $\Phi_g \neq 0$, the Wheatstone bridge becomes imbalanced and the TIB is changed to a 'transmit mode'. Transmission is shown in Fig. S7c for both an example transmit mode and an example reflect mode, corresponding to horizontal linecuts of the data in Fig. S7b, at the gradiometric flux bias specified by the black arrows. By changing $\Phi_g$, the on/off ratio can be tuned to greater than approximately 50 dB at any frequency between 4 and 7.3 GHz.

The bandwidth of the TIB is set by the octave bandwidth of its Marchand balun, here designed to have a center frequency near 6 GHz [41]. Bandwidth is also affected by the maximum bridge imbalances, $l_+/l_- \sim 4$ [35–37]. In general, a higher maximum imbalance improves bandwidth. The bandwidth and center frequency are also affected by choice of matching capacitors, here $\sim 0.5\,\mathrm{pF}$.

### 2. Power handling

For use in conjunction with a parametric amplifier, it is important for the TIB to remain linear when processing an amplified signal.

We characterize the power handling of a TIB by using the system described in Fig. S8: a single TIB operated near reflect mode is placed at the strongly coupled port of a readout cavity with a resonance frequency of approximately 6.1 GHz. The readout cavity contains a transmon qubit (a different qubit than used elsewhere in this work, at a frequency of approximately 8 GHz). We choose to characterize the TIB power handling in this manner because firstly, the TIB is measured by itself rather than integrated into the more complicated SIMBA. Secondly, inclusion of a qubit in this system allows for an expedient comparison between the separate power thresholds of a standard superconducting qubit-cavity system and a TIB. The separation between these power thresholds, apparent in Fig. S8, is *independent* of any calibration or estimation of loss and gain in the measurement chain.

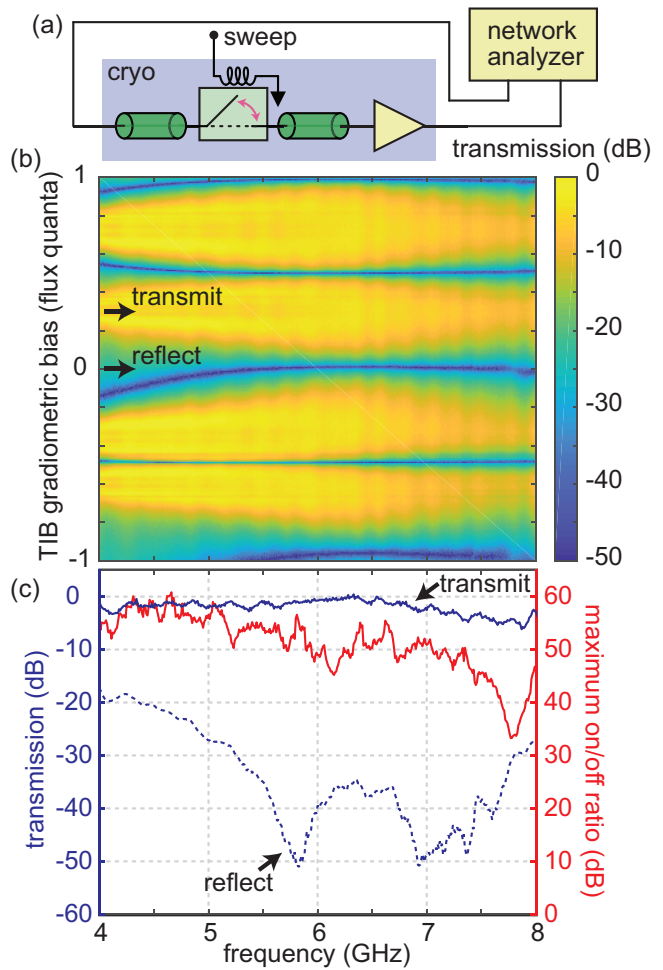

FIG. S7. **TIB transmission characterization.** (a) A single TIB is measured in transmission while sweeping its gradiometric flux bias, $\Phi_g$. (b) Measurement of transmission vs. $\Phi_g$ and frequency. Transmission is normalized to the maximum overall transmission. (c) Transmission vs. frequency at an example transmit mode ($\Phi_g \sim \pi/4$) and reflect mode ($\Phi_g \sim 0$), for uniform flux $\Phi_u \sim \pi/8$. At the reflect mode, transmission is lower than -30 dB over several GHz. The maximum on/off ratio that can be achieved at any given frequency is plotted in red, and is better than 50 dB between 4 to 7.3 GHz. The reflect mode data and on/off ratio are smoothed with a 100 MHz moving filter.

To characterize the TIB linearity, we measure transmission through the readout cavity while sweeping probe power. Doing so, we see a familiar low-power resonance which, at a certain power threshold, transitions into a high-power 'bright-state' peak [42]. As the probe power is further increased by about 10 dB, transmission begins to decrease due to the nonlinearity of the SQUID arrays in the TIB. We quote the power handling of the TIB as the power where this transmission decreases by 1-dB. Using an estimate of the attenuation between the network analyzer and TIB-chip, we determine that the TIB 1-dB compression point is approximately -98 dBm when

near reflect mode, Fig. S8c. This is similar to the power handling reported in previous superconducting switch designs which use similar arrays of SQUIDs [35, 36].

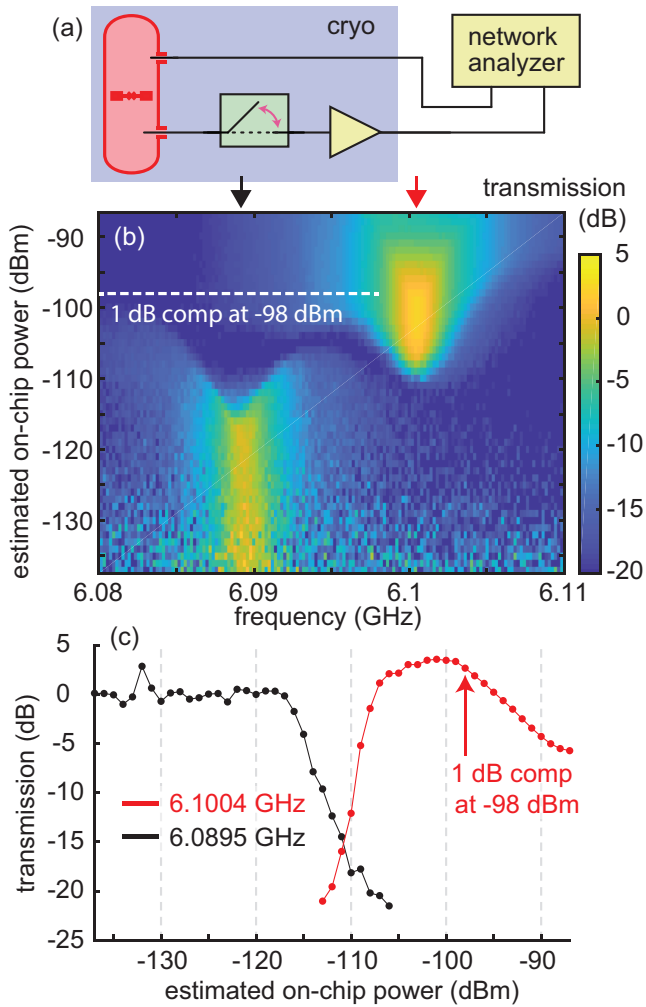

FIG. S8. **TIB power handling characterization.** (a) Transmission as a function of probe power is measured through a readout cavity + transmon qubit system, with a sample box containing a single TIB at its output. The TIB is biased near reflect mode (not fully in reflect mode, so that transmission is still easily measured). (b) Transmission vs. probe power and probe frequency. (c) Linecuts of transmission at the low power cavity frequency (black) and high power cavity frequency (red), specified by the arrows in (b).

### D. Parametric amplifier

Next, we discuss the Josephson parametric amplifier (JPA) contained within our SIMBA. The lumped-element schematic of this JPA is shown in Fig. S6d. A resonator is formed by an inductor (realized with a SQUID array similar to that in the TIBs), with 430 fF capacitors to ground on either side. When the SQUID array inductance is biased to be minimum, the array has an inductance of 0.66 nH. The geometric inductance of the resonator is 0.52 nH. To turn on parametric gain, flux through these SQUIDs is modulated at twice the JPA resonance frequency using a microwave bias line, which contains an on-chip capacitor to block dc-current. Current through a second flux bias line (containing low pass filters) can be used to change the uniform dc-flux through the SQUID array, tuning the JPA frequency. A fingered capacitor (80 fF) is placed between the JPA and each TIB. This limits the coupling rate into/out of the JPA to $\kappa_p^s/2\pi = 52$ MHz when a TIB is in transmit mode.

We characterize the JPA by setting TIB2 to transmit mode and measuring in reflection off of TIB2. Doing so, we find that the JPA frequency is tunable between approximately 4 and 7 GHz (Fig. S9), a similar range over which the TIB is designed to operate. We emphasize that the SIMBA may therefore be tuned to operate over a several GHz frequency range.

We note that the flux tuning curve shown in Fig. S9b was taken when sweeping current through an on-chip bias line rather than through an off-chip coil. The rest of the data reported in this work was taken in a separate cooldown, using an off-chip coil to tune the JPA frequency rather than the on-chip bias line. This functionality was temporarily removed out of concern for low-frequency noise in the on-chip bias line, which can cause the bifurcated amplifier state to become unstable, creating a source of readout error. With proper low-frequency filtering, however, no off-chip coil is needed to operate the SIMBA reported in this work.

## IV. BIFURCATION AMPLIFIER

In this section, we analyze a parametric amplifier pumped near twice its natural resonance frequency. When pumped hard enough, the steady-state field in the resonator bifurcates into one of two states, characterized by equal amplitude but opposite phase [43].

For simplicity, we model our amplifier as the parallel combination of a linear resistor, inductor and capacitor, along with an array of superconducting quantum interference devices (SQUIDs) also in parallel, Fig. S10 [44].

### A. SQUID array model

To model an array of SQUIDs, we first assume that both the self capacitance of the Josephson junctions and the geometric inductance of the SQUIDs themselves, is negligible. We also assume that all junctions in the array have equal critical currents. In these limits, each SQUID is equivalent to a single Josephson junction whose critical current is $2I_c|\cos(\varphi_{\text{ext}}/2)|$ [45]. Here, $I_c$ is the critical current of a single Josephson junction, and $\varphi_{\text{ext}}(t) = 2\pi\Phi_{\text{ext}}(t)/\Phi_0$ is the external magnetic flux applied through each SQUID loop, $\Phi_{\text{ext}}(t)$, scaled by the

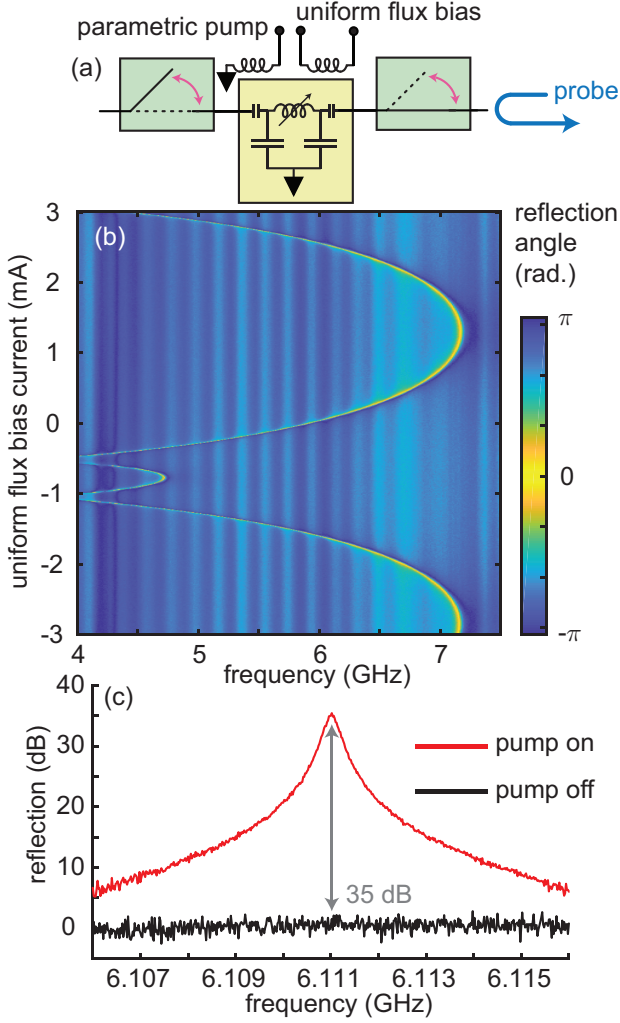

FIG. S9. **JPA characterization.** (a) The JPA within the SIMBA is characterized when TIB1 is set to reflect mode and TIB2 is set to transmit mode. Reflection off of TIB2 is measured. (b) The JPA resonance frequency is tuned by changing a uniform flux through its SQUID array, applied by current through its on-chip dc bias line (blue, Fig. S6). (c) When the JPA is pumped near resonance at the correct amplitude, it operates as a linear amplifier [39]. In the example shown here, the JPA is tuned to give greater than 20 dB of gain over a 2.2 MHz range, with a maximum of 35 dB of gain.

magnetic flux quantum $\Phi_0$. In general, this flux may be time-dependent.

Denoting the superconducting phase difference over an array of identical SQUIDs as $2\pi\Phi/\Phi_0$, the phase difference across each individual SQUID equals $2\pi\Phi/N\Phi_0$, where $N$ is the number of SQUIDs. From Kirchoff's current law, the common-mode current through each individual SQUID equals the current $I_\text{array}$ running through the entire SQUID array. The relationship between current flowing through the array and phase across it is given by the Josephson relation, therefore

$$I_\text{array} = 2I_c \left|\cos\left(\frac{\varphi_\text{ext}}{2}\right)\right| \sin\left(\frac{2\pi\Phi}{N\Phi_0}\right). \tag{S26}$$

From Eq. S26 we see that the effective nonlinearity of the SQUID-array is reduced relative to a single SQUID. In particular, the ratio of the leading nonlinear to the linear term in the expansion of the current-phase relation around $\Phi = 0$ scales with $1/N^2$.

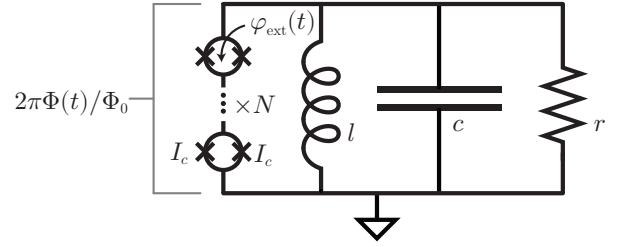

FIG. S10. **Parametric cavity model.** A parametric amplifier is modelled as a parallel combination of a resistor, capacitor, linear inductor, and array of $N$ SQUIDs. The phase difference across the SQUID array is $2\pi\Phi/\Phi_0$.

### B. Full equations of motion

The equations of motion for the circuit in Fig. S10 are two coupled first order differential equations governing the evolution of flux $\Phi(t)$ across the inductor $l$, and charge $Q(t)$ across the capacitor $c$. From Kirchoff's circuit laws, they are:

$$\dot{\Phi} = \frac{Q}{c}, \tag{S27}$$

$$\dot{Q} = -\frac{\Phi}{l} - \frac{Q}{rc} - I_\text{array}. \tag{S28}$$

It is useful to rewrite these equations in terms of the amplitude

$$a = \frac{1}{\sqrt{2\hbar}}\left(\frac{\Phi}{\sqrt{Z_0}} + i\sqrt{Z_0}Q\right), \tag{S29}$$

and its complex conjugate $a^*$ [46]. Here, $Z_0$ is the characteristic impedance of the resonator. The circuit equation of motion becomes,

$$\dot{a} = i\left[-\frac{a - a^*}{2cZ_0} - \frac{Z_0}{2l}(a + a^*) - \sqrt{\frac{Z_0}{2\hbar}}I_\text{array}\right] - \frac{\kappa}{2}(a - a^*), \tag{S30}$$

where $\kappa = 1/rc$.

Note that the linear inductance in Eq. S30 is comprised both of the inductor $l$ and the linear inductance of

the SQUID array $l_{\text{array}}$. We first consider a static external flux bias $\varphi_{\text{ext}}(t) = \varphi_s$. Combining Eq. S30 together with Eq. S26, it follows that the SQUID array adds a $\varphi_s$-dependent linear inductance $l_{\text{array}}$ in parallel to the inductor $l$. The resulting total linear inductance $l_{\text{res}}$ of the device becomes,

$$\frac{1}{l_{\text{array}}} = \frac{2\pi}{\Phi_0} \frac{2I_c \cos(\varphi_s/2)}{N}, \qquad \frac{1}{l_{\text{res}}} = \frac{1}{l} + \frac{1}{l_{\text{array}}}. \quad \text{(S31)}$$

The corresponding values for the resonator impedance and resonance frequency are,

$$Z_0 = \sqrt{\frac{l_{\text{res}}}{c}}, \qquad \omega_0 = \frac{1}{\sqrt{l_{\text{res}} c}}. \quad \text{(S32)}$$

The resonance frequency $\omega_0$ is a $2\pi$-periodic function of the static external flux bias $\varphi_s$. It assumes a maximum at $\varphi_s = 0$ and a minimum at $\varphi_s = \pm\pi$.

Next, we consider the parametric modulation of the SQUID critical current. This is achieved by a modulation of the external flux $\varphi_{\text{ext}}$ through each SQUID around its static bias value,

$$\varphi_{\text{ext}}(t) = \varphi_s + \varphi_m \cos(\Omega t), \quad \text{(S33)}$$
$$\Omega = 2\omega_0 + 2\delta. \quad \text{(S34)}$$

where $\varphi_m$ is the modulation amplitude and the pump detuning $\delta$ is assumed to be small, such that $\delta \ll \omega_0$.

### C. Simplified equations of motion

Eq. S30, along with Eqs. S33 and S34 substituted into Eq. S26, represent the full time-dependent equations of motion for the circuit in Fig. S10. These equations greatly simplify in certain realistic limits.

First, we consider only weak excitations, $2\pi\Phi/\Phi_0 \ll 1$, and in this limit we can expand the current-phase relation of the SQUID arrays in Eq. S26. We also assume that for the entire modulation period, the resonator remains detuned from its maximum frequency such that $0 < \varphi_{\text{ext}}(t) < \pi$, so that we may ignore the absolute value in Eq. S26. Finally, we assume weak parametric modulation amplitude $\varphi_m \ll 1$ so that $\cos(\varphi_{\text{ext}}/2) = \cos(\varphi_s/2) - (\varphi_m/2)\sin(\varphi_s/2)\cos(\Omega t) + \mathcal{O}(\varphi_m^2)$. In these limits, the circuit equation of motion then becomes,

$$\dot{a} = i\left[-\omega_0 a + \beta \cos(\Omega t)(a + a^*) + \frac{\zeta}{3}(a + a^*)^3\right]$$
$$- \frac{\kappa}{2}(a - a^*). \quad \text{(S35)}$$

Where we have defined an effective parametric drive amplitude $\beta$ and an effective nonlinearity $\zeta$,

$$\beta = \frac{\varphi_m \tan(\varphi_s/2) Z_0}{4 l_{\text{array}}}, \quad \text{(S36)}$$

$$\zeta = \frac{(2e)^2}{\hbar} \frac{Z_0^2}{8N^2 l_{\text{array}}}, \quad \text{(S37)}$$

where $e$ is the charge of an electron.

Finally, we transform Eq. S35 to the rotating frame such that $a = Ae^{-i\Omega t/2}$ and $\dot{a} = \left(\dot{A} - i\Omega A/2\right)e^{-i\Omega t/2}$, where $A$ is the cavity field amplitude in the rotating frame. Averaging over fast oscillations removes explicit time-dependence, and the equation of motion simplifies to,

$$0 = i\dot{A} + \delta A + \frac{1}{2}\beta A^* + \zeta|A|^2 A + \frac{1}{2}i\kappa A. \quad \text{(S38)}$$

### D. Phase diagram

We now examine stable solutions to Eq. S38. For absent or small parametric pumping, $\beta \leq \kappa$, Eq. S38 has only the trivial solution $A = 0$. However, for $\beta > \kappa$, Eq. S38 admits non-trivial steady-state solutions, a phenomenon known as parametric oscillation. One finds two stable, nontrivial steady states $A = |A|e^{i\theta_{\text{po}}}$ to Eq. S38, both with the same amplitude,

$$|A|^2 = \frac{1}{\zeta}\left(-\delta + \frac{1}{2}\sqrt{\beta^2 - \kappa^2}\right), \quad \text{(S39)}$$

but with $\pi$-shifted phases which are determined by $\sin(2\theta_{\text{po}}) = \kappa/\beta$. Here $|A|^2$ is the steady-state photon number in the resonator.

From Eq. S39, we can see that these states appear only for $\beta > \kappa$ and $\delta < \delta_{\text{th}} = \frac{1}{2}\sqrt{\beta^2 - \kappa^2}$. In the parameter regime $\beta > \kappa$ and $|\delta| < \delta_{\text{th}}$ the trivial state $A = 0$ is instable, while for $\beta > \kappa$ and $\delta < -\delta_{\text{th}}$ it coexists as a stable state with the parametric oscillation states [9, 47].

In general, noise and nonlinear terms in the resonator equations of motion will lead to more exotic behavior [48] which can limit the performance of the parametric cavity as a bifurcation amplifier. To understand any such effects, we map the region in parameter space where robust bifurcation is observed, Fig. S11. To do so, we make single-shot measurements of the pumped parametric cavity with TIB2 in transmit mode, TIB1 in reflect mode, and no readout or qubit pulses. The variance of the set of these measurements in the I/Q plane is plotted as the color axis in Fig. S11a,b. Variance above the background level indicates a pumped parametric cavity state which is no longer vacuum, including the presence of parametric oscillation. Stable bifurcation is observed when pumping near twice the bare cavity resonance frequency, Fig. S11b. The region of gain deviates from the parabolic region predicted by Eq. S39, however. This deviation is understood to result from higher order nonlinearities in the cavity equations of motion, which can occur when pumping at high enough amplitudes such that the cavity resonance frequency is no longer modulated linearly with flux [9]. Such nonlinearities can lower the steady state photon number [49]. The approximate operating point used to calibrate qubit readout is given by the star symbol, Fig. S11b,c,d, where we observe two stable steady states. At other detunings and pump powers we can measure

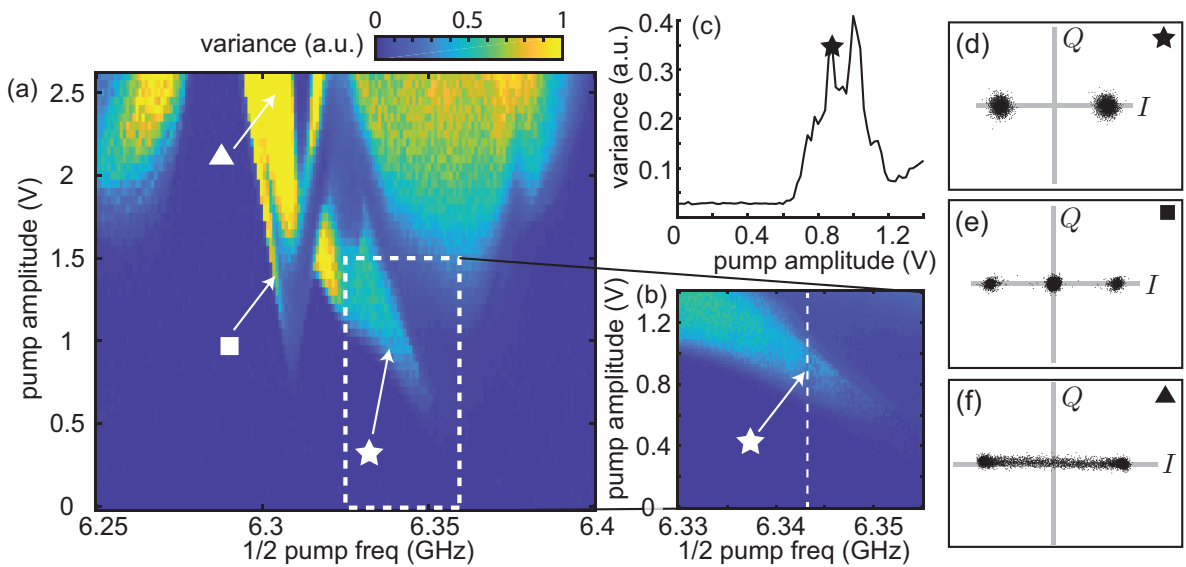

FIG. S11. **Phase diagram for the pumped parametric cavity**. (a) We examine the behavior of the pumped parametric cavity by computing the variance of the distribution of single-shot measurements in the I/Q plane, i.e. the sum of I and Q variances. This is plotted as a function of pump half-frequency and pump amplitude (in units of voltage bias on a double balanced mixer used to modulate the pump amplitude). (b) A higher resolution sweep of the boxed region in (a), and (c) a vertical linecut (the dashed white line in (b)) of variance vs. pump amplitude at the half-pump frequency used for qubit readout. The star indicates the pump amplitude used in this work. (d) At this operating point, the oscillator latches into one of two stable steady states. Each black dot is a single shot measurement. (e) At a negative detuning and higher pump amplitude, we can see three stable states, one at the origin [47]. (f) At even higher pump amplitudes, the oscillator jumps at random between two quasi-stable states, averaging to the distribution shown.

tri-stable states, e.g. Fig. S11e, or amplification but the lack of a stable steady state, e.g. Fig. S11f.

In conclusion, it is useful to understand the dynamics of the parametric cavity in order to use the SIMBA for qubit readout. The phase diagram in Fig. S11 shows that the parametric cavity used in this work deviates somewhat from the lowest-order model previously presented in this section, Eq. S39. This deviation can complicate calibration of readout using a SIMBA, as calibration is in general easier when the parametric cavity bifurcates over a wider range of pump amplitudes and detunings. In order to improve the robustness of bifurcation, loss $\kappa$ should be reduced and the effective nonlinearity $\zeta$ should also be reduced, while taking care that the amplified state does not exceed the TIB power handling.

### E. Photon number

#### 1. Model

We now estimate the pumped parametric cavity photon number $|A|^2$, at the operating point used for the qubit readout reported in the main text. We make this estimate by plugging realistic experimental parameters into Eq. S39. Our JPA has $c = 0.85\,\text{pF}$, $I_c = 5\,\mu\text{A}$ and $N = 20$, referring to the circuit in Fig. S10. At the operating point used in this experiment, the resonator has been detuned from its maximum frequency of $7.1\,\text{GHz}$ to $6.34\,\text{GHz}$ such that $l_{\text{array}} = 0.86\,\text{nH}$, $l = 5.5\,\text{nH}$ and $l_{\text{res}} = 0.74\,\text{nH}$. Eq. S37 therefore gives an effective nonlinearity of $\zeta/2\pi = 49\,\text{kHz}$. Finally, at the operating point where readout is calibrated in the main text, $\delta \sim 0$ and $\beta/\kappa \sim 1.3$ (seen by the star in Fig. S11b). The parametric cavity loss rate is approximately $\kappa/2\pi = 52\,\text{MHz}$ when TIB2 is in transmit mode such that $\beta = 1.3 \times \kappa$ and $\beta/2\pi = 68\,\text{MHz}$. Using these numbers, Eq. S39 predicts a steady-state amplitude of $|A|^2 \sim 450$ photons. This number is expected to be an overestimate, given that higher-order modifications to Eq. S39 are present in this system, seen by the non-parabolic region of bifurcation in Fig. S11b [9, 49].

#### 2. Measurement

We can alternatively measure the photon number in the bifurcated state by comparing the signal-to-noise of our qubit readout to the gain of our amplifier chain. The photon number in the pumped parametric cavity is equal to [47],

$$|A|^2 = \frac{P_s - P_n}{2\left(\kappa_p^s/2\pi\right)\hbar\omega_p 10^{G/10}}, \quad (S40)$$

where $P_s = 710\,\text{nW}$ and $P_n = 200\,\text{nW}$ are the measured signal and noise power levels of our readout, $\kappa_p^s =$

$2\pi \times 52$ MHz is the coupling rate of the parametric cavity to the measurement chain, $\omega_p = 2\pi \times 6.3432$ GHz is the parametric cavity frequency and $G = 69$ dB is the estimated gain of our amplifier chain between the parametric cavity and I/Q mixer. Our estimate of $G$ is based on the specified gain of the amplifiers and the specified loss in the cables/isolators in this line, at the parametric cavity resonance frequency. Using Eq. S40 we predict that the bifurcated state in the parametric cavity contains $|A|^2 \sim 148$ photons. The power the TIBs are exposed to is therefore $148 \times \hbar\omega_p \times \left(\kappa_p^s/2\pi\right) = -105$ dBm, relatively close to, but below, the 1-dB compression point of the TIBs at $-98$ dBm.

We note that this determination of photon number is approximately consistent with the excess backaction of our measurement and the isolation provided by TIB1, both of which we have independently measured. The ratio of excess backaction to bifurcated state amplitude is $10\log_{10}(0.66/148) = -24$ dB. For comparison, the isolation provided by TIB1 is $-26$ dB based on comparison of the swap times between the readout and parametric cavities with TIB1 in transmit mode or reflection mode, Fig. 2 and Fig. S14, respectively.

### F. Bifurcated state stability

Finally, we characterize the stability and reset time of the bifurcated state at the operating point used in qubit readout (Fig. S11d). Ideally, the pumped state in the bifurcation amplifier should be stable while pumped, and quickly decay when the pump is turned off. These qualities are tested by the measurements shown in Fig. S12. While TIB2 is set to transmit mode, the parametric cavity is pumped so that it latches into one of its two bistable states. While doing so, the state in the parametric cavity is digitized over two 100 ns intervals (the same duration used in our readout scheme). These intervals are separated by a delay in which the JPA pump is turned off and TIB2 is set to reflect mode. When sweeping this delay, the second measurement is initially highly correlated with the first. This indicates that the bistable state of the first measurement has yet to dissipate and is seeding the second measurement. After approximately 100 ns without pumping, however, correlation between the first and second measurements has nearly vanished, indicating that the pumped state in the JPA has decayed to near-vacuum. When the pump and TIB state are not changed during the variable delay however, the bifurcated state is extremely stable: the two measurements are greater than 99.9% correlated for any delay between them (with a maximum measured delay of 50 $\mu s$).

## V. EXPERIMENTAL PROCEDURE

We now describe details about how the SIMBA is used to measure a superconducting qubit. We begin by listing parameters of the qubit-cavity system in Table S2. We are using a transmon qubit with a frequency $\omega_q/2\pi$ coupled to a 3d aluminum readout cavity at frequency $\omega_r/2\pi$. The readout cavity dispersive shift $\chi$ is defined such that the readout cavity frequency is $\omega_r/2\pi \pm \chi/2\pi$ dependent on the qubit state [10]. When TIB1 is in reflect mode, the readout cavity is nearly degenerate with the parametric cavity frequency $\omega_p$.

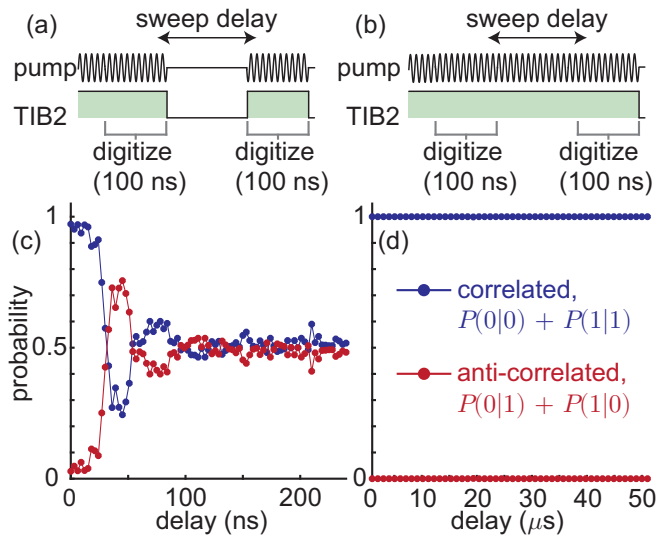

FIG. S12. **Bifurcation reset and stability.** The bifurcated state of the pumped parametric cavity is measured twice, with a variable delay between measurements. Here, the measurements of '0' and '1' correspond solely to the state of the pumped parametric cavity, with TIB1 set to reflect mode for the entire sequence. (a) During the delay, the pump is turned off and TIB2 is set to reflect mode. (b) During the delay, the pump remains on and TIB2 remains in transmit mode. (c) When the pump is turned off, the second measurement quickly becomes uncorrelated with the first. (d) When the pump remains on, however, the state in the cavity remains exceedingly stable. Data points in both plots are the mean of 1024 single-shot measurements.

### A. Calibration

The calibration procedure for superconducting qubit readout using a SIMBA is not significantly more complicated than other superconducting qubit readout schemes. This procedure is summarized below:

1. Tune the JPA frequency to the readout cavity frequency (Fig. 2a).

2. Sweep the JPA pump amplitude such that the JPA gives desired gain/bifurcation Fig. S11c).

3. Choose the readout pulse amplitude and frequency, and the qubit pulse amplitude and frequency. Because the SIMBA is a phase-sensitive amplifier, the

TABLE S2. **Qubit, readout cavity, and parametric cavity parameters.** See Table S3 for details on the loss in readout and parametric cavities.

| Qubit parameters | Value |
| --- | --- |
| probe frequency: $\omega_r/2\pi$ | 6.3432 GHz |
| qubit frequency: $\omega_q/2\pi$ | 4.5077 GHz |
| dispersive shift: $2\chi/2\pi$ | -1.93 MHz |
| qubit anharmonicity: $\alpha$ | -285.9 MHz |
| $T_1$ | $\sim 9\,\mu s$ |
| $T_2^*$ | $\sim 15\,\mu s$ |

| Readout cavity parameters | Value |
| --- | --- |
| external coupling rate (strong): $\kappa_r^s/2\pi$ | 13.8 MHz |
| external coupling rate (weak): $\kappa_r^w/2\pi$ | 8 kHz |
| loss rate, TIBs in reflect mode: $\kappa_r/2\pi$ | 440 kHz |

| Parametric cavity parameters | Value |
| --- | --- |
| external coupling rate: $\kappa_p^s/2\pi$ | 52 MHz |
| loss rate, TIBs in reflect mode: $\kappa_p/2\pi$ | 4.0 MHz |
| effective nonlinearity (Kerr constant): $\zeta/2\pi$ | 49 kHz |

| Readout/parametric cavity coupling rate | Value |
| --- | --- |
| $g/2\pi$, TIB1 in transmit mode | 12.5 MHz[a] |
| $g/2\pi$, TIB1 in reflect mode | 660 kHz[b] |

[a] Measured as the inverse of 4× the 20 ns swap time reported in Fig. 2a.
[b] Measured as the inverse of 4× the 380 ns swap time reported in Fig. S14.

phase difference between the readout tone at $\omega_r/2\pi$ and the pump tone at $2\times\omega_r/2\pi$ must be calibrated.

4. To optimize readout fidelity, sweep the duration for which TIB1 is set to transmit mode (Fig 2b).

5. Fine-tune TIB reflect modes to minimize backaction (Fig. S13c), and to maximize the measurement efficiency.

The first three steps are generally true of any readout scheme which uses a tunable, narrow band and phase-sensitive parametric amplifier. The final two steps are SIMBA-specific.

To maximize efficiency and minimize excess backaction, special care should be taken to determine the best reflect modes for TIB1 and TIB2. The reflect modes of both TIBs occur when current in their gradiometric bias lines is set near zero. This can be quickly checked by measuring transmission through the readout cavity while sweeping the gradiometric flux bias on either TIB1 or TIB2 with the other fixed, Fig. S13a,b. In practice, the optimal reflect mode may occur when this current is slightly offset from zero. Fig. S13c shows a calibration of TIB1 reflect bias by characterizing excess backaction $n_b$, Eq. 2, as a function of the reflect mode gradiometric bias of TIB1 when making a qubit measurement.

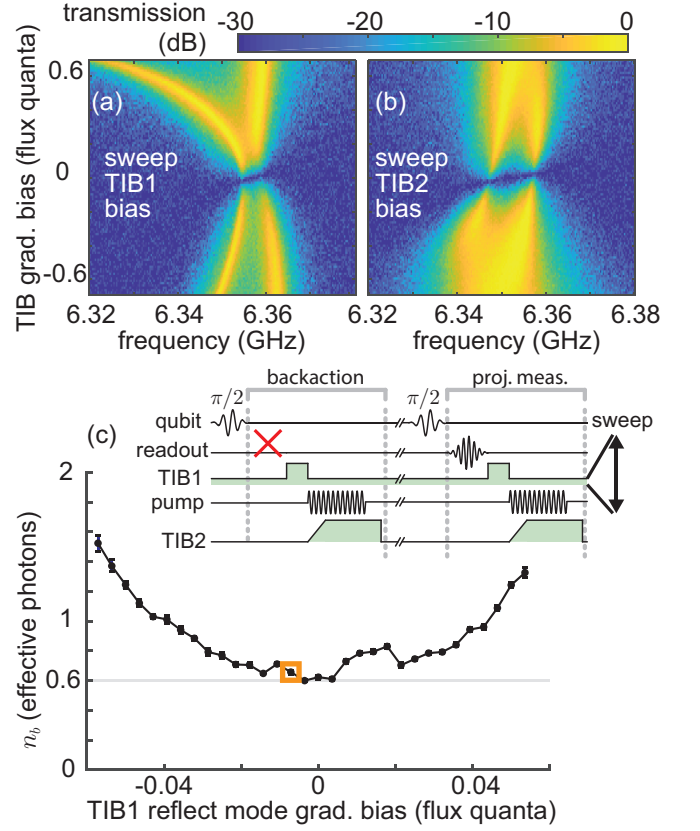

FIG. S13. **Calibration of TIB1 reflect mode.** Transmission through the readout cavity when the gradiometric bias on (a) TIB1 and (b) TIB2 is swept with the gradiometric bias on the other TIB set to $\sim 0.3$. The JPA is tuned near resonance with the readout cavity in both cases. The presence of a reflect mode (transmission $<$ -30 dB in this measurement) is seen by the null in transmission near zero gradiometric bias. (c) Excess backaction $n_b$ is measured by inserting a measurement sequence with zero readout amplitude into a Ramsey sequence, as in Fig. 3. Excess backaction is measured as a function of the gradiometric bias on TIB1, swept near reflect mode. The gold square indicates the operating point used to calibrate measurement efficiency in the main text, Fig. 3 and Fig. 4. Zero gradiometric bias on the $x$-axis is calibrated to correspond to approximately minimum excess backaction.

In the main text, we characterize measurement efficiency when the reflect mode of TIB1 is tuned to provide minimum backaction (gold square, Fig. S13c). Measurement of excess backaction at this operating point is a measure of the isolation provided by TIB1. This isolation can alternatively be measured by the procedure illustrated in Fig. S14: the qubit is prepared in the excited state, and then projectively measured after a delay placed between the readout pulse and the rest of the measurement procedure. The resulting oscillations correspond to

the readout pulse swapping back and forth between the readout and parametric cavities when TIB1 is in reflect mode. The average measured swap time is 380 ns.

The isolation provided by TIB1 can thus be expressed by comparing the ratio of the swap time when TIB1 is in reflect mode, Fig. S14 vs. when TIB1 is in transmit mode (20 ns, Fig. 2b): $T = 20\log_{10}(20/380) = -25.6\,\mathrm{dB}$. For comparison, one commercial cryogenic ferrite circulator provides $\sim -18\,\mathrm{dB}$ of isolation.

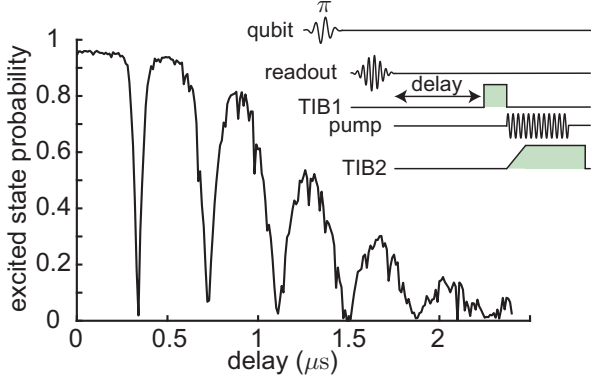

FIG. S14. **Measurement of TIB1 isolation.** The qubit is prepared in the excited state and then measured. A variable delay is swept after the readout pulse but before setting TIB1 to transmit mode. Oscillations show the readout pulse swapping between the readout and parametric cavities when TIB1 remains in reflect mode. These oscillations are much slower than if TIB1 were instead set to transmit mode, Fig. 2b.

### B. Dephasing due to TIB switching

Next, we further examine the backaction caused by switching a TIB. In the main text, Fig. 3c, we measure that connecting the TIBs as in a projective measurement causes $0.05 \pm 0.01$ effective photons of backaction (the left-most data point in Fig 3c, 'pump off' data, cyan). Dephasing due to TIB switching is further investigated by square wave modulating either TIB1 or TIB2 during a Ramsey sequence.

Fig. S15 shows the result of this experiment. The red data (line to guide the eye) results from square-wave modulating TIB1 between reflect and transmit modes during a Ramsey sequence. The gray data results from square wave modulating TIB2. Modulating TIB2 does little to change the qubit coherence time of $T_2^* \sim 15\,\mu\mathrm{s}$. Modulating TIB1, however, reduces the coherence time to several $\mu\mathrm{s}$. The reduction in coherence is abrupt with any square wave modulation frequency, and not highly dependent on the modulation rate. This indicates that perhaps the measured dephasing is not dominated by the act of switching a TIB itself, but is due to increased exposure of the qubit-cavity system to a noise source as a result of setting TIB1 to transmit mode.

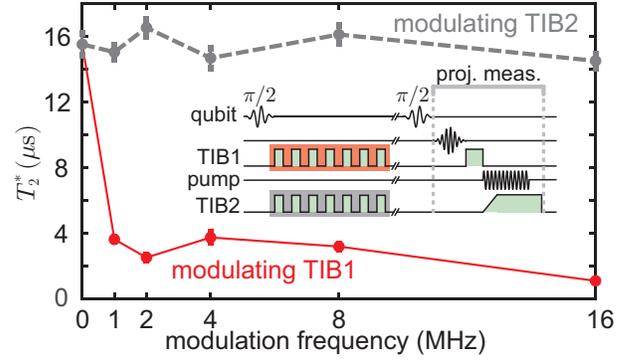

FIG. S15. **Backction from TIB modulation.** A reduction in coherence from the act of switching a TIB is measured by square wave modulating TIB1 (red) or TIB2 (gray) during a Ramsey sequence.

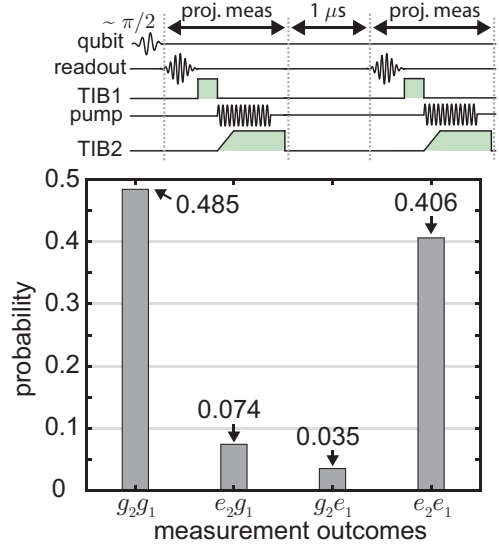

FIG. S16. **Quantum non-demolition characterization.** The qubit is prepared in a superposition state, projectively measured, and projectively measured again after a 1 $\mu$s delay. Measurement outcomes are far more likely to be correlated ($g_2g_1$ and $e_2e_1$) than anti-correlated ($g_2e_1$ and $e_2g_1$), indicating the measurement is largely quantum non-demolition. Reported numbers are the mean of 20480 single-shot measurements.

### C. Quantum non-demolition measurement

Finally, we characterize the degree to which qubit readout using a SIMBA is quantum non-demolition (QND). QND-ness is defined as the likelihood for a measured qubit to remain in its measured eigenstate [51]. It is important that a measurement is QND when a qubit must be repeatedly measured, for instance in measurement-based quantum error correction schemes [13, 52]. In practice, a measurement can be non-QND by kicking the

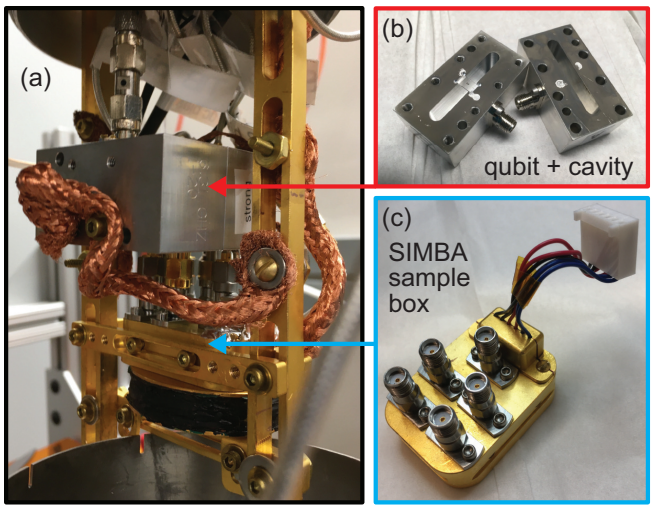

FIG. S17. **Photograph of the experiment.** (a) The qubit-cavity system is attached to the SIMBA sample box using an SMA connection. (b) Photograph of the transmon qubit + 3d readout cavity system, similar in design to the one used inf Ref. [50]. (c) Photograph of the SIMBA sample box. It has five SMA ports: the input/output ports of the SIMBA, two TIB bias ports, and a JPA pump port. The optional dc uniform bias lines on the SIMBA (not used in this work) are connected to the sample box using a 6-pin molex connector which is soldered to the circuit board. In this experiment, a uniform magnetic flux is provided to the SIMBA using an external superconducting coil.

qubit out of its two-level subspace. In general, these effects can become pronounced in readout schemes which require high amplitude readout pulses, or have too much backaction [53, 54].

Low power dispersive readout of superconducting qubits is understood to be QND [10]. We test if this remains true when using a SIMBA by making the measurement shown in Fig. S16: the qubit is prepared in a superposition state, and then projectively measured twice with a $1\,\mu$s delay between measurements (much longer than the reset time, Fig. S12a). QND-fidelity is defined as the extent to which these two measurement outcomes are correlated [55],

$$F_{QND} = \frac{1}{2}\left[P(e_2|e_1) + P(g_2|g_1)\right]. \quad (S41)$$

Where $P(e_2|e_1)$ and $P(g_2|g_1)$ are the probabilities that the second are first measurements both yield 'excited state' ($e_2$ and $e_1$) or both yield 'ground state' ($g_2$ and $g_1$), respectively. Plugging the data shown in Fig. S16 into Eq. S41 gives $F_{QND} = 89\%$ [56].

QND-infidelity is reflected by the probability of outcomes $g_2 e_1$ and $e_2 g_1$ corresponding to decay from the excited to ground state, and excitation from the ground to excited state, respectively. Probability of decay from the excited to ground state is consistent with the qubit relaxation time of $T_1 = 9\,\mu$s, and the $1\,\mu$s delay between measurements. The rate of spontaneous excitation from the ground to excited state may be reflective of the cavity bath temperature when the TIBs are disconnected, potentially related to the dephasing measured when modulating TIB1, S15. (In this work, prior to any qubit measurement both TIBs are set to transmit mode for a duration much longer than $T_1$, as this improved readout fidelity seemingly by reducing the residual qubit excited state population).

### D. Full experimental schematic

The qubit, readout cavity and SIMBA are placed inside of a cryoperm can at the base temperature stage of a dilution refrigerator, Fig. S17. The complete experimental schematic for qubit readout using a SIMBA is shown in Fig. S18. Eccosorb filters are placed on the lines running in and out of the qubit + readout cavity + SIMBA system, in order to shield the qubit from high-frequency radiation.

We note that the SIMBA should be placed as close as possible to the readout cavity in order to minimize the electrical length between them. If any mode formed by this electrical length falls close in frequency to the readout/parametric cavity frequency, a significant fraction of the readout pulse can also couple into it. This lowers the measurement efficiency and can complicate the calibration procedure. In this work, the strongly coupled port of the readout cavity is constructed using an SMA connector, which is then screwed directly into another SMA connector on the SIMBA sample box. This results in approximately 3 cm of waveguide between the readout cavity and SIMBA chip. This length may be significantly shortened in future designs by engineering a more compact connection mechanism.

## VI. FUTURE PERSPECTIVES

In conclusion, using a SIMBA we demonstrate superconducting qubit readout with state-of-the-art measurement efficiency and low excess backaction. The combination of these features is achieved without any ferrite circulator or isolator placed between the qubit and parametric amplifier. Readout is also fast, high fidelity and largely quantum non-demolition. As such, we believe the SIMBA is a promising platform for scalable superconducting qubit measurement. We conclude with an analysis of the limitations of this experiment and opportunities for future improvement.

### A. Measurement efficiency limitations

#### 1. Measurement efficiency model

We seek to understand the various sources of loss which limit measurement efficiency using this SIMBA. To do so,

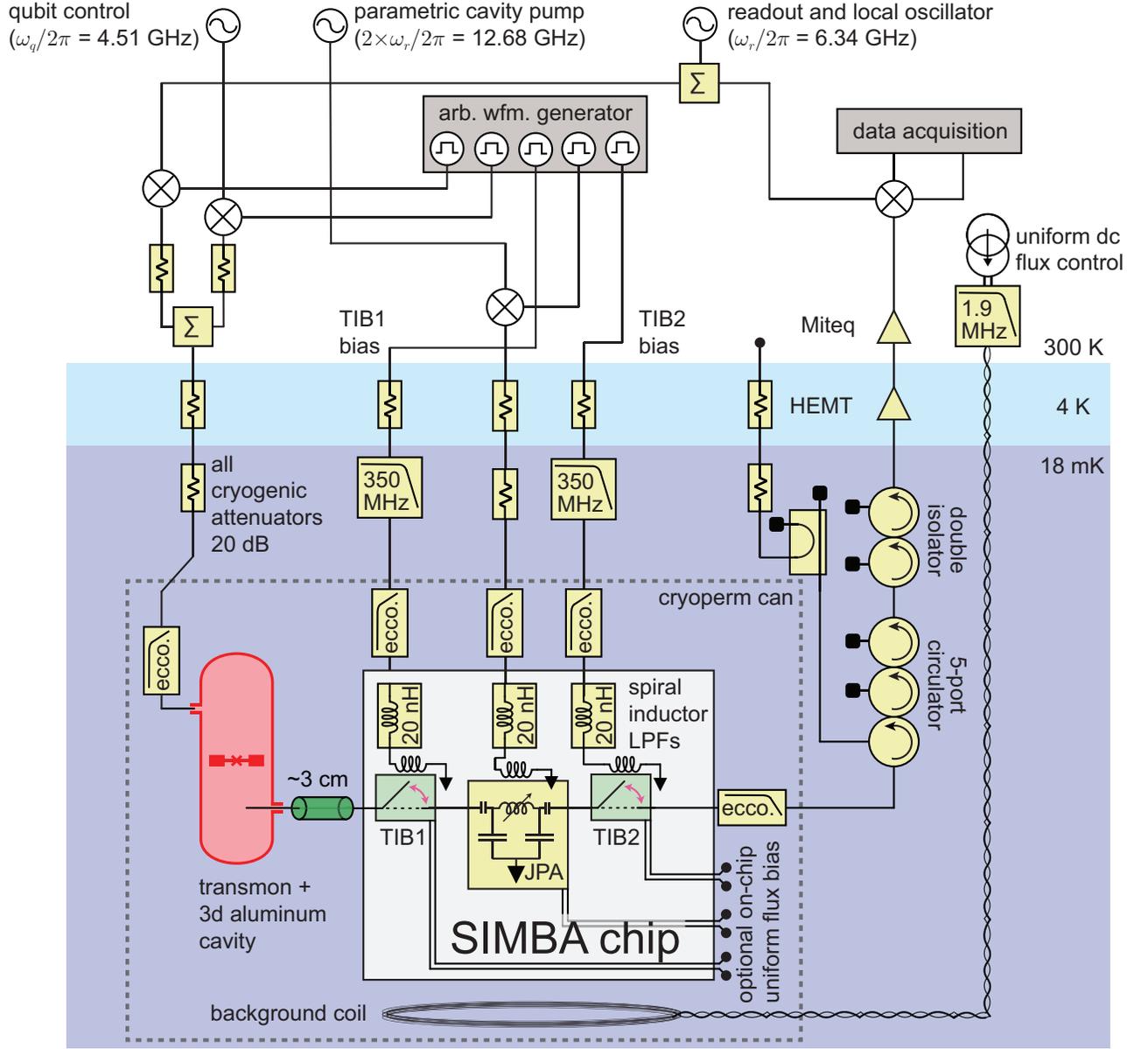

FIG. S18. **Experimental schematic.**

we begin with a model for the combined system of the readout and parametric cavity.

Consider the classical, time-dependent readout cavity and parametric cavity fields, $\mathbf{a}e^{i\omega_0 t}$ and $\mathbf{b}e^{i\omega_0 t}$, respectively, which rotate at bare cavity resonance frequency $\omega_0$ (assuming no detuning between the cavities). These cavities are coupled with a time-dependent coupling $g(t)$. The total loss rate in the parametric cavity is $\kappa_p$, and the total loss in the readout cavity is $\kappa_r$. Loss in the readout cavity is the sum of an internal loss rate $\kappa_r^{\text{int}}$ along with loss from a weakly coupled port $\kappa_r^w$ such that $\kappa_r = \kappa_r^{\text{int}} + \kappa_r^w$. Using input-output formalism [57], this port interacts with an input field $\mathbf{a_I}e^{i\omega_0 t}$ and output field $\mathbf{a_O}e^{i\omega_0 t}$. In a frame rotating at $\omega_0$, the Heisenberg-Langevin equations of motion for this system simplify to,

$$\sqrt{\kappa_r^w}\mathbf{a} = \mathbf{a_I} + \mathbf{a_O}, \tag{S42}$$

$$\dot{\mathbf{a}} = -\frac{\kappa_r}{2}\mathbf{a} + ig(t)\mathbf{b} + \sqrt{\kappa_r^w}\mathbf{a_I}, \tag{S43}$$

$$\dot{\mathbf{b}} = -\frac{\kappa_p}{2}\mathbf{b} + ig(t)\mathbf{a}. \tag{S44}$$

Eqs. S42-S44 are solved with the initial conditions $\mathbf{a} = \mathbf{b} = 0$ and $\dot{\mathbf{a}} = \dot{\mathbf{b}} = 0$ at time $t = 0$, and with the cavities initially decoupled, i.e. $g(t) = 0$. Beginning at $t = 0$, an input field with a Gaussian profile of standard deviation $s$ and amplitude maximum at time $\tau_1/2$ is incident on the readout cavity until time $\tau_1$. At $\tau_1$, the input field is turned off and a coupling $g(t) = g_0$ is turned on until $\tau_2$. These boundary conditions and tunable interactions are

described by the piecewise functions,

$$\mathbf{a_I}(t) = \begin{cases} e^{-(t-\tau_1/2)^2/2s^2}, & 0 < t < \tau_1 \\ 0, & \tau_1 < t \end{cases} \quad \text{(S45)}$$

$$g(t) = \begin{cases} 0, & 0 < t < \tau_1 \\ g_0, & \tau_1 < t < \tau_2. \end{cases} \quad \text{(S46)}$$

Time $\tau_2$ corresponds to the time at which the coupling $g(t)$ is turned off, and when phase-sensitive parametric gain is turned on in the parametric cavity. In this model, gain is assumed to instantaneously exceed loss.

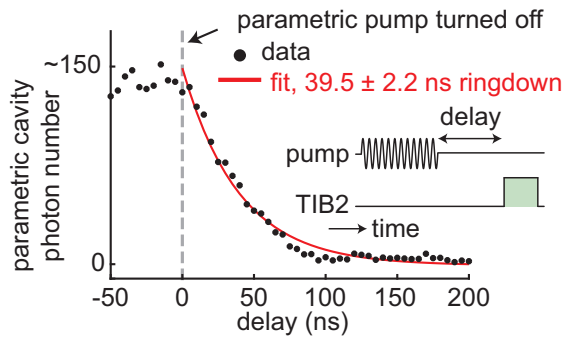

FIG. S19. **Parametric cavity loss measurement.** Measurement of the loss rate in the parametric cavity with TIB1 in reflect mode. The parametric cavity is pumped to bifurcation and at 'delay' = 0, the pump is turned off. After a variable delay, TIB2 is briefly connected to the measurement chain and the output signal is digitzed. Energy decays at a rate $\kappa_p/2\pi = 4.0 \pm 0.2$ MHz when both TIBs are set to reflect mode.

To estimate the fraction of readout power lost before $\tau_2$, we first numerically solve Eqs. S42-S44 using the definitions in Eq. S45 and Eq. S46, while setting the cavity internal loss rates to zero ($\kappa_r - \kappa_r^w = 0$ and $\kappa_p = 0$). Doing so, we compute the energy in the parametric cavity immediately before amplification to be $E_0 = \hbar\omega_0 |\mathbf{b}(\tau_2)|^2$. Next, we choose $\kappa_r \neq 0$ and $\kappa_p \neq 0$, and re-simulate Eqs. S42-S44 to compute the energy in the parametric cavity before amplification *including* this loss, $E'$. The estimated measurement efficiency is the ratio of these two numbers, $\eta_{model} = E'/E_0$.

The simulated efficiency depends on the parameters $\kappa_p$, $\kappa_r$ and $g_0$. We experimentally measure the parametric cavity loss rate to be $\kappa_p/2\pi = 4.0 \pm 0.2$ MHz with both TIBs in reflect mode, Fig. S19. We also measure $g_0 = 12.5$ MHz from the data shown in Fig. 2a, and $\eta = 0.70 \pm 0.01$ from the measurement in Fig. 3. We compute $\eta_{model} = 0.70$ by using these parameters and estimating the loss rate of the readout cavity to be $\kappa_r/2\pi = 440$ kHz. Note that this efficiency is approximately equal to $g_0/(g_0 + \kappa_r + \kappa_p) = 0.74$.

2. *Sources of loss*

Estimates of the loss sources in both the readout and parametric cavities are given in Table S3. Loss in the parametric cavity is a combination of the imperfect isolation provided by TIB2 (coupling the parametric cavity to the 50 Ohm measurement chain, even when TIB2 is in reflect mode), on-chip dissipation, and coupling to spurious modes in the $\sim 3$ cm connection between the readout and parametric cavities. These effects can all be mitigated in future designs. In particular, any coupling to cable modes can be greatly reduced by shortening the connection between the readout and parametric cavities to less than 1 cm. An effective way to achieve this would be to construct a chip-scale qubit + readout cavity + SIMBA system using through-silicon-via technology [63], or more simply, using a 2-dimensional transmon + readout resonator wire bonded to a SIMBA chip.

Estimates for measurement efficiency using several different loss models are shown in Fig. S20c. Model (1), blue bar, represents the measurement efficiency computed in Figs S20a,b. The other models are the simulated efficiency with lower amounts of loss. Making all improvements suggested in Table S3 predicts $\eta_{model} > 0.9$ (e.g. model (4) in Fig. S20). Loss may also be reduced by simply increasing the coupling rate between the readout and parametric cavities for a faster swap time (simulations assuming a 10 ns swap time is shown by the red bars in Fig. S20c), and also by using a faster readout pulse. In-principle, a measurement efficiency of $\eta \geq 0.99$ can be achieved by making all of these changes and using lower loss Josephson junctions within the SIMBA.

### B. Excess backaction limitations

Excess backaction, indicative of the finite isolation provided by TIB1, is limited to $n_b = 0.63 \pm 0.01$ effective photons for the data shown in Fig. 3 of the main text. In this experiment, turning down the parametric pump amplitude could reduce excess backaction to as low as $n_b \sim 0.1$ effective photons. However, doing so reduced the maximum readout fidelity to $F_0 \sim 80\%$, now limited by failure of the parametric cavity to latch due to low pump amplitude.

We determine TIB1 transmission to be $-26$ dB when set to reflect mode from comparison of the readout cavity to parametric cavity swap time with TIB1 in transmit mode (Fig. 2b) and reflect mode (Fig. S14), Table S2. This is worse than the 50 dB on/off ratio of the TIB measured in isolation, Fig. S7. The discrepancy between these numbers may result from an alternative on-chip transmission path within the SIMBA, which can in-principle be engineered-away in future designs. It could also result from too low a power handling of the TIB compared to the amplified state in the parametric cavity; this can be improved by increasing the number of SQUIDs per array in the TIBs, and the TIB junction

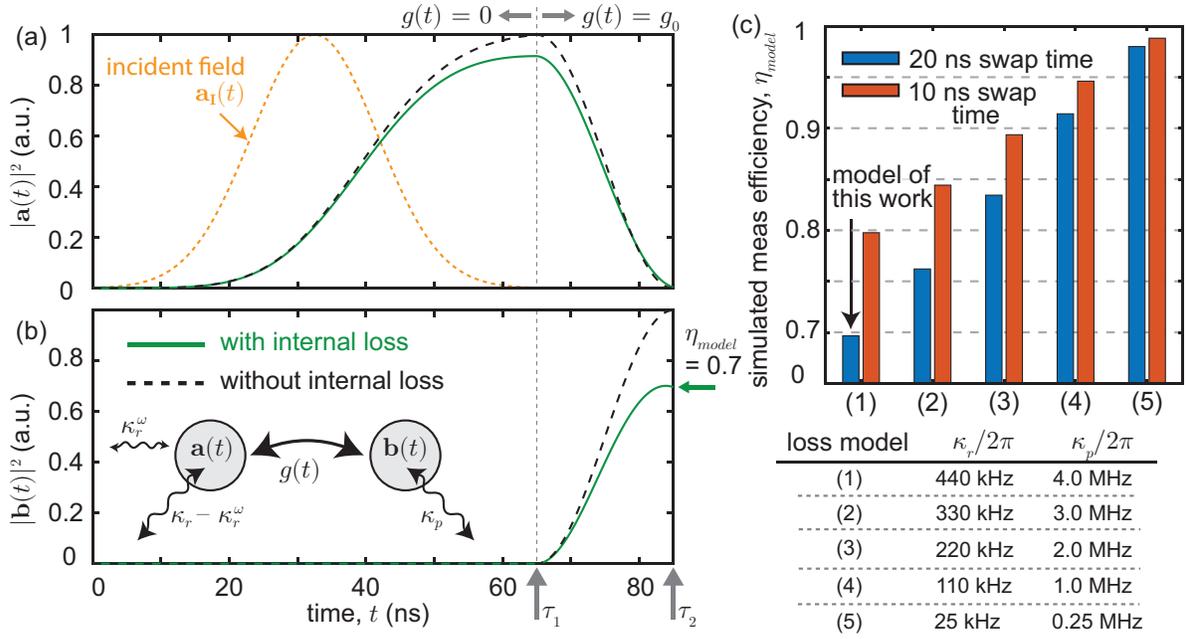

FIG. S20. **Measurement efficiency model.** (a,b) Model of energy vs. time in the readout cavity (a), and parametric cavity (b), which is proportional to the square of their field amplitudes $|\mathbf{a}(t)|^2$ and $|\mathbf{b}(t)|^2$, respectively. This simulation uses Eqs. S42-S44 with the parameters $\kappa_r^w/2\pi \times 8\,\mathrm{kHz}$, $\kappa_r/2\pi \times 440\,\mathrm{kHz}$, $\kappa_p/2\pi = 4.0\,\mathrm{MHz}$, $s = 13\,\mathrm{ns}$, $\tau_1 = 65\,\mathrm{ns}$, $\tau_2 = \tau_1 + 20\,\mathrm{ns}$ and $g_0/2\pi = 12.5\,\mathrm{MHz}$, chosen to emulate experiment. The solid green lines are the results of a simulation which includes the internal loss in both cavities, while the dashed black lines are the results of a simulation with the internal loss turned off, $\kappa_r - \kappa_r^w = \kappa_p = 0$. Simulated measurement efficiency $\eta_{model}$ is taken to be the ratio of energy remaining at time $\tau_2$ with/without internal loss included in the model. The time $\tau_2$ corresponds to when the parametric pump is turned on in experiment. (c) Simulated measurement efficiency for five different models of loss in the readout and parametric cavities. The blue bar in model (1) is the simulation of the current experiment (subplots a,b). Measurement efficiency is expected to improve to above $\eta \geq 0.9$ after making the improvements suggested in Table S3 (model (4)), and can be improved to $\eta \geq 0.99$ if the internal loss rate of the parametric cavity can be substantially reduced (model (5)).

critical currents (see Section IV.A).

In principle, excess backaction could also result from the $\sim 12.68\,\mathrm{GHz}$ pump signal coupling into higher order modes of the readout cavity. To the extent that the qubit is still dispersively coupled to these modes, such coupling would dephase the qubit. We note that the readout cavity used in this work has a fundamental mode at 6.34 GHz, the next lowest frequency mode at 10.44 GHz, and remaining modes above 14 GHz. We therefore expect the vast majority of any JPA pump signal incident on the readout cavity to be reflected by the cavity. In general when using a SIMBA, however, care must be taken to prevent higher order readout cavity modes from falling near the JPA pump frequency — or at least, to carefully isolate the pump signal from these modes.

Rather than speculate further on the minimum excess backaction which can be achieved using a SIMBA, we briefly offer a proof-of-principle demonstration of minimal excess backaction $n_b \ll 1$, Fig. S21, using a different SIMBA than discussed elsewhere in this work. This existence-proof uses an earlier version of the SIMBA constructed from a one-port JPA connected to two TIBs via a microwave T-junction. The JPA and TIBs are on separate chips, connected together on the same printed circuit board. The measurement efficiency and readout fidelity of qubit readout demonstrated in this setup compare unfavorably to the readout demonstrated in the main text. Additionally, the calibration of this device was complicated both by the significant presence of trapped flux vorticies near the Josephson parametric amplifier, and by the microwave T-junction between the TIBs and JPA. However, the excess backaction of qubit readout was significantly lower than reported in the main text, with $n_b = 0.018 \pm 0.002$ effective photons of excess backaction compared to $n_r^{\mathrm{proj}} \sim 4$ effective photons in a projective readout pulse. We therefore conclude that it is indeed possible to reach a limit of negligible excess backaction $n_b \ll 1$ using a SIMBA.

### C. Readout fidelity limitations

We believe readout infidelity to result from a combination of state preparation errors along with a probability of latching error in the parametric cavity. These error mechanisms are summarized in Table S4.

TABLE S3. **Sources of loss**. Contributions to loss are determined by either measurement (M), simulation (S), referencing a similar system in the literature (L), or inferring a value (I) to be similar to related measurements, or so that total loss adds up to a measured value.

| Parametric cavity parameter | Estimate | Method | After improvement |
| --- | --- | --- | --- |
| Transmission rate through TIB2 when in reflect mode | 2.5 MHz | I[a] | < 200 kHz[b] |
| On-chip dissipation rate | 1 MHz | L[c] | < 1 MHz[d] |
| Coupling to cable modes | 400 kHz | I[e] | negligible[f] |
| Transmission rate out of the JPA pump bias line | 80 kHz | S[g] | < 1 kHz[h] |
| Transmission rate out of a TIB gradiometric bias lines | < 1 kHz | S[i] | |
| Total loss rate, $\kappa_p/2\pi$ | 4.0 MHz | M[j] | |

| Readout cavity parameter | Estimate | Method | After improvement |
| --- | --- | --- | --- |
| Coupling to cable modes | 400 kHz | I[k] | negligible |
| Internal dissipation rate: | 30 kHz | L[l] | < 1 kHz[m] |
| Weak port coupling rate, $\kappa_r^w/2\pi$ | 8 kHz | M | |
| Transmission rate out of TIB1 gradiometric bias line | < 1 kHz | S | |
| Total loss rate, $\kappa_r/2\pi$ | 440 kHz | M/S[n] | |

| Connector dissipative loss | Estimate | Method | After improvement |
| --- | --- | --- | --- |
| Single pass absorption in 3 cm of cryogenic Cu waveguide | 20 kHz | L[o] | negligible[p] |

[a] Making the assumption that transmission through TIB2 when in reflect mode is similar to that measured through TIB1 when in reflect mode, Fig. S14.
[b] We expect that improving the TIB1 reflect mode from $-26$ dB (as demonstrated by TIB1 within this SIMBA) to $-50$ dB (as demonstrated in a single TIB measured in isolation) should equivalently improve the reflect mode of TIB2.
[c] Estimated to be similar to the dissipation measured in Ref. [58].
[d] We believe the greatest source of on-chip dissipation within this SIMBA to be the $SiO_2$ dielectric used within the SQUID arrays [34]. This dielectric has a loss tangent of $2.8 \times 10^{-3}$ [59]. Without changing the junction fabrication process, dissipation may be reduced by lowering the participation ratio of the SQUID array within the resonator (i.e. decreasing $l_{array}/l_{res}$). Doing so will reduce the effective nonlinearity $\zeta$ of the parametric cavity, and also decrease the frequency range over which it can be tuned. Note that the dissipation rate within the Si substrate is expected to be only several kHz [60].
[e] In general, this coupling depends on the phase of the signal reflected from the TIB, e.g. whether it reflects like an open, short, or something in-between (as is generally the case, according to simulations). Moreover, this reflection also depends on the value of the uniform bias flux seen by the TIBs. We estimate cable loss to be approximately the same as that of the readout cavity.
[f] Note that a 1 cm length of coaxial transmission line has a half-wavelength resonance at 10 GHz. Reducing the connection between the readout and parametric cavities to below this length is desirable to ensure there are no cable modes near the readout/parametric cavity frequency.
[g] This coupling is simulated using a finite-element model of the parametric cavity, with both of its microwave ports connected to 50 Ohms, both of its dc bias ports open as in this experiment, and its pump port connected to 50 Ohms through its on-chip coupling capacitor.
[h] Improved high-pass filtering on this bias line can significantly reduce this coupling.
[i] Loss is simulated using a finite-element model of the parametric cavity connected to TIBs. The simulated loss rate is 0.9 kHz with both TIBs simulated in transmit mode, and even smaller when the TIBs are simulated in reflect mode.
[j] Obtained from the measurement in Fig. S19.
[k] We model coupling to all (lossy, detuned) cable modes as simply an effective loss rate. Here we infer this loss to account for the remaining fraction of readout cavity loss not accounted for by other known sources of dissipation.
[l] Assumed to be approximately the same as in the similar cavities used in Ref. [50].
[m] By using a high-Q cylindrical cavity, for instance as done in Ref. [61].
[n] Readout cavity loss is determined so that given the measurement of $\kappa_p$ obtained in Fig. S19, the simulated measurement efficiency in Fig. S20a,b is $\eta_{model} = 0.70$, therefore matching its measured value in Fig. 3c.
[o] Based on the measurement of -56 dB single pass absorption in Ref. [62], and taking and $g_0 = 12.5$ MHz.
[p] This resistive loss can be significantly decreased by engineering an all superconducting connection.

In the main text we report a maximum readout fidelity of $F_0 = 95.5 \pm 0.3\%$. The best readout fidelity we measured during the same cooldown is $F_0 = 96.5 \pm 0.2\%$ (measured by fitting Eq. 4 to a measurement of readout fidelity $F_r$ vs. readout amplitude $\epsilon$ as in Fig. 3). We attribute the 1% difference between these numbers to a parametric cavity latching failure probability, which can be eliminated by increasing pump amplitude or more precisely calibrating the time at which the parametric pump is turned on.

Other sources of infidelity include decay before/during state preparation and measurement, and a residual ex-

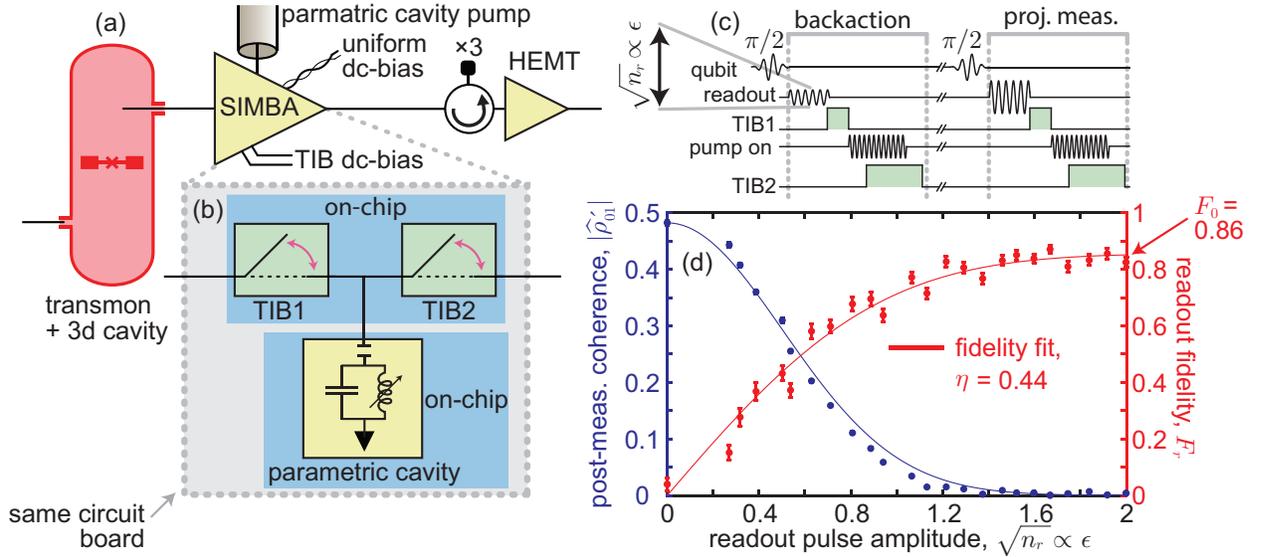

FIG. S21. **Negligible backaction proof-of-principle demonstration.** (a) Alternate SIMBA experiment. Rather than a 2-port JPA with TIBs on each port as discussed elsewhere in this work, a SIMBA is constructed from a one port JPA connected by a T-junction to two TIBs. (b) The TIBs are on a separate chip from the parametric cavity, but share the same circuit board. (c) Excess backaction, maximum fidelity, and measurement efficiency are characterized by inserting a variable measurement into a Ramsey sequence, as in Fig. 3. (d) Results of this procedure. Note that in the data shown, the JPA pump remains on during the variable measurement. Fitting data to the models in Eq. 3 and Eq. 4 yield measurement efficiency $\eta = 0.44 \pm 0.03$ and maximum readout fidelity $F_0 = 0.86 \pm 0.01$. The post-backaction coherence at zero readout amplitude is $\rho_b = 0.482 \pm 0.004$ corresponding to $n_b = 0.018 \pm 0.002$ effective photons of backaction.

TABLE S4. **Sources of readout infidelity.**

| Error mechanism | Estimated probability |
| --- | --- |
| State preparation | 2.3% |
| Qubit decay before measurement | 1.3% |
| Parametric cavity latching failure | 1.1% |

cited state population which causes preparation infidelity. Note that for a projective measurement (e.g. the last red data point in Fig. 3), readout error probabilities are $P(g|\pi) = 2.9\%$ and $P(e|0) = 1.6\%$. The difference between these numbers $P(g|\pi) - P(e|0) = 1.3\%$ estimates the likelihood of the qubit to have decayed to the ground state before/during measurement. This is consistent with $T_1 = 9\,\mu$s, the 65 ns readout pulse duration, and the 100 ns duration $\pi$-pulse used to prepare the qubit in the excited state.

After accounting for latching failure and qubit decay, the remaining 2.3% infidelity is attributed to a residual thermal excited state population. This can also be estimated by the measured difference between $T_2^*$ and twice $T_1$. The cavity average thermal photon number $\bar{n}_{\rm th}$ yields a dephasing rate of [64],

$$\frac{1}{T_\varphi^{\rm th}} = \frac{4\kappa\chi^2}{\kappa^2 + 4\chi^2}\bar{n}_{\rm th}, \tag{S47}$$

where $1/T_\varphi^{\rm th} = 1/T_2^* - 1/2T_1$, and taking the limit $\bar{n}_{\rm th} \ll 1$. Plugging in the experimental values given in Table. S2 and solving for the readout cavity thermal occupancy gives $\bar{n}_{\rm th} = 0.0042$ corresponding to a cavity temperature of $T = 56$ mK given that $\bar{n}_{\rm th} = 1/\left(e^{\hbar\omega_r/k_B T} - 1\right)$. Assuming the qubit to be at the same temperature as the readout cavity, the Maxwell-Boltzmann distribution for a two-level system gives the qubit residual excited state population to be $1/\left(1 + e^{\hbar\omega_q/k_B T}\right) = 2\%$. In general the qubit can be at a different temperature than the cavity and also may not be in a thermal distribution [65], effects which will modify this estimate.

In conclusion, there are no indications of inherent limits to readout fidelity when using a SIMBA. State preparation errors can be improved by better thermalization at base temperature and optimized $\pi$-pulses. Fidelity should also improve using a qubit with a longer $T_1$ time and a faster readout sequence. Finally, any latching failure can be eliminated by increasing the parametric cavity pump amplitude, although potentially at the cost of additional excess backaction.

### D. Measurement time limitations

Readout using a SIMBA can be improved to be significantly faster than the 265 ns measurement time reported in this work without detriment to the readout performance. Dispersive readout using a SIMBA is different

from standard dispersive readout schemes because the external coupling rate is now tunable. Advantageously, the readout cavity external coupling can be made large during the measurement allowing for a fast readout, but is otherwise tuned close to zero so that the qubit $T_1$ time is not limited, obviating the need for a Purcell filter [66]. For optimal readout using a SIMBA, it is desirable to minimize loss in the readout cavity such that $2\chi \gg \kappa_r$. Then, to turn on a large external coupling $g_0$ to the parametric cavity in order to quickly and efficiently swap the readout signal.

The 265 ns readout reported in the main text is divided into four steps (Fig. 1c): sending a readout pulse into the readout cavity (65 ns), swapping the pulse into the parametric cavity (20 ns), ringing up the parametric cavity (80 ns), and data acquisition of the pumped state of the parametric cavity (100 ns). There is significant room for improvement in the speed of all these steps:

1. *Readout pulse* (65 ns): because in this experiment the readout cavity external coupling rate is tunable, the rate at which the readout pulse acquires a qubit state-dependent phase shift is set *only* by $\chi$, rather than the ratio of $\chi$ to the external coupling rate (assuming this as the dominant source of loss) as in conventional dispersive readout [10, 30, 67]. As such, increasing $\chi$ will allow for a proportionally faster readout pulse. Additionally, we note that a 50 ns readout pulse has been shown to achieve readout fidelity of greater than 98% in an experiment using a ferrite circulator and Josephson parametric amplifier [17]. Finally, careful pulse shaping of the readout pulse [17, 68], not yet done in this work, has been shown to increase readout fidelity for a given pulse duration.

2. *Swap time* (20 ns): the 20 ns swap speed is controlled by the coupling rates of readout and parametric cavities, Table. S2. These coupling rates may be increased in order to decrease the swap time. In general, measurement efficiency increases as the swap time decreases, but, excess backaction also increases.

3. *Digitization time* (100 ns): The digitization time is chosen in order to be able to clearly distinguish two well separated bifurcated states (the histograms in Fig. 1c). In this experiment, effectively no readout infidelity is introduced by the inability to distinguish these states, and seen by the separation of the histograms in Fig. 1c. The digitization time required to do this depends on the amplitude of these states which depends on the JPA pump amplitude, pump detuning and resonator nonlinearity, Eq. S39. In particular, decreasing the resonator nonlinearity can lead to a much higher bifurcated state amplitude and thus a much shorter digitization time. Increasing $\kappa_p^s$, the maximum coupling rate between the JPA and measurement chain, can also reduce the digitization time.

In conclusion, all steps of the SIMBA readout pulse sequence may be sped up. Complicated trade-offs arise between the length of different readout steps and the various metrics for readout performance. With further optimization, we anticipate a total measurement time of approximately 100 ns is achievable without significant detriment to the performance reported in this work.

### E. Ease of calibration

Finally, ease of calibration is an important factor to consider with regards to scalability. Calibration of qubit readout when using a SIMBA is not particularly more complicated than use of other parametric amplifiers, however it can still be improved.

In particular, in this experiment the parametric amplifier performance (for example, the region where it bifurcates as a function of pump frequency and amplitude), is somewhat dependent on the uniform bias flux seen by both TIBs. This is because the SQUID arrays within the TIBs participate to some extent in the parametric amplifier mode. In general this is undesirable; it requires extra care when choosing the uniform magnetic flux applied to the device. This participation can be easily lowered by reducing the coupling between the JPA and TIBs, but at the cost of a slower swap time. More difficult but perhaps more useful, the superconducting switch element can be redesigned to participate minimally in the parametric cavity resonance. Or, the parametric cavity parameters (such as maximum resonance frequency and nonlinearity), can be changed so that it bifurcates over a wider region in parameter space.

[1] A. A. Clerk, M. H. Devoret, S. M. Girvin, Florian Marquardt, and R. J. Schoelkopf, "Introduction to quantum noise, measurement, and amplification," Reviews of Modern Physics **82**, 1155 (2010).

[2] C. C. Bultink, B. Tarasinski, N. Haandbæk, S. Poletto, N. Haider, D. J. Michalak, A. Bruno, and L. DiCarlo, "General method for extracting the quantum efficiency of dispersive qubit readout in circuit qed," Applied Physics Letters **112**, 092601 (2018).

[3] A. Eddins, S. Schreppler, D. M. Toyli, L. S. Martin, S. Hacohen-Gourgy, L. C. G. Govia, H. Ribeiro, A. A. Clerk, and I. Siddiqi, "Stroboscopic qubit measurement with squeezed illumination," Phys. Rev. Lett. **120**, 040505 (2018).

[4] A. Eddins, J. M. Kreikebaum, D. M. Toyli, E. M. Levenson-Falk, A. Dove, W. P. Livingston, B. A. Levitan, L. C. G. Govia, A. A. Clerk, and I. Siddiqi, "High-efficiency measurement of an artificial atom embedded in

[5] S. Touzard, A. Kou, N. E. Frattini, V. V. Sivak, S. Puri, A. Grimm, L. Frunzio, S. Shankar, and M. H. Devoret, "Gated conditional displacement readout of superconducting qubits," Phys. Rev. Lett. **122**, 080502 (2019).

[6] F. Lecocq, L. Ranzani, G. A. Peterson, K. Cicak, X. Y. Jin, R. W. Simmonds, J. D. Teufel, and J. Aumentado, "Efficient qubit measurement with a nonreciprocal microwave amplifier," Phys. Rev. Lett. **126**, 020502 (2021).

[7] A. A. Clerk, S. M. Girvin, and A. D. Stone, "Quantum-limited measurement and information in mesoscopic detectors," Phys. Rev. B **67**, 165324 (2003).

[8] Rui Han, Gerd Leuchs, and Markus Grassl, "Residual and destroyed accessible information after measurements," Phys. Rev. Lett. **120**, 160501 (2018).

[9] Waltraut Wustmann and Vitaly Shumeiko, "Parametric resonance in tunable superconducting cavities," Phys. Rev. B **87**, 184501 (2013).

[10] Alexandre Blais, Ren-Shou Huang, Andreas Wallraff, S. M. Girvin, and R. J. Schoelkopf, "Cavity quantum electrodynamics for superconducting electrical circuits: An architecture for quantum computation," Physical Review A **69**, 062320 (2004).

[11] Johannes Heinsoo, Christian Kraglund Andersen, Ants Remm, Sebastian Krinner, Theodore Walter, Yves Salathé, Simone Gasparinetti, Jean-Claude Besse, Anton Potočnik, Andreas Wallraff, and Christopher Eichler, "Rapid high-fidelity multiplexed readout of superconducting qubits," Phys. Rev. Applied **10**, 034040 (2018).

[12] Christian Kraglund Andersen, Ants Remm, Stefania Lazar, Sebastian Krinner, Johannes Heinsoo, Jean-Claude Besse, Mihai Gabureac, Andreas Wallraff, and Christopher Eichler, "Entanglement stabilization using ancilla-based parity detection and real-time feedback in superconducting circuits," npj Quantum Information **5**, 69 (2019).

[13] Christian Kraglund Andersen, Ants Remm, Stefania Lazar, Sebastian Krinner, Nathan Lacroix, Graham J. Norris, Mihai Gabureac, Christopher Eichler, and Andreas Wallraff, "Repeated quantum error detection in a surface code," Nature Physics **16**, 875–880 (2020).

[14] T. Peronnin, D. Marković, Q. Ficheux, and B. Huard, "Sequential dispersive measurement of a superconducting qubit," Phys. Rev. Lett. **124**, 180502 (2020).

[15] Baleegh Abdo, Oblesh Jinka, Nicholas T. Bronn, Salvatore Olivadese, and Markus Brink, "On-chip single-pump interferometric josephson isolator for quantum measurements," arXiv preprint arXiv:2006.01918 (2020).

[16] B. Abdo, N. T. Bronn, O. Jinka, S. Olivadese, A. D. Córcoles, V. P. Adiga, M. Brink, R. E. Lake, X. Wu, D. P. Pappas, and J. M. Chow, "Active protection of a superconducting qubit with an interferometric josephson isolator," Nature communications **10**, 3154 (2019).

[17] T. Walter, P. Kurpiers, S. Gasparinetti, P. Magnard, A. Potočnik, Y. Salathé, M. Pechal, M. Mondal, M. Oppliger, C. Eichler, and A. Wallraff, "Rapid high-fidelity single-shot dispersive readout of superconducting qubits," Phys. Rev. Applied **7**, 054020 (2017).

[18] C. Macklin, K. O'Brien, D. Hover, M. E. Schwartz, V. Bolkhovsky, X. Zhang, W. D. Oliver, and I. Siddiqi, "A near–quantum-limited Josephson traveling-wave parametric amplifier," Science **350**, 307–310 (2015).

[19] M. Hatridge, S. Shankar, M. Mirrahimi, F. Schackert, K. Geerlings, T. Brecht, K. M. Sliwa, B. Abdo, L. Frunzio, S. M. Girvin, R. J. Schoelkopf, and M. H. Devoret, "Quantum back-action of an individual variable-strength measurement," Science **339**, 178—181 (2013).

[20] Another important readout metric is how quantum-nondemolition (QND) a measurement is. Dispersive readout of superconducting qubits is understood to be highly QND [10]. The hardware requirements for any qubit readout scheme (size/footprint of the detector, number of bias controls, etc) are of practical importance, also.

[21] T. Thorbeck, S. Zhu, E. Leonard Jr., R. Barends, J. Kelly, John M. Martinis, and R. McDermott, "Reverse isolation and backaction of the slug microwave amplifier," Physical Review Applied **8**, 054007 (2017).

[22] The measurement efficiency is not characterized in Ref. [21]. Also, note that Ref. [16] demonstrates qubit readout using the combination of both a superconducting isolator and a ferrite circulator before a parametric amplifier.

[23] F. Lecocq, L. Ranzani, G. A. Peterson, K. Cicak, A. Metelmann, S. Kotler, R. W. Simmonds, J. D. Teufel, and J. Aumentado, "Microwave measurement beyond the quantum limit with a nonreciprocal amplifier," Phys. Rev. Applied **13**, 044005 (2020).

[24] K. Kraus, "General state changes in quantum theory," Ann. Phys. **64**, 311–335 (1971).

[25] The partial trace $\text{Tr}_2(\cdot)$ is defined as the unique linear operator satisfying $\text{Tr}_2(X \otimes Y) = X\text{Tr}(Y)$ for any two operators $X$ and $Y$. When applied to a bipartite state, the partial trace $\text{Tr}_2$ yields the marginal of the first constituent system.

[26] Rodney Loudon, *The Quantum Theory of Light*, 3rd ed. (Oxford Science Publications, 2000).

[27] Christopher Gerry and Peter Knight, *Introductory Quantum Optics* (Cambridge University Press, 2004).

[28] Note that measurement efficiency has also been referred to as 'quantum efficiency' or simply 'efficiency' in the superconducting qubit literature [1–5, 11].

[29] Note that when instead using a phase-preserving amplifier with no additional added noise, $\eta_{\text{amp}} = 1/2$ such that $\eta = \frac{1}{2}\eta_{\text{loss}}$ [69, 70].

[30] Alexandre Blais, Arne L. Grimsmo, and Andreas Wallraff, "Circuit quantum electrodynamics," arXiv:2005.12667 (2020).

[31] Note that Ref. [2] derives an expression which is written identically to Eq. S12, but whose derivation contains two differences from ours. First, Ref. [2] defines a measurement efficiency of one to correspond to an ideal phase-*preseving* amplifier, such that $0 \leq \eta \leq 1/2$ accordingly to our definition. Second, SNR is defined in Ref. [2] to equal $(d_1 + d_0)/w$ (taking $w_1 = w_0 = w$, referring to our Fig. S1), which differs from our definition in Eq. S9. Our definition of SNR is chosen to keep with the convention in Refs. [5, 30], and references therein. In practice, such conflicting definitions can be a source of confusion, and it is important to make sure that the definitions for $\eta$ and SNR are consistent with the formula being used.

[32] Jay Gambetta, W. A. Braff, A. Wallraff, S. M. Girvin, and R. J. Schoelkopf, "Protocols for optimal readout of qubits using a continuous quantum nondemolition measurement," Phys. Rev. A **76**, 012325 (2007).

[33] C.W. Helstrom, *Quantum detection and estimation theory* (Academic Press: New York, 1976).

[34] J. A. B. Mates, G. C. Hilton, K. D. Irwin, L. R. Vale, and K. W. Lehnert, "Demonstration of a multiplexer of

dissipationless superconducting quantum interference devices," Applied Physics Letters **92**, 023514 (2008).
[35] Benjamin J. Chapman, Bradley A. Moores, Eric I. Rosenthal, Joseph Kerckhoff, and K. W. Lehnert, "General purpose multiplexing device for cryogenic microwave systems," Applied Physics Letters **108**, 222602 (2016).
[36] Benjamin J. Chapman, Eric I. Rosenthal, Joseph Kerckhoff, Leila R. Vale, Gene C. Hilton, and K. W. Lehnert, "Single-sideband modulator for frequency domain multiplexing of superconducting qubit readout," Applied Physics Letters **110**, 162601 (2017).
[37] Benjamin J. Chapman, Eric I. Rosenthal, Joseph Kerckhoff, Bradley A. Moores, Leila R. Vale, J. A. B. Mates, Gene C. Hilton, Kevin Lalumière, Alexandre Blais, and K. W. Lehnert, "Widely tunable on-chip microwave circulator for superconducting quantum circuits," Phys. Rev. X **7**, 041043 (2017).
[38] M. Malnou, D. A. Palken, Leila R. Vale, Gene C. Hilton, and K. W. Lehnert, "Optimal operation of a josephson parametric amplifier for vacuum squeezing," Phys. Rev. Applied **9**, 044023 (2018).
[39] M. Malnou, D. A. Palken, B. M. Brubaker, Leila R. Vale, Gene C. Hilton, and K. W. Lehnert, "Squeezed vacuum used to accelerate the search for a weak classical signal," Phys. Rev. X **9**, 021023 (2019).
[40] F. Lecocq, L. Ranzani, G. A. Peterson, K. Cicak, R. W. Simmonds, J. D. Teufel, and J. Aumentado, "Nonreciprocal microwave signal processing with a field-programmable Josephson amplifier," Phys. Rev. Applied **7**, 024028 (2017).
[41] A Marchand balun [71] consists of a half-wave resonator connected to the unbalanced port, coupled to two quarter-wave resonators which are each coupled to one node of the balanced port. The balun bandwidth is controlled by the characteristic impedance of these resonators compared to the port impedance (typically 50 Ohms).
[42] M. D. Reed, L. DiCarlo, B. R. Johnson, L. Sun, D. I. Schuster, L. Frunzio, and R. J. Schoelkopf, "High-fidelity readout in circuit quantum electrodynamics using the jaynes-cummings nonlinearity," Phys. Rev. Lett. **105**, 173601 (2010).
[43] A second type of bifurcation, distinguished by two stable states of different amplitude, is not discussed in this work. This bifurcation occurs in a Duffing oscillator pumped near resonance [9], and has also been studied in Josephson junction based circuits [55, 72, 73]. Such an oscillator is an example of four-wave mixing, as two pump photons, near the oscillator resonance frequency, are converted to one signal and one idler photon, both also near the oscillator resonance frequency. The amplifier used in this work, however, is an example of three wave mixing since one pump photon is converted to one signal and one idler photon, each near half the pump frequency.
[44] Note that the topology of the circuit in Fig. S10 is different than in Fig. S6d. We choose to model the parametric cavity using the circuit in Fig. S10 because its equations of motion are simpler to write down, but capture the same essential physics.
[45] Theodore Van Duzer and Charles William Turner, *Principles of superconductive devices and circuits*, 2nd ed. (Prentice Hall, 1981).
[46] This definition is chosen so that in the quantum regime, these fields become operators which have the commutation relation $[\hat{a}, \hat{a}^*] = 1$.
[47] P. Krantz, A. Bengtsson, M. Simoen, S. Gustavsson, V. Shumeiko, W. D. Oliver, C. M. Wilson, P. Delsing, and B. Bylander, "Single-shot read-out of a superconducting qubit using a josephson parametric oscillator," Nature communications **7**, 11417 (2016).
[48] Waltraut Wustmann and Vitaly Shumeiko, "Parametric effects in circuit quantum electrodynamics," Low Temperature Physics **45**, 848 (2019).
[49] P. Krantz, Y. Reshitnyk, W. Wustmann, J. Bylander, S. Gustavsson, W. D. Oliver, T. Duty, V. Shumeiko, and P. Delsing, "Investigation of nonlinear effects in josephson parametric oscillators used in circuit quantum electrodynamics," New J. Phys. **15**, 105002 (2013).
[50] Hanhee Paik, D. I. Schuster, Lev S. Bishop, G. Kirchmair, G. Catelani, A. P. Sears, B. R. Johnson, M. J. Reagor, L. Frunzio, L. I. Glazman, S. M. Girvin, M. H. Devoret, and R. J. Schoelkopf, "Observation of high coherence in josephson junction qubits measured in a three-dimensional circuit qed architecture," Phys. Rev. Lett. **107**, 240501 (2011).
[51] V. B. Braginsky and F. Ya. Khalili, "Quantum nondemolition measurements: the route from toys to tools," Rev. Mod. Phys. **68**, 1–11 (1996).
[52] Austin G. Fowler, Matteo Mariantoni, John M. Martinis, and Andrew N. Cleland, "Surface codes: Towards practical large-scale quantum computation," Phys. Rev. A **86**, 032324 (2012).
[53] D. H. Slichter, R. Vijay, S. J. Weber, S. Boutin, M. Boissonneault, J. M. Gambetta, A. Blais, and I. Siddiqi, "Measurement-induced qubit state mixing in circuit qed from up-converted dephasing noise," Phys. Rev. Lett. **109**, 153601 (2012).
[54] Daniel Sank, Zijun Chen, Mostafa Khezri, J. Kelly, R. Barends, B. Campbell, Y. Chen, B. Chiaro, A. Dunsworth, A. Fowler, E. Jeffrey, E. Lucero, A. Megrant, J. Mutus, M. Neeley, C. Neill, P. J. J. O'Malley, C. Quintana, P. Roushan, A. Vainsencher, T. White, J. Wenner, Alexander N. Korotkov, and John M. Martinis, "Measurement-induced state transitions in a superconducting qubit: Beyond the rotating wave approximation," Phys. Rev. Lett. **117**, 190503 (2016).
[55] A. Lupaşcu, S. Saito, T. Picot, P. C. De Groot, C. J. P. M. Harmans, and J. E. Mooij, "Quantum nondemolition measurement of a superconducting two-level system," Nature Physics **3**, 119–123 (2007).
[56] Note that in this definition of QND fidelity, $F_{\mathrm{QND}} = 0.5$ when the results of the two subsequent measurements are uncorrelated (e.g. if $g_2 g_1$, $g_2 e_1$, $e_2 g_1$ and $e_2 e_1$ occur with equal probably of $1/4$).
[57] D. F. Walls and G. J. Milburn, *Quantum Optics* (Springer, Berlin, 1994).
[58] M. A. Castellanos-Beltran and K. W. Lehnert, "Widely tunable parametric amplifier based on a superconducting quantum interference device array resonator," Applied Physics Letters **91**, 083509 (2007).
[59] Aaron D. O'Connell, M. Ansmann, R. C. Bialczak, M. Hofheinz, N. Katz, Erik Lucero, C. McKenney, M. Neeley, H. Wang, E. M. Weig, A. N. Cleland, and J. M. Martinis, "Microwave dielectric loss at single photon energies and millikelvin temperatures," Applied Physics Letters **92**, 112903 (2008).
[60] G. Calusine, A. Melville, W. Woods, R. Das, C. Stull,

V. Bolkhovsky, D. Braje, D. Hover, D. K. Kim, X. Miloshi, D. Rosenberg, A. Sevi, J. L. Yoder, E. Dauler, and W. D. Oliver, "Analysis and mitigation of interface losses in trenched superconducting coplanar waveguide resonators," Applied Physics Letters **112**, 062601 (2018).

[61] Matthew Reagor, Hanhee Paik, Gianluigi Catelani, Luyan Sun, Christopher Axline, Eric Holland, Ioan M. Pop, Nicholas A. Masluk, Teresa Brecht, Luigi Frunzio, Michel H. Devoret, Leonid Glazman, and Robert J. Schoelkopf, "Reaching 10 ms single photon lifetimes for superconducting aluminum cavities," Applied Physics Letters **102**, 192604 (2013).

[62] Philipp Kurpiers, Theodore Walter, Paul Magnard, Yves Salathe, and Andreas Wallraff, "Characterizing the attenuation of coaxial and rectangular microwave-frequency waveguides at cryogenic temperatures," . EPJ Quantum Technology **4**, 8 (2017).

[63] D. Rosenberg, D. Kim, R. Das, D. Yost, S. Gustavsson, D. Hover, P. Krantz, A. Melville, L. Racz, G. O. Samach, S. J. Weber, F. Yan, J. L. Yoder, A. J. Kerman, and W. D. Oliver, "3d integrated superconducting qubits," npj Quantum Inf **3**, 42 (2017).

[64] Fei Yan, Simon Gustavsson, Archana Kamal, Jeffrey Birenbaum, Adam P. Sears, David Hover, Ted J. Gudmundsen, Danna Rosenberg, Gabriel Samach, S. Weber, Jonilyn L. Yoder, Terry P. Orlando, John Clarke, Andrew J. Kerman, and William D. Oliver, "The flux qubit revisited to enhance coherence and reproducibility," Nature communications **7**, 12964 (2016).

[65] Fei Yan, Dan Campbell, Philip Krantz, Morten Kjaergaard, David Kim, Jonilyn L. Yoder, David Hover, Adam Sears, Andrew J. Kerman, Terry P. Orlando, Simon Gustavsson, and William D. Oliver, "Distinguishing coherent and thermal photon noise in a circuit quantum electrodynamical system," Phys. Rev. Lett. **120**, 260504 (2018).

[66] M. D. Reed, B. R. Johnson, A. A. Houck, L. DiCarlo, Chow J. M., D. I. Schuster, L. Frunzio, and R. J. Schoelkopf, "Fast reset and suppressing spontaneous emission of a superconducting qubit," Applied Physics Letters **96**, 203110 (2010).

[67] Jay Gambetta, Alexandre Blais, M. Boissonneault, A. A. Houck, D. I. Schuster, and S. M. Girvin, "Quantum trajectory approach to circuit qed: Quantum jumps and the zeno effect," Phys. Rev. A **77**, 012112 (2008).

[68] Evan Jeffrey, Daniel Sank, J. Y. Mutus, T. C. White, J. Kelly, R. Barends, Y. Chen, Z. Chen, B. Chiaro, A. Dunsworth, A. Megrant, P. J. J. O'Malley, C. Neill, P. Roushan, A. Vainsencher, J. Wenner, A. N. Cleland, and John M. Martinis, "Fast accurate state measurement with superconducting qubits," Phys. Rev. Lett. **112**, 190504 (2014).

[69] Carlton M. Caves, "Quantum limits on noise in linear amplifiers," Phys. Rev. D **26**, 1817–1839 (1982).

[70] Alexander N. Korotkov, "Quantum bayesian approach to circuit qed measurement with moderate bandwidth," Phys. Rev. A **94**, 042326 (2016).

[71] Nathan Marchand, "Transmission-line conversion transformers," Electronics **17**, 142–145 (1944).

[72] I. Siddiqi, R. Vijay, F. Pierre, C. M. Wilson, M. Metcalfe, C. Rigetti, L. Frunzio, and M. H. Devoret, "Rf-driven josephson bifurcation amplifier for quantum measurement," Phys. Rev. Lett. **93**, 207002 (2004).

[73] V. E. Manucharyan, E. Boaknin, M. Metcalfe, R. Vijay, I. Siddiqi, and M. Devoret, "Microwave bifurcation of a Josephson junction: Embedding-circuit requirements," Phys. Rev. B **76**, 014524 (2007).






\end{document}